# EXPLORING COST-BENEFIT ANALYSIS OF RESEARCH, DEVELOPMENT AND INNOVATION INFRASTRUCTURES: AN EVALUATION FRAMEWORK

Massimo Florio[1], Stefano Forte[2], Chiara Pancotti[3], Emanuela Sirtori[3], Silvia Vignetti[3]

[1] Dipartimento di Economia, Management e Metodi Quantitativi,
Università di Milano , via Conservatorio 7, I-20122 Milano, Italy
[2] TIF Lab, Dipartimento di Fisica, Università di Milano and
INFN, Sezione di Milano, Via Celoria 16, I-20133 Milano, Italy
[3] CSIL, Centre for Industrial Studies,
Corso Monforte 15, I-20122 Milano, Italy

## Abstract

Governments, funding agencies and policy makers have high expectations on research, development and innovation (RDI) infrastructures in the context of science and innovation policies aimed at sustaining economic growth in the long term. The stakes associated with their selection and evaluation are therefore high.

Cost-benefit analysis of RDI infrastructures is a new field. The intangible nature of some benefits and the uncertainty associated to the achievement of research results have often discouraged the use of a proper CBA for RDI infrastructures. Recently, some attempts to develop a CBA theoretical framework for RDI infrastructures have been made in the context of the use of Structural Funds by the Czech government and JASPERS. Moreover, the new Guide for the CBA of investment projects in the context of Cohesion Policy, recently adopted by the European Commission (2014) provides guidelines to appraise RDI projects, but also admits that – due to lack of experience and best practices – further steps are needed to improve the evaluation framework.

This paper presents the results and the lessons learned on how to apply ex-ante CBA for major RDI infrastructures by a team of economists and scientists at the University of Milan and CSIL during a three-year research project supported by a EIBURS grant of the European Investment Bank Institute. Albeit the comprehensive conceptual framework presented in the paper builds on principles firmly rooted in CBA tradition, their application to the RDI sector is still in its infancy. So far, the model has been applied on two cases in physics involving particle accelerators (the Large Hadron Collider (LHC) at CERN and the National Centre for Oncological Treatment (CNAO) in Italy)).

In a nutshell, the model presented break down benefits into two broad classes: i) use benefits, held by different categories of infrastructure's users such as scientists, firms, students and general public visitors, and ii) non-use benefits, denoting the social value for the discovery potential of the RDI infrastructure regardless of its actual or future use. We argue that the social value of discovery can be estimated with contingent valuation techniques. Another significant feature of our approach is the stochastic nature of the CBA model, intended to deal with the uncertainty and risk of optimism bias in the estimates.

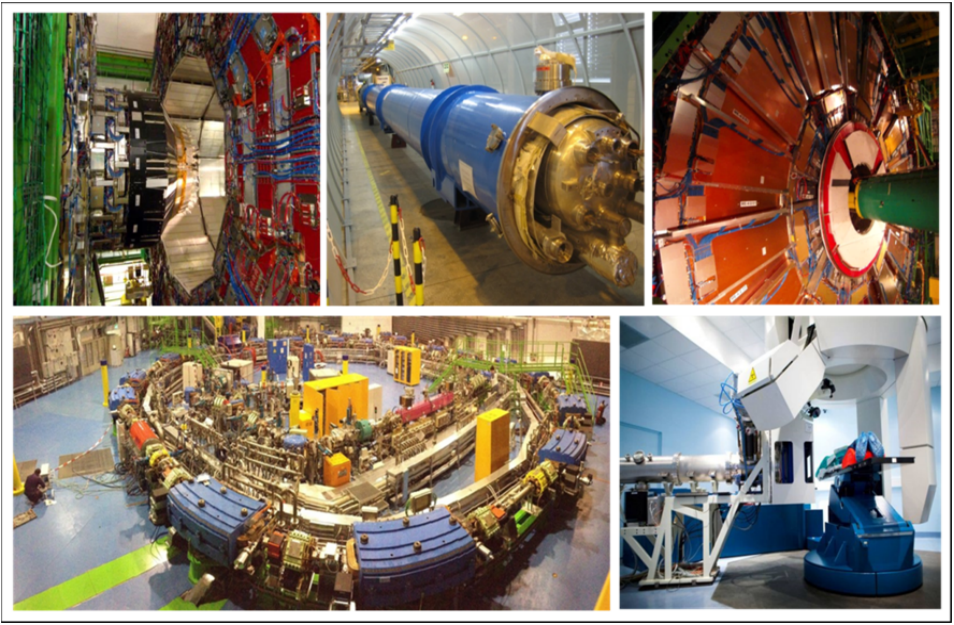

**Acknowledgments**: The paper has been produced in the frame of the research project '*Cost/Benefit Analysis in the Research, Development and Innovation Sector*' sponsored by the EIB University Research Sponsorship programme (EIBURS), whose financial support is gratefully acknowledged. Further details on this research project can be found at: www.eiburs.unimi.it.

The authors are grateful for specific inputs to other members of the Milan EIBURS Team including particularly Giuseppe Battistoni, Chiara Del Bo, Stefano Carrazza, Gelsomina Catalano, Donatella Cheri, Mario Genco, Lucio Rossi, Silvia Salini, and for specific comments on earlier drafts to Gianni Carbonaro, Andres Faíña, Ugo Finzi, Per-Olov Johansson, Mark Mawhinney, Davide Sartori, and Alessandro Sterlacchini. Also, the authors are thankful for feedbacks and suggestions from the participants at the EIBURS-UNIMI Workshop hosted by DG Research - European Commission Brussels, November 13, 2015.

**Disclaimer**: This Working Paper should not be reported as representing the views of the EIB. Any errors remain those of the authors. The findings, interpretations and conclusions presented in this article are entirely those of the author(s) and should not be attributed in any manner the EIB or other institutions.

**Photo credits**: 1) The Large Hadron Collider, CMS detector. Source: Authors. 2) The Large Hadron Collider. Source: Authors. 3) The Large Hadron Collider, CMS detector. Source: Authors. 4) The CNAO synchrotron. Source: Authors. 5) The CNAO treatment room, Source: CNAO Foundation.

European Investment Bank · Institute

# Table of contents

# Abbreviations

| | |
|---|---|
| CBA | Cost-benefit analysis |
| CERN | European Organization for Nuclear Research |
| CNAO | National Hadrontherapy Centre for Cancer Treatment |
| DG Regio | Directorate General for Regional and Urban Policy |
| EBITDA | Earnings before interest, taxes, depreciation and amortisation |
| EC | European Commission |
| EIB | European Investment Bank |
| EIRR | Economic Internal Rate of Return |
| ENPV | Economic Net Present Value |
| ESA | European Space Agency |
| ESF | European Science Foundation |
| ESFRI | European Strategy Forum on Research Infrastructures |
| EU | European Union |
| EUR | Euro |
| EXV | Existence Vale |
| FIRR | Financial Internal Rate of Return |
| FNPV | Financial Net Present Value |
| GHG | Greenhouse gas |
| HEATCO | Harmonised European Approaches for Transport Costing and Project Assessment |
| ICT | Information and Communications Technology |
| LHC | Large Hadron Collider |
| LRMSC | Long Run Marginal Social Cost |
| NACE | Statistical Classification of Economic Activities in the European Community |
| NASA | National Aeronautics and Space Administration |
| NOAA | National Oceanographic and Atmosphere Administration |
| OECD | Organization for Economic Cooperation and Development |
| PDF | Probability Distribution Function |
| Pr | Probability |
| QOV | Quasi-Option Value |
| QUALY | Quality-Adjusted Life Year |
| RDI | Research, Development, Innovation |
| TCM | Travel Cost Method |
| VOSL | Value of Statistical Life |
| WTP | Willingness-to-pay |

# Main notation

| | |
|---|---|
| $A$ | Market equilibrium |
| $B_n$ | Non-use benefits |
| $B_u$ | Use-benefits |
| $C_u$ | Use-Costs |
| $D$ | Demand curve |
| $d$ | Market distortion |
| $\mathbb{E}$ | Expected value |
| $ex$ | Externality of patent |
| $F$ | Value of knowledge spillover |
| $\Delta$ | Variation |
| $\Delta G$ | Project demand |
| $H$ | Value of human capital formation |
| $i$ | Index of goods/entities/individuals ranging from 1 to I |
| $m$ | Multiplier of impact for knowledge output |
| $MSV$ | Marginal social value of patents |
| $n$ | Index for non-use values |
| $O$ | Value of knowledge outputs |
| $P$ | Social value of patents |
| $p$ | Supply price |
| $pv$ | Private value of patent |
| $q$ | Demand price |
| $R$ | Value of recreational benefits |
| $r$ | Social discount rate |
| $ref$ | Average number of references included in patents |
| $S$ | Supply demand |
| $s_t$ | Discount factor |
| $t$ | Index for years, ranging from 0 to $\mathcal{T}$ |
| $U(\cdot)$ | Utility function |
| $u$ | Index for use values |
| $use$ | Average rate of usage of granted patents |
| $v$ | Shadow price |
| $X$ | Vector of project inputs, ranging from x to X |
| $W$ | Willingness-to-pay for a visit |
| $Y$ | Social cost of producing knowledge outputs |
| $Z$ | Value of developing new/improved products, services and technologies |
| $\alpha$ | Set of multiplicative parameters representing the characteristics of the scientific community |
| $\Pi$ | Profit |
| $\beta$ | Parameter determining the shape of the curve that change according to the papers weights, distinguishing between excellent and mediocre papers |

# Foreword

This paper summarises the main findings and lessons learned on how to apply cost-benefit analysis (CBA) for research, development, and innovation (RDI) infrastructures drawing from a research project carried out by the University of Milan in partnership with the CSIL Centre for Industrial Studies. The study was supported by an EIB Institute grant (*University Research Sponsorship Programme – EIBURS)* and involved more than twenty researchers from a broad range of scientific fields[1].

After three years of research, the team developed a conceptual framework and applied it to two selected cases (i.e. the Large Hadron Collider (LHC) at CERN[2] and the National Centre for Oncological Treatment (CNAO) in Pavia (IT)[3]) to explore methodological challenges and further potential applications. A special issue of the journal, 'Technology Forecasting and Social Change'[4], is expected to be published in 2016 with articles by economists, physicists and other experts from several EU countries and from China.

In distilling and communicating the main lessons learned during the three years of research, it is important to stress from the beginning that this terrain is mostly uncharted. Although the conceptual model builds on principles firmly rooted in CBA tradition, its application to the RDI sector is still in its infancy. Hence, the approach adopted in this discussion paper is explorative and heuristic.

The concepts presented in the paper would benefit from further applications and testing beyond what has been possible to do in the specific framework of this research project. Future versions of this paper will take into account as best as possible any comments or new findings.

The structure of the paper is as follows. After presenting the rationale for using CBA and the main specificities of RDI infrastructures in the CBA perspective (Section 1), suggestions on how to perform the financial analysis of the RDI projects as a preliminary step (Section 2) and the socio-economic analysis (Section 3) are given (Section 4). Then, section 5 provides suggestions on how to perform a proper risk analysis and indications on how to effectively present the results of the analysis (because communication of results is important in a new field) and concludes.

---

[1] The full list of researchers is available on the project website at www.eiburs.unimi.it
[2] See Florio *et al.* (2015).
[3] See Pancotti *et al.* (2015).
[4] www.journals.elsevier.com/technological-forecasting-and-social-change

# 1. Executive Summary

## 1.1 Introduction

Governments, funding agencies and policy makers have high expectations on research, development and innovation (RDI) infrastructures in the context of science and innovation policies aimed at sustaining economic growth in the long term. The construction and operation of these often large and complex RDI facilities is increasingly costly and there is competition for public support. A wide range of approaches and methods are used in different institutional settings for the selection and prioritisation of such infrastructures. These include peer review of the scientific case, the development of national, international or sectoral roadmaps, and a number of quali-quantitative indicators of results. Yet, a consensus on the most appropriate methodology to assess the socio-economic long-term impact of major RDI projects is still missing. The objective of this discussion paper is to explore the applicability of the cost-benefit analysis (CBA) approach to this issue.

CBA is grounded in welfare economics, and its application to traditional infrastructures, such as transport, water, energy is firmly established, as revealed for example by a recent survey of OECD countries (OECD, 2015). More recently CBA have also been successfully applied to environment, education, cultural investment. Until now, however, the use of CBA to evaluate RDI infrastructures has often been hindered by the intangible nature and the uncertainty associated to the achievement of research results.

Recently, some attempts to develop a CBA theoretical framework for RDI infrastructures have been made in the context of the use of Structural Funds by the Czech government and JASPERS.[5] Moreover, for the first time after its previous four editions, the new Guide for the CBA of investment projects in the context of Cohesion Policy, recently adopted by the European Commission (2014) provides guidelines to appraise RDI projects, but also admits that – due to lack of experience and best practices – further steps are needed to improve the evaluation framework.

This discussion paper presents the results and the lessons learned on how to apply ex-ante CBA for major RDI infrastructures by a team of economists and scientists at the University of Milan and CSIL during a three-year research project supported by a EIBURS grant of the European Investment Bank Institute. Albeit the comprehensive conceptual framework presented in the paper builds on principles firmly rooted in CBA tradition, their application to the RDI sector is still in its infancy. The model has been applied on two cases in physics involving particle accelerators (the Large Hadron Collider (LHC) at CERN[6] and the National Centre for Oncological Treatment (CNAO) in Italy)[7], but further applications and testing are needed to fine tune and expand the proposed appraisal techniques as well as to contribute to building a larger information base.

## 1.2 A CBA model for RDI projects

The core of CBA is an evaluation (ex-ante or ex-post) of the project intertemporal socio-economic benefits and costs, all expressed in units of a welfare *numeraire* (usually money in present value terms). The net effect on society is finally computed by a quantitative performance indicator (the net present value, or the internal rate of return, or a benefit/cost ratio). In line with the general CBA fundamentals, a CBA model of RDI should make use of: 1) shadow prices to capture social costs and benefits beyond the market or other observable values; 2) a counterfactual scenario to ensure that all costs and benefits are estimated in incremental terms relative to a 'without project' world; 3) discounting to convert any past and future value in their present equivalent; and 4) a consistent framework to identify social benefits by looking at the different categories of agents (producers, consumers, employees, tax-payers).

For the purpose of RDI project evaluation, it is convenient to divide social intertemporal benefits in two broad classes. On the one hand there are *use benefits*, accruing to different categories of direct and indirect users of the infrastructure services, such as e.g. scientists (insiders and outsiders), students starting their career within the facility, firms benefitting of technological spillovers, consumers benefitting of innovative services and products, and general public visitors of the facility or those enjoying outreach activities. The identification of use-beneficiaries is project specific and must carefully avoid omissions or double counting of impacts. On the other hand, there are *non-use benefits*, reflecting the social value of the discovery potential of the RDI infrastructure, regardless its ex-ante predictable actual use. This is a measure of the social preference for pure discovery, akin to social preferences for culture, or environment protection *per se*. The sum of these use and non-use benefits is then compared with costs. Whenever possible, the risks of forecasting errors should be taken into account by attaching probabilities to the values of each critical variable entering the model.

---

[5] See Czech Ministry of Education, Youth and Sport and JASPERS (2009) and JASPER (2013).
[6] See Florio *et al.* (2015).
[7] See Pancotti *et al.* (2015).

A significant feature of a CBA approach to RDI is the stochastic nature of the model. Consistently with best practice in the field, the project performance is assessed in probabilistic terms using a Monte Carlo simulation that approximates the probability distribution functions of the socio economic net present value (or other indicators), their cumulative distribution functions, the expected value, etc.

In a nutshell, the CBA model presented in this paper is intended to predict socio-economic benefits and costs of RDI projects in measurable form. While the model includes the measurable forecasting errors, it deliberately leaves aside what is intrinsically non-measurable. The model should be seen as a complement, not as a substitute, of the scientific case evaluation, financial and budgetary issues, managerial and strategic considerations, or political dimensions of investment in RDI, which are also obviously important. A preliminary financial analysis of the project in long-term cash-flow terms is helpful, but should not be confused with the social CBA.

## 1.3 Social Costs

As for projects in other fields, the main categories of costs associated to RDI infrastructure relate to the present value of capital, labour cost (including the labour cost of scientific personnel and the labour costs of other administrative and technical staff), other operating costs, such as materials, energy, communication, maintenance, etc., negative externalities, like air pollution or noise during construction and operations, and decommissioning.

However, when the infrastructure is designed to perform a range of different experiments and activities or the project consists of several inter-related but relatively self-standing components, delimiting the project borders and, in turn, its costs is challenging. If the analysis focuses on assessing a single experimental facility part of a larger complex, the costs that are shared by other experiments out of the scope of the analysis should be duly apportioned to the infrastructure under examination. Sunk costs should not be included in the computation of costs.

## 1.4 Social benefits

Traditionally, in economics, agents are classified according to their roles: firm owners, consumers, employees, tax-payers. These classifications are flexible, as in some cases producers of goods are also consumers (e.g. small farmers), employees are also tax-payers, shareholders are also firm managers, and so on. In CBA it is crucial to identify the beneficiaries in a way consistent with first principles of welfare economics.

Having identified the main beneficiaries of an RDI infrastructure, a list of typical benefits can be attached to each group. Depending on the project's nature, some of these benefits may accrue to different types of target groups. Also, the intensity of each benefit may be highly variable across the different typologies of RDI infrastructures. Hence, only a case by case study can design the appropriate research strategy.

## 1.5 Social benefits to firms

The following benefits to firms can often be identified and evaluation methods applied in relation to RDI infrastructure projects:

- The *development of new/improved products, services or technologies* has a socio-economic value measurable by the *expected incremental shadow profits* (i.e. after using shadow prices when needed for inputs and outputs) expected from their sale as compared to the without-the-project scenario;

- The *grant of a patent* has a marginal social value taking into account both the private value, i.e. the value from the patent holder point of view, and the externality, i.e. the knowledge spillover brought about by patents in generating a cascade of innovation;

- The creation of *start-ups and spin-offs or (and) an increase in their survival rate*: the economic value of this benefit is valued as the *expected shadow profit* gained by the created business during its overall expected lifetime compared with the without-the-project scenario. Whereas, if the RDI infrastructure contributes to increasing the survival rate of start-ups, then the benefit is valued as the *incremental expected shadow profit* attained by businesses that survive longer than in the without-the-project scenario

- The occurrence of *knowledge spillovers from the RDI project to third-parties* (businesses, professionals, public organisations), can be valued using alternative approaches (or a combination of them provided that double counting is carefully avoided) depending on the category of beneficiaries (i.e. incremental shadow profit; avoided costs, willingness-to-pay (WTP) for time saving);
- The *learning-by-doing benefit* for firms in the supply chain of a major RDI infrastructure, is valued through the incremental shadow profit expected by supplier companies thanks to the fact they have collaborated with the scientific and technical staff of the infrastructure and, in turn, have acquired new knowledge and technological skills;

## 1.6 Social Benefits to researchers and human capital formation

In general, employment is a social cost (except when there is very large unemployment). Therefore, the social benefit of employment has to be taken into account by exclusively using shadow wages, i.e. by considering that the opportunity cost of employing a person in the project under assessment is lower than that from using the same person for any alternative use.

However, RDI infrastructure projects have the mostly unique peculiarity that some producers of services are also their beneficiaries. Namely:

- Students and young scientists who will spend a period working within a major RDI infrastructure will earn higher human capital relative to their peers. The socio-economic value of this benefit is expressed as the expected *incremental lifelong salary* earned by such individuals over their entire careers compared with the without-the-project scenario;
- Scientists at the RDI facility produce knowledge, but are also users of such knowledge. The process is embodied in the production of knowledge outputs (i.e. technical reports, preprints, working papers, articles in scientific journals and research monographs) and their degree of influence on the scientific community in form of citations. The socio-economic benefit related to the production of scientific outputs can be valued using their *marginal production cost,* which is common practice in CBA for certain types of services, when market prices are not relevant and when WTP is not the appropriate empirical approach. Instead, the degree of influence of such outputs is reflected in the number of people that would cite it and valued through the *opportunity cost of time* employed by a scientists to download, read and understand someone else's output and decide to cite it. An important consequence of valuing scientific outputs at their marginal cost, which is mainly labour cost, is that to a certain extent scientific work pays for itself (an analogy is self-employment in subsistence farming where the benefit of the output – i.e., food – net of other costs, is exactly the value of the labour input). It is important to avoid the confusion, however, between the value of knowledge *outputs* (publications) and the value of *knowledge per se* embodied in such publications. The former is usually predictable, while the latter is often unmeasurable (the social value of producing and selling a book is unrelated to the social value of understanding its content and elaborating on it by the readers).

## 1.7 Social benefits to consumers of services produced by the RDI project

Benefits to consumers are highly project specific.

- They may derive from the use of the infrastructure's equipment and/or the provision of specific services to *external users* (e.g. industries, governmental bodies and other research teams). In this case, the socio-economic benefit is valued by either using the l*ong run marginal cost* of the services provided or estimating *external-users' WTP for the service*. Alternatively, when market prices are available and are supposed to be non-distorted, i.e. they reflect economic prices, the *nominal (market) price*s can be used.

- Also, benefits to consumers may derive from the practical application of a research effort (e.g. reduction of GHG and air pollutant emissions; improved energy efficiency; reduction of vulnerability and exposure to natural hazards; improved health conditions, or simply lower production cost and sale price, etc.). The methods to quantify and value these benefits depend on the types of new services or products made available by the infrastructure. However, these methods are generally based on the willingness to pay or avoided cost approaches, and are often well established in CBA.

- Other use benefits include the cultural effects enjoyed by both on-site and virtual visitors of the RDI project because of outreach activities. The expected marginal social value of this benefit is valued using the visitors' implicit willingness-to-pay for a visit. Concerning in-person visits, the standard way to estimate the WTP is using the travel cost method, while a broadly used method to attach a monetary value to non-market goods such as virtual visits is contingent valuation;

## 1.8 Non-use benefits and the tax-payers: discovery as a public good

There are other social benefits to be considered, more elusive but nevertheless important. These are non-use benefits. While for applied research, development and innovation most benefits accrue to direct and indirect users (firms, consumers, researchers and students) for fundamental research it is usually impossible to identify who will be the ultimate beneficiaries of a discovery.

- If there is a potential but largely unknown future use-benefit, this can be defined a *quasi-option value* and while it is conceptually important to acknowledge its role, CBA methods are often unable to quantitatively determine it, even if research on the topic is ongoing. It is suggested to conservatively set to zero such value, except when the evaluator is confident of being able to make predictions on the economic value of applications of fundamental research.

- As the tax-payers ultimately foot the bill of some government-supported research infrastructures, it is appropriate to know their willingness to pay for their discovery potential. This is a non-use value of a public good, similar to the notion of existence value in environmental CBA. In principle, the social preference for pure knowledge *per se*, regardless the fact that it might find some use in the future, is empirically testable by stated preference techniques. Similarly to what is done in environmental or cultural economics for estimating the economic value of endangered species protection, preservation of natural resources, or conservation of cultural heritage assets, the contingent valuation methodology can be exploited to elicit the taxpayers' WTP for having a discovery, regardless of its actual or potential use.

## 1.9 The probability distribution of the economic net present value

Once the socio-economic benefits (including both use and non-use benefits) and costs associated with an RDI infrastructure have been identified, valued in monetary terms and discounted using a social discount rate, the effect on society is finally computed by a quantitative performance indicator (the net present value, or the internal rate of return, or a benefit/cost ratio).

Whenever possible, the risks of forecasting errors should be taken into account by attaching probabilities to the values of each critical variable entering the model. Hence, consistently with best practice in the field, the project performance is assessed in probabilistic terms using a Monte Carlo simulation that approximates the probability distribution functions of the Economic Net Present Value (or other indicators), their cumulative distribution functions, the expected value, etc.

## 1.10 Conclusions

The CBA model presented in the discussion paper provides a comprehensive framework for ex-ante assessment of major RDI infrastructures, consistent with the general applied welfare economics fundamentals, but innovating the field in several ways.

The model is also novel and heuristic because it intends to apply principles firmly rooted in the CBA tradition into a new uncharted field. The application of the model on two pilot case studies in physics involving particle accelerators, respectively in pure science and medical research, has contributed to explore its empirics. Therefore, this discussion paper is intended to sow the seed of further applications and testing to fine tune and expand the currently methodologies and techniques.

Experience in other fields of CBA, such as environmental and cultural economics, suggests that several years of practical testing are needed before new ideas are embodied in an accepted paradigm by practitioners. It is hence needed to be experimented in different RDI domains, with their specificities, the empirical analysis.

# 2. Motivation and Principles

## 2.1 Increasing need for accountability

Policy makers have growing expectations that RDI infrastructures are essential components of technological and scientific progress (EC, 2010; ESFRI, 2010; Technopolis, 2011; ESF, 2013). Hence, the stakes associated with the selection and evaluation of such infrastructures are high.

Traditionally, the selection and appraisal process of RDI projects relies on science's own internal quality control mechanisms and the policy context. Typically, these mechanisms involve a peer review process that assesses the 'science case', sometimes complemented by a 'business case' or considerations related to the socio-economic impact (Feller, 2013). Although this approach is usually efficient and fair, it is not suitable for appropriately assessing the socio-economic effects of a project. The scientific or business cases are complementary evaluation tools but are not necessarily correlated to the socio-economic impact of a project. For this reason, specific evaluation tools and methods are needed.

Recently, a more strategic approach to RDI investments has been promoted by international practice, with the development of roadmap exercises to prioritise RDI infrastructures at the national or European level (ESFRI, 2008; OECD, 2008; Research Council UK, 2010). Typically, roadmaps assess RDI infrastructures according to a set of quali-quantitative criteria ranging from scientific and technological excellence to socio-economic impact indicators and governance and financial aspects. In some cases, the approach also includes the consideration of risk factors. However, no consensus exists on a unique evaluation model; instead, a variety of different experiences exist (Pancotti, *et al.*, 2014). This lack of consensus hinders the possibility to systematically compare the impact of different projects that may compete for scarce funding, or to compare ex-ante, on-going, and ex-post evaluations, as suggested by the best international practice for project appraisals of major infrastructures. Although roadmaps are relevant strategic and planning tools, they are not designed to provide a socio-economic impact evaluation framework, consistent with applied welfare economics principles.

Against the demand for credible methodologies to assess RDI infrastructures, cost-benefit analysis (CBA) is considered a promising candidate. CBA emerged from more than one hundred years of intellectual history (Dupuit, 1844 and 1853; Pigou, 1920; Little and Mirrlees, 1974) and is now a recognised evaluation technique (Florio, 2014). Currently, CBA is widely adopted by international institutions and governments to assess the socio-economic profitability of investment projects in many fields.[8] Although already advocated by some international and national organisations, even with a number of caveats and possible adaptations (ESA, 2012; EIB, 2013; OECD, 2015), up to now CBA has not been systematically adopted on a broad scale in the RDI sector.

Until recent years, the development of CBA in the RDI field was hindered by the perception of the unpredictability of the long-term benefits of knowledge (Martin and Tang, 2007). Actually, the nature of knowledge creation – the typical output of RDI projects – is such that the effects of a discovery may appear in the very distant future, long after the decommissioning of the RDI infrastructure. The uncertain impacts of the RDI infrastructure on social welfare, combined with the difficulties in measuring them, have probably slowed down the diffusion of CBA in the RDI sector. However, a renewed interest is observable in recent years because the stakes associated with RDI infrastructure selection and ex-ante appraisal have increased[9].

Earlier attempts to develop a CBA theoretical framework in the field of RDIs was initiated by the Czech government in 2009 and further expanded by the JASPERS team at the European Investment Bank[10]. On the basis of the experience gathered in the 2007–2013 programming period, JASPERS produced a staff working paper as preliminary guidance for the application of the CBA approach into practice in the RDI sector[11]. Another recent contribution is provided by the European Space Agency (2012), which proposed a methodology called 'SCBA-plus' to establish the impact of space programmes. The methodology consists of a combination of social cost benefit analysis and multi-criteria analysis. A recent survey by the OECD (OECD, 2015) showed that some governments are using CBA in the RDI field.

---

[8] See, for instance, Adler (1987); Atkinson *et al.* (2006); EC-EIB (2005); EC (2007); Economics and Development Resource Center (1997); Pearce *et al.* (1994); WHO (2006); and Asian Development Bank (2013).
[9] For example, in the UK, the Science and Technology Facilities Council is committed to mobilising a methodology for to make a difficult decision on where to discontinue funding in a context of a 'flat-cash' budget (Technopolis, 2013).
[10] Czech Ministry of Education, Youth and Sport and JASPERS (2009). Background methodology for preparing feasibility and cost-benefit analysis of R&D infrastructure projects in Czech Republic, supported by the Cohesion Fund and the European Regional Development Fund in 2007–2013.
[11] JASPERS (2013). Project Preparation and CBA of RDI Infrastructure Project, Staff Working Papers, JASPERS Knowledge Economy and Energy Division.

The approach proposed in this discussion paper draws from and further develops the mentioned previous work and is consistent with the general methods suggested in the updated EC Guide for CBA (European Commission, 2014). Hence, the paper takes advantage of the substantial experience gained by researchers and practitioners worldwide on the evaluation of infrastructures in a range of sectors and tries to apply the lessons learned in other contexts to the specific challenges posed by RDI infrastructures. The proposed approach should be intended to complement and not substitute for other evaluation methods, including peer review assessments, road mapping, qualitative evaluation of socio-economic impact and monitoring of performance indicators.

## 2.2 The perspective of social cost-benefit analysis

Social CBA is grounded in welfare economics, according to which the welfare of a society depends on the aggregate individual utility of all of its members. In a welfare economic frame CBA arises as the solution for the government's planning problem of the constrained optimisation of a social welfare function (Drèze and Stern, 1990; Florio, 2014).

CBA theory and application have evolved over time and have undergone different phases of experimentation, consolidation, and diffusion in a variety of institutional settings and sectoral traditions. Yet, a number of key features and principles offer a good framework for a solid and systematic approach to RDI project appraisal and selection. In particular, these features and principles are as follows.

- Social CBA is a tool aimed at informing decision making on the economic viability of investment decisions by quantitatively expressing all of the costs and benefits to society. The net economic benefit to society is used as the performance criterion. Costs and benefits are expressed through a monetary metric, but any welfare *numeraire* or appropriate accounting unit also works. Against the existing evaluation approach for RDI projects, CBA offers a tool to systematically compare both costs and benefits on a unique accounting basis.

- A long-run timeframe is adopted to assess the social welfare change attributable to it, implying the identification of a proper time horizon and the consideration of long-term sustainability.

- CBA makes a clear distinction between social welfare effects and financial effects. The former are expressed in accounting prices that convey the social opportunity costs of a project's inputs and outputs ('shadow prices'), whereas the latter are prices adopted and observed on the market. CBA makes use of shadow prices and uses market prices only to assess the financial viability of a project (see below).

- Costs and benefits are considered incrementally, which requires a systematic comparison between the project option and a proper counterfactual ('with' and 'without the project' scenarios) (see *Box 1*).

- The key strength of CBA is that it produces information on the project's net contribution to society's welfare, synthesised into simple indicators, such as net present value. This approach leads to the possibility of comparing several investment options or past expectations with actual outcomes.

> Box 1. Choosing a proper counterfactual scenario
>
> The incremental approach requires that the costs and benefits of the proposed project are estimated in incremental terms within a counterfactual scenario (without-the-project). This method serves to grasp the 'net' change, i.e. the change specifically attributed to the intervention analysed. The choice of the counterfactual scenario requires a careful examination and implies defining what would happen in the absence of the project. In-depth discussions with science specialists involved in the project design and with independent professionals capable of providing sufficiently disinterested judgments are fundamental to choosing a proper counterfactual scenario. The following two broad options are available.
>
> ➡ For a completely new facility ('green field'), the without-the-project scenario is usually a 'zero-based' scenario. In other words, the incremental scenario coincides with the 'with-the-project' scenario.
> ➡ For investments aimed at improving an already existing RDI facility, the counterfactual should include the costs and benefits to operate and maintain it at a level that keeps it operable (this scenario is referred to as 'business as usual' or 'do-nothing') or even small adaptation investments that were programmed to occur anyway (do-minimum).
>
> Examples of green field RDI facilities are the National Hadrontherapy Centre for Cancer Treatment (CNAO) based in Pavia, the Jules Horowitz Reactor (JHR) built on the Cadarache site in France. Examples of incremental RDI facilities are the High Luminosity Large Hadron Collider at CERN and the upgrade of the European Synchrotron Radiation Facility in Grenoble.

## 2.3 Key features of RDI infrastructures

Given the wide variety of facilities that are generally referred to in this field, no established and agreed definition of RDI infrastructures exists in the literature and in policy documents[12]. However, a number of constituent features of typical RDI infrastructures exist that lend themselves particularly well to be assessed using CBA. These features are as follows.

- <u>Based on tangible assets</u>: The assets can be either single-sited, mobile or distributed. Most RDI infrastructures, such as particle accelerators, telescopes, and technological platforms, are single-sited, i.e. a unique facility or a combined set of infrastructures and equipment located in a single physical location, as defined by ESFRI (2008). However, mobile[13] and geographically distributed facilities also exist. The latter includes for instance grid computing systems or seismographic stations consisting of a network of infrastructures located in different areas but with a strong functional relationship among all of their parts.[14] Finally, some RDI infrastructures provide their services in electronic form, i.e. through storage, transmission, and elaboration of coded information. High performance ICT-based technologies can be essential for an RDI infrastructure to making particularly complex computations and simulations, thus actually producing new knowledge and, for this reason, are to be distinguished by traditional ICT infrastructure.[15]

- <u>High-capital intensity facilities</u>: Capital expenditures[16] overcome operating costs, i.e. they represent a large share of the total present value of the project cost. Hence, different from RDI programmes, once the financing decisions are made, discontinuing such facilities before the full materialisation of their benefits becomes expensive.[17]

- <u>Major facilities</u>: They require substantial capital investments in infrastructure. For example, in the field of EU cohesion policy, a major project is conventionally defined as requiring a total investment cost in excess of EUR 50 million[18].

- <u>Long-lasting facilities</u>: The economic life of RDI infrastructures is not different from that of more standard infrastructures. Generally, these facilities remain operational for more than two decades after their construction. In some domains, research infrastructures only develop their full scientific potential if a long-term data series can be generated and recorded (Wissenschaftsrat, 2013). The adoption of a long time horizon is necessary for a comprehensive appreciation of their performance.

---

[12] See, for instance, ESFRI (2008); Horlings, E. and Versleijen, A. (2008); JRC-IPTS (2002); Research Council UK (2010); and Technopolis (2011).
[13] An example is provided by research vessels.
[14] Conversely, a network of mutually-independent RDI infrastructures, each providing its service without depending on the service provided by another facility in the same network, is not accounted for as a single distributed RDI facility but instead as a number of single-sited infrastructures.
[15] Examples are supercomputers and grid computing, which consists of computer resources specifically developed to process big data and produce outputs for scientific use. On the contrary, it can be argued whether collaborative ICT infrastructures where large volume of algorithms for data pre-processing, statistical analysis and annotation are integrated and chained to build ad hoc workflows for users, qualify as RDI infrastructures per se. ICT tools for integration of datasets can be a component or an output of a RDI infrastructure. The collaborative aspect can be important in terms of spillover effects since it allows an easier exchange of research and innovation outputs among the scientific communities; at the same time, the collaborative nature of different RDI infrastructures could pose some challenges in terms of benefits appropriation, since the boundaries of research and innovation outputs become blurred and their ownership difficult to be clearly attributable.
[16] They include both fixed capital and initial human capital formation expenditures.
[17] As a general remark, in the paper we consider the investment as a one-shot decision but in practice decision makers face options of expanding, deferring or abandoning the investment. RDI infrastructure projects are actually embedded in a sequential process with multiple stages and mile-stones associated with certain risk. Along a decision tree risks vary according to the likelihood of achieving different end points, taking these risks into account involves managing different discount factors. The theory of option prices provides a solution to manage this situation by means of a trick based on adjusting each state's probabilities in accordance with the 'risk neutral' discount factor. For a discussion of this issue see EIB (2013), Jägle (1999), Luehrman (1998), Courtney et al. (1997).
[18] As per article 100 of Regulation (EU) No 1303/2013.

- Facilities with the objective of producing social benefits through a generation of new knowledge and innovation for a variety of users: RDI infrastructures are realised with the main purposes of acquiring new knowledge in a given scientific or technological field and/or using the stock of new knowledge to devise new applications or produce innovation. A typical distinction is made between the following infrastructures.

    o Infrastructures for fundamental research, such as a large telescope, are meant to support basic research, i.e. undertaking theoretical or experimental work primarily to acquire new knowledge on the underlying foundations of phenomena and observable facts, without any direct practical application or use in view ('curiosity-driven').

    o Infrastructures for applied research and technological development are meant to support the acquisition of new knowledge for a potentially well-identified practical purpose, i.e. the development of new products, processes, or services (e.g. quantum computing or human genomics).

    o Innovation infrastructures, such as a laboratory within a pharmaceutical firm or working for a consortium of firms, aim to combine new knowledge and technology for the future commercial exploitation of newly developed applications.

- <u>Potentially supporting multiple experiments or testing</u>: In the RDI sector, some facilities are destined to remain unique at a regional, national, or even global level because a second facility is too expensive or because the number of users is not large enough. In these cases, the preferred setting usually entails arrangements for different teams of users of the same infrastructure and exploiting its features to perform different experiments or tests, either over time or at the same time.

Although such features distinguish RDI Infrastructures from traditional RDI programmes and initiatives, they are instead shared by traditional infrastructures in other sectors (e.g. environment and transport). For this reason, the use of a CBA framework seems particularly appropriate[19].

---

[19] For the same reason, examples of research infrastructures for which the justification of a CBA framework is less robust are the construction or modernisation of buildings with primarily educational purposes; knowledge-based resources such as collections, archives, or surveys, for which the service they provide – i.e. the collection and elaboration of data *per se* – is usually more labour than capital intensive; and relatively small RDI projects with capital costs of millions of Euros.

Figure 1. Examples of major RDI infrastructures

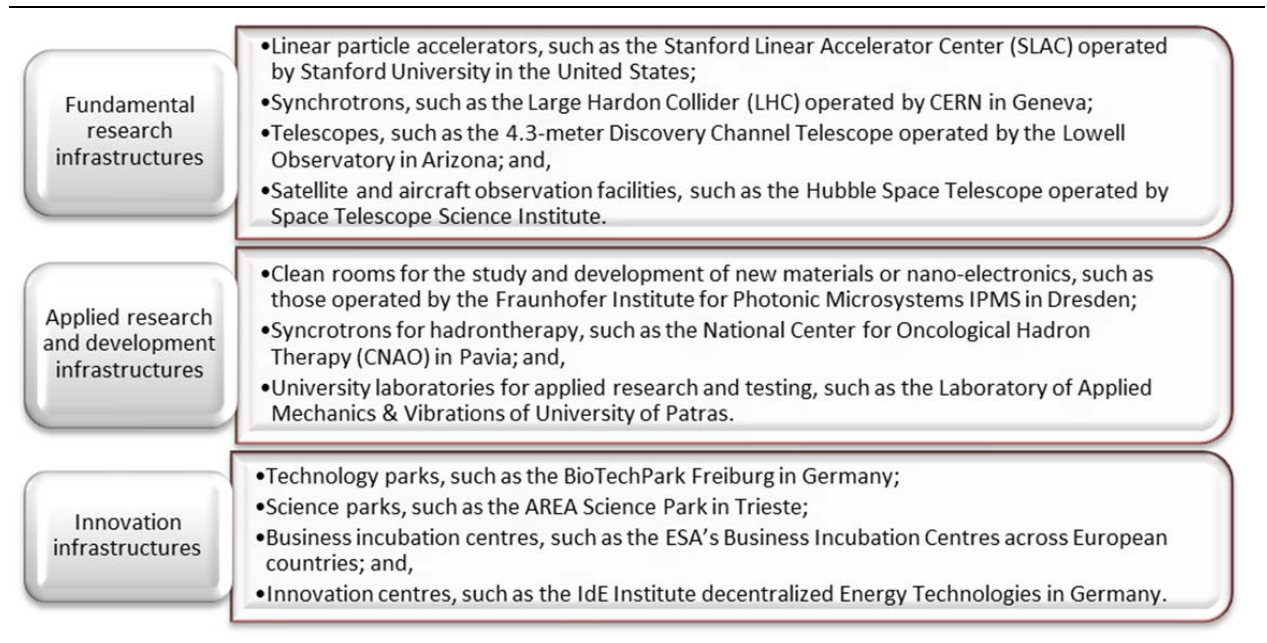

Source: Authors

## 2.4 Categories of potential beneficiaries

In economics, agents are traditionally classified according to their roles: firm owners, consumers, employees, tax-payers. These classifications are flexible, as in some cases producers of goods are also consumers (e.g. small farmers), employees are also tax-payers, shareholders are also firm managers, and so on.

From the CBA perspective, it is crucial to identify the potential beneficiaries (or losers) of a project. Consistently with first principles of welfare economics and because of the variety of possible services delivered by major RDI infrastructures, at least six groups of social agents exist whose welfare is potentially affected by a RDI project.

Table 1. Classification of potential beneficiaries of an RDI infrastructure

| Type of agent | Description |
| --- | --- |
| Businesses | Businesses include spin-offs and start-ups, small and medium enterprises, and large enterprises that directly enjoy the services provided by the project and/or benefit from indirect spillover effects, particularly through procurement and supply chain learning effects. |
| Employees | Scientists and researchers produce knowledge, but are also are direct users of the RDI facility. They encompass both inside staff and outsiders, who are the rest of the research community, including those working in other fields that may use the evidence provided by the experiments to produce further knowledge and innovations.<br>Young professionals, junior researchers, and students who will spend a period working within the RDI infrastructure. They include, for example, post-doctoral researchers, early career researchers who use the RDI facility to carry out their own studies/tests, and students, usually at graduate level, involved in training or the preparation of their dissertation or who have access to the facility through a training programme. |
| Consumers | Consumers of goods or services produced by RDI projects. They may include, for example, patients associated with medical treatment provided at a health research infrastructure and residents of a region in which major risks such as floods, earthquakes, and fires are better monitored/forecasted because of the research developed by observatories, stations, or satellites, among others.<br>The general public involved in outreach, which includes onsite visitors to the facility and virtual visitors on websites and social networks; media exposure is also the effect of outreach activities and people derive utility from being informed about research and technological progress; |
| Taxpayers | Non-use beneficiaries, such as most tax payers who fund RDI infrastructures without directly using it and people who do not plan to visit (personally or virtually) the infrastructure. Such individuals may derive some utility from the mere fact that they appreciate that scientific discoveries and technological progress are possible because of the infrastructure; these potential beneficiaries of knowledge create a global public good with an intrinsic non-use value. |

Source: Authors

The conceptual model developed stems from the consideration that the recognition and proper identification of actual and potential beneficiaries is an essential ingredient of the assessment. Indeed, different categories of target groups are associated with different types of benefits and externalities, thus estimating the present and future demand for the infrastructure. The infrastructure's outputs for each of the identified groups of beneficiaries are the starting point of the analysis. Regardless of the estimation technique chosen, the underlying assumptions (parameters, coefficients, and values) that show that a critical mass of users (and non-users) exists need to be stated transparently and tested in the risk assessment.

In most cases, the previous list of six groups (see Table 1) likely covers the universe of potential beneficiaries of RDI projects. The list is based on the consideration that the 'demand side' of the evaluation is correlated with benefits, whereas the 'supply side' is correlated with social costs. Because most RDI services are not marketed, some confusion may exist over the agents and determinants of supply and demand. However, these mechanisms are built in any project. In a CBA framework, the willingness to pay for the service by agents ultimately determines its social value, and each of the six groups that we mention 'has standing' in the project from this perspective: firms because RDI potentially creates value to their owners; scientists and young researchers because of the reputational effects and human capital formation; service beneficiaries because of the avoided costs and/or better quality of life; and the general public because of the direct cultural effects or the willingness to pay for a public good. Instead, employment, procurement, use of land, and others are items on the cost side. In the rest of this paper, we focus on the main items that enter into an evaluation, even though additional impacts may need to be considered in specific projects.

## 2.5 The model

The model on which the rest of the paper is built takes the form of a simple yet comprehensive equation (for slightly more technical details see Florio and Sirtori, 2014):

$$\mathbb{E}(ENPV_{RDI}) = \mathbb{E}(EPV_{B_u}) + \mathbb{E}(EPV_{B_n}) - \mathbb{E}(EPV_{C_u})$$

In this frame, the social CBA exercise consists of forecasting, in incremental terms, the expected economic net present value of the RDI infrastructure projects ($\mathbb{E}(ENPV_{RDI})$), defined as the sum of the expected net present value of economic benefits associated with any actual or predictable practical use of the infrastructure services ($\mathbb{E}(ENPV_{B_u})$) and the additional expected value of discovery (new knowledge) for which a possible use is not yet identified ($\mathbb{E}(ENPV_{B_n})$) but for which a social value can be empirically estimated (non-use value), minus the expected net present value of the costs. In other words, our approach breaks down intertemporal benefits into two broad classes – use and non-use benefits – and compares these benefits with costs, taking into account the probability density functions attached to the determinants of each critical variable entering into the model.

In our framework, both fundamental and applied research in principle can be addressed, with the non-use benefit being often negligible for the more applied innovation projects at one extreme, and significant benefits coming from non-use value at the other extreme of fundamental research.

Three distinct concepts are included in our approach:

- the *expectation* operator implies that all critical variables are treated as stochastic;
- *economic value* indicates that our valuation uses shadow prices to capture social benefits beyond their market or financial value; and,
- the *net present operator* implies that any past or future value is converted into its present equivalent and costs are treated as negative benefits.

Different from prior literature, one contribution of the present approach is the consistent identification of the social benefits of the RDI infrastructure and the provision of a comprehensive framework to evaluate and combine them to obtain a synthetic quantitative measure of social benefit. In a CBA frame, carefully identifying different types of beneficiaries is necessary to apply concepts and empirical methods appropriate for each of such types and to avoid double counting the benefits.

Another advantage of relying on a CBA framework to assess RDI infrastructure is that, when focusing on social benefits, the analysis is often developed starting from a preliminary financial analysis (European Commission, 2014). The financial analysis is a useful management tool for verifying the long-term financial sustainability of the project.

As mentioned, the proposed CBA model suggests that all critical variables are expressed in terms of expected values. This suggestion implies conjecturing on the probability distribution functions of quantities and accounting prices (i.e. empirical proxies of unknown shadow prices) rather than using their punctual 'best guess' values. Given the large risks implicit in RDI projects, framing ENPV in terms of probabilities draws from the consideration that risk is measurable, whereas radical uncertainty is not (common sense often confuses the two concepts). From a heuristic perspective, a bold but useful step is to set to zero the value of what is radically uncertain.

The presentation of the approach consists of providing a structured discussion around three building blocks – CBA theory, empirical approaches, and examples. This presentation structure ensures a balance of conceptualisation and practical shortcuts (purely illustrative) when presenting an experimental framework.

The following sections describe in detail the steps of a CBA for RDI infrastructures. Although reference is made to the DG REGIO Guide (European Commission, 2014) for the standard approach to CBA, the aim of this paper is to illustrate how individual steps need to be adjusted to account for the specificities of RDI infrastructures.

# 3. Financial analysis in the RDI field

A clear difference exists between the socio-economic impact of an RDI project and its financial performance. The latter is carried out from the point of view of the project promoter and aims primarily to assess the project's sustainability. Financial performance is preparatory to the economic appraisal, which instead assesses the project's worthiness to society (at the global, country, and regional levels as appropriate). More specifically, the financial analysis is useful to determine the costs and revenues (if any) arising from the project over the reference period and to verify whether the projected cash flow ensures adequate operation of the infrastructure and its financial sustainability in the long term.

In this section, after briefly discussing the proper unit of analysis, the typical cash inflows and outflows to be considered for RDI projects are presented. In closing, the financial performance and the sustainability criteria are considered.

## 3.1 Unit of analysis, project borders, and cost apportionment

A useful step before carrying out the financial (and economic) analyses of a RDI infrastructure is to identify the object of the analysis. In principle, the appraisal should focus on all of the components that are logically connected to the attainment of the intended objectives[20]. In the appraisal phase, the unit of analysis is typically related to the financing decision; however, in some cases, a mismatch may occur between funding and project identification (for example, when funding is broken down in different decisions over time or across parts of a project). Delimiting the borders of a RDI infrastructure project is challenging when the infrastructure is designed to perform a range of different experiments and activities or when the project consists of several interrelated but relatively self-standing components. For example, the LHC is a combination of an accelerator system (itself composed of several accelerators) and detectors, each managed by international collaborations of scientists and laboratory staff in various combinations.

Depending on the specific nature of the infrastructure and the scope of the analysis, for multi-purpose RDI facilities, the project analyst could either focus on assessing the costs and benefits of a single experimental facility or take into account the costs and benefits related to both the hosting infrastructure and all hosted experiments. If the former approach is adopted (as in Florio *et al.* 2015), the costs related to the common facilities and that are shared by other experiments out of the scope of the analysis should be duly apportioned to the infrastructure under examination. Similarly, for distributed facilities, the project analyst should ascertain whether synergies and functional relationships among the facility components are such that they justify the assessment of the entire infrastructure as a single unit of analysis[21]. Otherwise, each project component should be appraised independently.

## 3.2 Typology of costs and revenues

Costs are defined as cash out-flows directly paid to build, operate, maintain, upgrade the RDI infrastructure.[22] Although costs disaggregation is project-specific, common categories of financial costs, including investment and operating costs that are generally related to RDI infrastructures, are presented in Investment and operating costs. The typical spending profile and cost distribution over time of different categories of the RDI infrastructure show a double peak, as illustrated in Illustrative spending profile and cost distribution over time of a single-sited RDI infrastructure.

---

[20] For a further discussion on the issue, see Belli et al. (2001); European Commission (2014); European Investment Bank (2013); Jenkins et al. (2011); and Florio (2014).
[21] See OECD (2014a) to further explore the topic of internationally distributed research infrastructures.
[22] In line with the economic theory, every factor of production (capital, labor, knowledge) has an associated cost. More specifically, in accounting terms, each resource used in a project, enterprise, etc. is associated with a monetary value. This monetary market value is taken into account in the financial analysis. Instead, in economic terms, each resource has an opportunity cost, which is the value of the next best economic alternative foregone due to the chosen use a determined resource. The opportunity cost of resources is considered in the economic analysis.

Box 2. Investment and operating costs

| Investment costs | Operating costs |
|---|---|
| ➡ Planning and design | ➡ Scientific, technical, and administrative personnel |
| ➡ Land acquisition | ➡ Ordinary maintenance |
| ➡ Construction of buildings and plant | ➡ Material for the operation and repair of assets |
| ➡ Construction of plant and machinery | ➡ Utilities consumption |
| ➡ Machinery and equipment purchase | ➡ Services purchased from third parties |
| ➡ Utilities consumed during the construction phase (e.g. energy and waste disposal) | ➡ Rental of machinery |
| ➡ Start-up costs | ➡ Quality control |
| ➡ Licenses acquisition | ➡ Environmental protection measures |
| ➡ Replacement costs | ➡ General management and administration |
| | ➡ Property rights |
| | ➡ Promotional campaigns and other outreach expenditures |
| | ➡ Decommissioning |

The design phase of a RDI infrastructure can be very long[23]. In addition, new facilities are sometimes developed in the same location as that of previous infrastructures and experiments, to some extent taking stock of the existing assets. Costs incurred before the start of the appraisal period, such as costs for feasibility studies undertaken at an earlier date or construction costs already sustained for a previous project, are *sunk costs* and excluded from the investment costs in an ex-ante project analysis. Similarly, *in-kind contributions*, i.e. goods, services, and staff provided in kind by external parties for the construction or operation of the project, are not considered in the financial analysis from the promoter's perspective because they do not represent actual cash flows. However, in some cases, a consolidated financial analysis across different funding or management bodies may be helpful.

Box 3. Illustrative spending profile and cost distribution over time of a single-sited RDI infrastructure

The overall spending pattern shows a relatively large investment peak during construction, a quasi-flat spending period during operation, and followed by a new peak for decommissioning. For a major upgrade during operation, another investment peak would have been visible. The distribution of costs during construction shows that civil engineering and technical hardware costs represent the vast majority of spending during the investment phase. Salaries, consumables, and utilities are the main expenditures during the operation period.

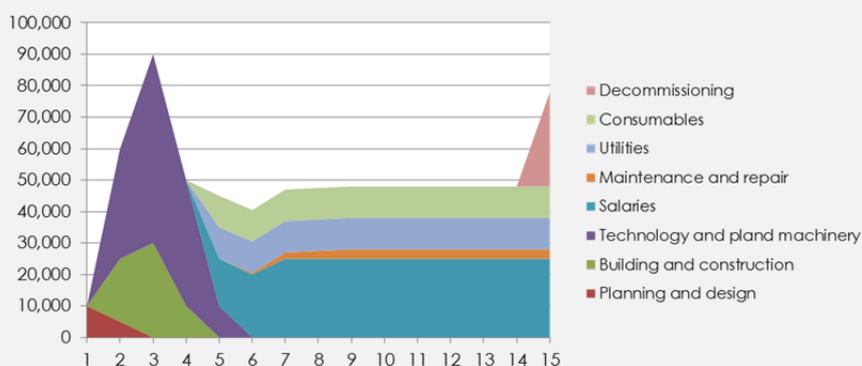

Source: Authors based on RAMIRI online handbook (http://www.ramiri-blog.eu/)

Revenues are defined as cash in-flows directly paid by project users for the services from which they benefit. Revenues can vary significantly from one project to another in relation to the specific type of services delivered by the infrastructure. Possible revenues are listed in *Errore. L'origine riferimento non è stata trovata.*.

---

[23] For instance, the Future Circular Collider study at CERN started in 2014 (see www.fcc.web.cern.ch) for an infrastructure proposed to follow the LHC around 2040.

Box 4. Revenues

- Sales of material, products, and equipment
- Sales of consultancy services
- Licence revenues gained from patent commercialisation
- Revenues from industrial research contracts and pre-commercial procurement contracts
- Research grants involving a transfer of ownership of a specific research output
- Entry fees to the laboratory and to use research equipment charged to researchers and businesses
- Rent of space
- Spin-off equity return to investor
- Student/Master/PhD fees
- Revenues from individuals using the research outputs (e.g. patients receiving innovative treatment)
- Revenues from outreach activities to the broader public (e.g. bookshop sales, entrance tickets)

The prediction of costs and revenues is an important step to prevent budgetary imbalances, which may affect the project's ability to generate the desired socio-economic effects. For instance, a technology park, for which forecasted operating revenues are inadequate to recover the initial investment and cover the expected operating costs, may risk halting its activities at a given moment. Unless other sources of financing are found, such a stoppage would imply that the desired effects on local businesses in terms of knowledge and technological progress will not materialise.

## 3.3 Dealing with inflows from research contracts and grants

Different from development and innovation activities that are expected to recoup their investments through future profits – particularly if carried out by businesses – infrastructure and experiments in fundamental research are typically supported by government funds or donations. This broad dualism involves different approaches to accounting for financial inflows. In particular, whether a financial inflow, particularly if granted by a public institution or agency, represents a source of financing or operating revenue for the project should be carefully assessed.

According to the DG Regio Guide (EC 2014, Chapter 7), public research contracts or contributions granted through either competitive or non-competitive arrangements should be considered operating revenues only if they are payments against a service directly rendered by the project promoter. Typically, this condition is verified when the ownership of the expected research output is transferred to the contracting public entity and does not remain with the RDI institution. Otherwise, these financing sources should be considered as transfers from state or regional budgets. As such, they should not be included as revenues but should account for the verification of the project's financial sustainability. This approach has consequences for the computation of financial performance indicators; see the next section.

Box 5. Operating revenues vs financing sources

| Examples of flows considered operating revenues for the project: | Examples of flows to be considered financing sources for the project: |
|---|---|
| - grant awarded by a national/regional public agency to a public research body, but in fact directed to the development/delivery of a new product/service commissioned by the agency; and, <br> - contributions paid by technology-based companies involved in co-development of equipment, software, and services to be able to use them as out-of-the-box products in the future. | - grants from European/national/regional research funding frameworks (e.g. Horizon, 2020); <br> - loans from banks or financial institutions acting as intermediaries of public bodies; <br> - regular or exceptional donations from state agencies; and, <br> - donations from charitable entities and philanthropic organisations or individuals. |

## 3.4 Financial profitability

An investment's financial profitability is the ability of a project to generate returns on the resources invested regardless of the sources of financing (loans, private equity, or grants).

The financial return on an investment is calculated using simple performance indicators, the financial net present value (FNPV), and the financial internal rate of return (FIRR). The former is expressed in monetary terms and is the discounted sum of the net financial flows for the entire time horizon. The latter is defined as the financial discount rate[24] that produces a zero FNPV.

A project with *positive financial performance* is associated with a positive FNPV, meaning that the total discounted inflows exceed the total discounted outflows. Under certain technical conditions[25], a FIRR higher that the reference financial discount rate provides the same information. Conversely, a project with *negative financial performance* is associated with a negative FNPV (and usually with a FIRR lower than the reference discount rate).

Frequently, financial indicators are used to set the correct volume of public support to be committed to welfare-improving projects (Florio, 2014), which require the contribution of public funds. For example, the European Commission allows for co-financing through grants only if the proposed major project is not financially profitable[26], i.e. the FNPV is negative and the FIRR is lower than the discount rate used for the analysis[27]. Errore. L'origine riferimento non è stata trovata. presents in anonymous form the financial performance indicators of a sample of RDI infrastructure projects co-financed by the European Commission.

Table 2. Examples of financial performance indicators of a sample of major projects co-financed by the European Commission during 2007–2013

| Country | Field | FIRR | FNPV | Reference period |
|---|---|---|---|---|
| Germany | Innovation business Incubator centre | -63.0 | -16,171,681 | 15 |
| Poland | Materials and biomaterials | 3.9 | -2,800,501 | 15 |
| Czech Republic | Laser infrastructure | -45.1 | -171,530,005 | 22 |
| Czech Republic | Biotechnology and biomedicine | -30.0 | -124,941,750 | 15 |
| Poland | Biological and chemical sciences | -3.9 | -12,349,562 | 15 |
| Lithuania | Physical and technological sciences | -12.5 | -29,878,183 | 15 |
| France | Advanced engineering materials | -33.0 | -102,161,236 | 15 |

Note: The reference financial discount rate adopted is 5% for all projects except for the Lithuanian project, which used an 8% rate.
Source: Authors based on EC Major Project database 2007–2013. Data extraction concerns a selected category of investments (i.e. RDI infrastructure and centres of competence in a specific technology), as per annex IV of EC regulation 1023 (2006).

RDI projects with potential profitability are regularly assessed by venture capitalists.[28] Clarifying the difference between CBA and the role of venture capital project analysis is useful (see

Venture capitalists' investment decision).

---

[24] This rate reflects the opportunity cost of capital from the perspective of financial investor(s) and is used to discount financial flows to estimate the investment's profitability indicators. The financial discount is valued as the loss of income from an alternative investment with a similar risk profile and is estimated by considering the return on an appropriate portfolio of financial assets lost from the best alternative investment, the real return on government bonds, or the long-term real interest rate on commercial loans.
[25] The condition is that the net benefits do not change sign during the life of the project (e.g. because of high decommissioning costs). Otherwise, more than one interest rate value may make the NPV equal to zero. Additionally, the IRR cannot be calculated when time-varying discount rates are used.
[26] We do not consider the case of financial instruments.
[27] In contrast, economic performance should be positive (see section 5.1).
[28] It must be acknowledged that, in some cases, the RDI infrastructures are involved in a transitioning process from the public sector to the private one or the venture capital arena.

Box 6. Venture capitalists' investment decisions

To screen investment opportunities, venture capitalists use a broad range of accounting and non-accounting information. Examples of information sources are business proposal, contracts with other venture capitalists, interviews with the entrepreneur, interviews with potential investors, and statistical information services (see Wright and Robbie, 1996; Manigart et al., 1997). The collection and analysis of information – the due diligence process – is needed to gain a thorough understanding of all business aspects.

The principal aspects considered by venture capitalists when looking for promising investments include[29]:

- viability of the product or service;
- potential for sustained growth of the company;
- efficient management team for efficient control and operation of the company;
- a balance between risk and expected profits;
- and justification of venture capital investment and investment criteria.

Additionally, the screening activity involves a variety of valuation techniques to determine the profitability of venture capitalists' investments, ranging from standard valuation methods based on discounted cash flow analysis (e.g. Brigham et al., 1999; Brealey and Myers, 2000) or the earnings multiple and the value of a company's assets, to the most innovative approaches based on option pricing theory (e.g. Seppä and Laamenen, 2001). The latter with respect to more traditional methods seems to better handle uncertainty. For a review, see Manigart et al. (2000), which examined the valuation methods used by venture capitals in five different countries[30].

## 3.5 Financial sustainability

A project is financially sustainable when the financial sources (including both operating revenues and any other sources of financing) are able to cover the expenditures (including investment costs, operating costs, reimbursements and interests on loans, taxes, and other disbursements) year-by-year. Hence, if the cumulated net cash is negative even for one year, the project is not financially sustainable. In this case, the project promoter is expected to demonstrate the capacity to raise additional sources of financing to cover the costs in each year of the time horizon.

In the RDI context, a number of factors influence long-term sustainability, which is only partially related to a sustainable funding profile. In particular, attracting scientific talent or operating in a field gaining scientific relevance is usually the underpinning for long-term sustainability. Thus, financial sustainability should be considered together with other sustainable criteria, as discussed in Long-term sustainability of the RDI infrastructure.

Box 7. Long-term sustainability of the RDI infrastructure

According to EIROforum (2015), the following five criteria are key to ensuring the long-term sustainability of RDI infrastructures.

- An infrastructure must be relevant to its scientific community and able to generate scientific excellence. Hence, before establishing a new RDI infrastructure, clearly defining its added value to the scientific community and its complementarity with respect to already existing facilities is regarded as essential.
- The governance model and legal framework should be sustainable. An infrastructure's managers should ensure that adequate programmes/projects are implemented to rapidly respond to the needs and ambitions of all member states/funders and to enable full exploitation of the research results.
- The funding model should be sustainable. The necessary on-going investments needed for optimal operation should be guaranteed to ensure that the infrastructure can continuously carry out its cutting-edge research activities.
- The infrastructure must attract scientific talent and develop a critical mass of scientific expertise. The ability to attract and retain talented researchers, which in turn builds scientific excellence and allows the infrastructure to maintain high standards, is closely related to the potential of the infrastructure to enable cutting-edge science.
- The infrastructure must drive major socio-economic changes and must play a crucial role in the development of society. The expected long-term changes led by a RDI infrastructure may concern two different levels: society as a whole and the immediate local environment.

---

[29] See www.capital-investment.co.uk
[30] Specifically, the United States, Great Britain, France, Belgium, and the Netherlands.

# 4. Forecasting and valuing social costs and benefits

Although financial analysis uses observed prices, for economic analysis (used synonymously with socio-economic impact, i.e. welfare analysis), such flows must be converted into shadow prices. In addition, the CBA needs to account for the positive and negative additional effects that are relevant to society but that have not been considered from the financial perspective (i.e. costs or benefits that spill over from the project towards other parties without monetary compensation)[31].

Externalities are particularly relevant for RDI infrastructure projects given the imperfect market in which they operate. By definition, knowledge creation is characterised by the fact that ex-ante information is imperfect because users literally only know that some probabilities are associated with different research outcomes when they embark on studying something unknown. Moreover, knowledge *per se* is an intangible public good and has a number of special features, namely:

- It is *non-rival*: a discovered fact does not prevent anyone else from potentially using the same knowledge; in other words, the benefits derived from knowledge may extend to mankind in general; and,

- To a certain extent, knowledge may be *non-excludable* because some knowledge cannot be patented or otherwise protected; thus, knowledge created by RDI projects is often a public good, which creates a market failure[32].

After illustrating the concept of shadow prices and their use, this section provides a detailed discussion of the social benefits associated with RDI infrastructure projects.

## 4.1 Shadow prices and main approaches for their estimation

The project's welfare impact is assessed by comparing its costs and benefits expressed as their social opportunity cost, rather than their observed market prices. Markets are typically distorted[33] and, thus, market prices are not signals of the social value of goods and are driven by different economic or political factors (Florio, 2014). Therefore, correcting for these distortions means identifying the marginal social (or shadow) value of goods, i.e. their opportunity cost to society of producing or consuming more or less of any good (see *Errore. L'origine riferimento non è stata trovata.*).

---

[31] According to standard practice, economic analysis involves three steps. First, financial costs are transformed into accounting prices using suitable conversion factors. Second, direct benefits to users are converted by replacing financial revenues with an estimation of users' WTP for project outputs less changes in supply costs. Third, the monetary valuation of externalities is added (EC, 2014).

[32] In economics, a market failure implies that the quantity of public goods demanded by consumers does not equate the quantity supplied by suppliers. This imbalance creates a case for public intervention.

[33] Distortions refer to taxes, domestic and international constraints on capital and labour flows, monopoly or oligopoly price setting, suboptimal distribution of assets and income, and information asymmetries.

Box 8.  Shadow price of experimental equipment

Consider a major RDI project that uses as input x special experimental equipment purchased in the market in which p is the supply price (marginal cost) of the equipment and q = p+d is the demand price before the project, where d is the distortion of the input price, created e.g. by an import duty. The new project is funded by a public sector grant and its effect can be seen as a shock affecting the previous equilibrium. The figure below shows the market for x. The supply curve S exhibits the quantity supplied at various supply prices, and the demand curve D shows the quantity demanded at various demand prices. By adding the distortion d to the original supply curve, a new supply curve S+d is obtained.

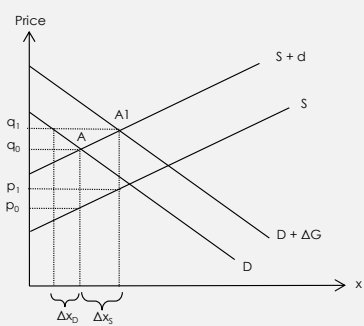

Market equilibrium occurs in A, with $p_0$ being the supply price and $q_0$ the demand price ($q_0 = p_0+d$). If the new RDI project purchases an amount ΔG from the market, the demand curve shifts right (recall that the new demand is supported by a public sector transfer), causing the demand and supply prices to rise to $q_1$ and $p_1$, respectively. Therefore, the project demand of new equipment ΔG is satisfied partially from an increase in supply ($Δx_S$) and partially from a reduction in demand ($Δx_D$) by the other users of the equipment because of the higher price. This change in the previous equilibrium caused by the project is associated with an opportunity cost. On the one side, there is an increase of suppliers' costs since they have to produce an additional amount of $x_S$: this opportunity cost can be approximated by $p_0Δx_S$. On the other side there is the reduction of users' benefits due to the foregoing purchases of $x_D$ which can be approximated by $q_0Δx_D$. The sum of these two effects gives the shadow price ($v$) of the equipment $x$ used in the project:

$$v = \frac{pΔX_S}{ΔG} - \frac{qΔX_D}{ΔG}$$

This equation shows that $v$ is different from both p and q, as it is a linear combination of both, as $\frac{ΔX_D}{ΔG} < 0$. The same formula holds if the project outputs displace the market supply and induce market demand. Finally, two special cases exist, namely:

- when the demand price is constant (i.e. the demand is infinitely elastic), the shadow price is q = p+d; and,
- when the supply price is constant (i.e. the elasticity of supply is infinite), the shadow price is p.

These simple examples ignore general equilibrium effects, i.e. consequences to other markets.

Source: Authors based on Boadway (2006).

Shadow prices can be empirically estimated using several approaches (reviewed *inter alia* by Boardman *et al.*, 2006; Brent, 2006; De Rus, 2010; Florio, 2014; Potts, 2002; and Potts, 2012a). In this section, two main approaches for their estimation are mentioned: users' marginal willingness-to-pay (WTP) and the long-run marginal social cost of production (LRMSC).

- The concept of (marginal) WTP refers to the maximum amount of money that the consumer is willing to pay to have an additional unit of a good[34]. This concept is primary used for the empirical valuation of a project's direct benefits, i.e. those related to the use of the goods or services rendered by the project, and for externalities. However, in some specific cases, WTP can also be used to proxy the opportunity cost of a project's inputs, such as land, whose use in the project leads to an adjustment in the net demand of other consumers of that good. The importance to using WTP is particularly evident for estimating externalities for which no monetary compensation is paid.

- The concept of the LRMSC of a good refers to the increase in the total cost to society as a whole, i.e. private costs plus external costs, required to increase the production of the good by one unit, keeping constant the production levels of all other goods. Typically, the LRMSC measures the economic value of non-tradable inputs, for which an increase in demand results in increased production. However, when the WTP approach is not possible or relevant, the LRMSC can be used to evaluate the output of some projects.

The example in *Box 8* shows that, in some cases, the combination of WTP and LRMSC is a proxy for the shadow price. However, the following sections treat cases in which either one or the other is appropriate, leaving the issue of combined estimation to specific applications.

---

[34] In a CBA framework, a good stands for a benefit or avoided costs. Hence, the WTP refers to the amount of money people are willing to pay to enjoy a benefit or to avoid a cost (Boadway, 2006).

## 4.2 Conversion factors for inputs other than labour

A simplified operational approach for expressing the costs of RDI infrastructures in shadow prices consists of applying suitable conversion factors to the main cost items considered in the financial analysis (e.g. materials, land, building construction, electricity) possibly retrieved from already existing benchmarks developed by the national public authorities for CBA in other fields.

A conversion factor is defined as the ratio between shadow prices and market prices. Thus, it represents the factor by which market prices have to be multiplied to obtain the shadow price[35].

Different approaches exist to calculate conversion factors. In general, if inputs are tradable goods, border prices are used[36]. Regarding non-tradable goods (i.e. procured domestically), a different approach is used depending on whether they are minor or major project items. For minor items, the standard conversion factor is adopted[37]. For major items (e.g. land, civil works, machinery, equipment), *ad hoc* assumptions should be made depending on the specific hypotheses adopted for market conditions.

For instance, consider land used for RDI infrastructures. As long as the real estate market operates under competitive conditions and no distortions occur, the financial cost of land can be assumed to be a reasonable proxy for its economic cost. However, in some RDI projects, land is provided free of cost by universities, donors, or public sector entities. Although no financial cost is included as part of the investment cost, corrections are needed to reflect the opportunity cost, which is the net benefit lost from the best possible alternative use of that land.

## 4.3 Conversion factors for labour

Special attention should be given to the opportunity cost of labour, measured using the shadow wage. The shadow wage rate is the social opportunity cost of labour and may differ from the observed wage because of distortions related to labour (e.g. unemployment, migration, taxes, minimum wages) and in the product markets. Following the opportunity cost concept, the shadow wage should reflect the social benefit of employing a person in a region/country and sector characterised by certain labour market conditions, rather than in others[38].

The application of a shadow wage to the labour cost is particularly important because it is the recommended way to capture a project's effects on employment (European Commission, 2014; Del Bo et al., 2011). Against the conventional argument (primarily referred to by politicians and project managers) that jobs created are a direct benefit of an infrastructure project (with the consequences that salaries of newly employed scientists are sometimes added as such to other economic benefits), the economic reasoning points to the consideration that wages are rather a share of the total costs of the project. Therefore, the social benefit of employment has to be taken into account by exclusively using shadow wages, i.e. by considering that the opportunity cost of employing a person in the project under assessment is lower than that from using the same person for any alternative use (including, possibly, unemployment).

According to the DG Regio Guide (European Commission, 2014), the shadow wage can be assumed:

- equal to or typically not less than the value of unemployment benefits (or other proxies when unemployment benefits do not exist) for unskilled workers previously employed in similar activities (in principle, if unemployment benefits are high, the shadow wage can be lower);

- equal to the value of the output forgone in previous informal activities for unskilled workers drawn to the project from such activities; and,

- equal or close to the market wage for skilled workers previously employed in similar activities.

---

[35] If the conversion factor for a good is higher than one, the opportunity cost of that good is higher than that captured by the market. Conversely, if the conversion factor is lower than one, then the observed price is higher than the shadow price. For instance, a conversion factor of 0.9 implies that the shadow price is 10% below the market price or that the market price is 11.1% higher than the shadow price (1/0.9 = 1.111) given taxes or other market distortions that add to the marginal social value of a good and determine a higher market price.

[36] Empirically, FOB (free on board, before insurance and freight charges) prices are retained through the best guess of the economic value of exported outputs, whereas that of imported inputs is captured by CIF (cost, insurance, and freight) prices. This approach relies on Little and Mirrlees (1974).

[37] Following Little and Mirrlees (1974), the standard conversion factor is a proxy of the average distance between world prices and domestic prices. The formula is SCF = (M+X)/(M+X+TM), where M is the total value of imports at shadow prices, i.e. CIF prices; X is the total value of exports at shadow prices, i.e. FOB prices; and TM is the total value of duties on import.

[38] The CBA literature offers different shadow wage formulae on the basis of the different hypothesis on labour and product market conditions. Recent theoretical contributions include Potts (2002); Londero (2003); de Rus (2010); and Potts (2012b). Recent empirical contributions include Honohan (1998); Saleh (2004); Picazo-Tadeo and Reig-Martinez (2005); and Del Bo et al. (2011). The latter presents a new, simple framework for the empirical computation of shadow wages at the regional level and empirical estimations for EU regions.

As a result, the conversion factor should generally be computed for unskilled workers, whereas no conversion factor is normally needed for scientists assuming an international non-distorted labour market in their case. However, when a shortage of a skilled workforce exists in a country, a suitable conversion factor must be computed even for such workers. This anomaly may lead to a conversion factor higher than 1.

In some cases, migrant workers are employed by RDI facilities. The economic cost of their labour should reflect the opportunity cost from the country of origin, not from the country in which the project is located, unless wages are the same in the two countries. When wages are higher in the country in which the infrastructure is located than in the country of origin, the appropriate conversion factor for labour is equal to the sum of the opportunity cost in the country of origin and the migration costs divided by the project wage (Asian Development Bank, 2013).

### 1. Shadow wages

The appraisal of a greenfield materials science and engineering laboratory of national relevance includes selecting its location among possible alternative sites. Region ALFA is affected by high unemployment of unskilled workers, but not of other labour force. The labour market in Region BETA is under full employment. The laboratory will hire twenty senior researchers; forty-five young researchers; thirty unskilled workers. The real market wage forecasted is EUR 50,000 per year for senior researchers, EUR 30,000 for young researcher, and EUR 24,000 for unskilled workers and is the same in both regions.

The shadow wages have been previously estimated by a national authority in compliance with government CBA guidelines :

- equal to market wage for the forty-five young researchers since their labor market is assumed to be open to international competition, with easy mobility across countries. Hence, the cumulated annual shadow labor cost for young researchers is EUR 1.35 million in both regions;
- 25% higher than market wage for the twenty senior researchers as their reference labour market is not fully competitive since for them the international mobility is hindered by linguistic barriers and high relocation costs. Hence, the cumulate annual shadow labor cost for senior researchers is (EUR 50,000 * 1.25 * 20) = EUR 1.25 million;
- 25% lower than market wage for the thirty unskilled workers due to high unemployment in region ALFA, but equal to market wage in region BETA. Hence, the cumulated annual shadow labor cost for unskilled workers in region ALFA is (EUR 24,000 * 0.75 * 30) = EUR 540,000, while it is 720,000 in region BETA. As a consequence, ceteris paribus, the Net Present Value of the project will be higher by EUR 180,000 in region ALFA.

The laboratory is expected to recur to some local enterprises as suppliers of the infrastructure during the construction phase. These firms will increase the number of unskilled employees in both regions, but only in region ALFA this fact will decrease local unemployment. This additional effect on regional employment is already captured through the adoption of the different shadow wages in the two regions. No additional benefit calculations are needed. Other employment effects are expected as the result of adoption of innovations in materials by a range of industries in the country. However, as the laboratory is of national relevance, such impact is not relevant in the selection of the location.

# 5. Social benefits

Once the main beneficiaries (either users or non-users) of an RDI infrastructure have been identified (see section 1.4), a list of typical benefits can be attached to each group. Depending on the project's nature, some of them recur for different types of target groups. For instance, the value of patents as a potential benefit may accrue to large businesses, SMEs, academics, or inventors outside academia.

This section reviews the possible approaches to forecasting the quantities of benefits over the project's time horizon and giving them an economic value.

The intensity of each benefit may be highly variable across the different typologies of RDI infrastructures. For instance, the social benefit of human capital formation is highly relevant for basic or applied research infrastructures through which students are often involved in research activities. However, this benefit is less relevant for technological development and innovation infrastructures. Only a case-by-case appraisal can determine the category of benefit that is more or less important for a specific project.

Table 3. Navigator table of typical benefits associated with RDI infrastructure projects

| Benefit | Marginal social value | Estimation method | Beneficiary target group(s) | Page |
|---|---|---|---|---|
| Development of new/improved products, services and technologies | Incremental shadow profits | Survey of business; statistical inference from company data | Businesses | 29 |
| Patents | Marginal Social value of patents | Inventors' survey; statistical inference from data on decision to renew patents or on economic terms of patent transactions; stock market valuation of market patent portfolio | Businesses, Academics; Researchers | 31 |
| Establishment of more numerous or more long-lived start-ups and spin-offs | Incremental shadow profits | Survey of start-ups and spin-offs; statistical inference from start-ups and spin-offs data; benefit transfer | Start-ups and spin-offs | 34 |
| Knowledge spillovers (not protected by patents) | Incremental shadow profits; avoided costs, willingness-to-pay for time saving | Survey of businesses; avoided cost for the production or purchase of a technology; avoided cost thanks to the exploitation of a new technology; Benefit transfer | Businesses; Professionals; Citizens; Organisations | 36 |
| Learning-by-doing benefits for the supply chain | Incremental shadow profits; avoided costs, | Survey of business; statistical inference from company data; Benefit transfer | RDI suppliers | 38 |
| Human capital formation | Incremental lifelong salary | Survey to former students; Benefit transfer | Young professionals, researcher; Students | 41 |
| Knowledge outputs and their impact | Marginal production cost | Gross salary of scientists; Value of time | Academics; Researchers | 44 |
| Provision of services | Long-run-marginal cost (or observed price) or WTP for the service | Cost incurred by the infrastructure to make the services available; Contingent valuation | Businesses; Professionals; Organisations; Government; Third research teams | 48 |
| Social benefits of RDI services for target groups | Avoided costs, Willingness-to-pay | Avoided economic cost of emissions; Opportunity cost of avoided energy sources; Avoided damage of capital stocks; Travel cost Method; Opportunity cost of land; Contingent valuation, Cost of illness; revealed preference approach; human capital approach; Benefit transfer | Businesses; Target groups of population | 49 |
| Recreational benefits for the general public | Willingness-to-pay | Travel cost method; Contingent valuation; Choice modelling; Benefit transfer | General public | 54 |
| Non-use benefits of new knowledge as a public good | Willingness-to-pay | Contingent valuation; Benefit transfer | Tax-payers | 56 |

## 5.1 Development of new/improved products, services, and technologies

The development of new/improved products, services, or technologies is the expected direct benefit of innovation infrastructures. These developments may accrue to either the RDI infrastructure itself (e.g. the research centre of a large manufacturing company that directly sells the new products on the market) or external users (e.g. users of a technological park or incubators). In minor cases, they may be side effects of fundamental or applied research infrastructures.

When a project entails the development of innovative products, services, and technologies, the social value of these goods is expressed using the *incremental shadow profits* expected from their sale. In particular: 'incremental' means that profits expected from the sale of new/improved products, services, and technologies generated by the project must be compared with the profits in the without-the-project scenario; and, 'shadow' means that market distortions should be duly considered; for instance, the shadow profit is higher than the gross financial profit if the infrastructure is located in an area of high unemployment.

Given that $i$ is the number of innovations (products, services, and technologies) over time t, $\mathbb{E}(\Pi_{it})$ represents the expected incremental profits directly imputable to these innovations and $s_t$ represents the discount factor. Then, the expected present value of developing new/improved products, services, and technologies ($Z$) is expressed as:

$$\mathbb{E}(Z) = \sum_{i=1}^{I} \sum_{t=0}^{T} s_t \cdot \mathbb{E}(\Pi_{it}).$$

*Empirics*

The ex-ante estimation of these benefits involves the following calculations.

- The benefits must be quantified by forecasting the demand for new/improved products/services/technologies over time. This forecast depends on the project's objective and, as best as possible, should rely on benchmarking with similar RDI infrastructures and interviews with experts in the considered sector;

- The marginal value of new/improved products/services/technologies should be estimated. However, if the expected profit from the new/improved products/services/technologies has been included among the project's financial revenues (i.e. when the new/improved products/service/technology is directly sold by the infrastructure), the economic estimation of their value should not be included in the economic analysis, provided that a suitable conversion factor is used to convert the financial flows from direct commercial exploitation of innovations.

The following different possible approaches exist to predicting expected profits.

- Information on profitability, average costs, and sales can be retrieved from databases in the public domain or may be granted by data providers. Typically, earnings before interest, taxes, depreciation, and amortisation (EBITDA) can be used to proxy companies' profits because interest and taxes are effectively transfers between agents and depreciation is inconsistent with the discounted cash flow approach that supports the computation of the NPV. For instance, the Amadeus Database, maintained by the consultancy Bureau van Dijk[39], which provides balance sheet data reported to national registries and statistical offices by European companies, has been used by Florio *et al.* (2015) to calculate average sector-specific values for profitability up to the four-digit NACE level.

- Benchmarking with similar RDI infrastructures in other contexts could also offer some inputs to forecasting future profits, and access to systematic project datasets at national or supranational levels may be helpful.

- Interviews with experts in the sector can assist in conjecturing on the possible changes in the profitability of businesses under different scenarios.

For the purpose of exemplification, the ten-year average EBITDA margins associated with companies whose primary activity falls within a selected list of NACE codes[40] are presented in Table 4 and are broken down by country (Italy, France, Germany, and the United Kingdom). Data were gathered from the ORBIS world database of companies' financial information[41] and refer to sectors often involved in the procurement of major RDI infrastructures.

---

[39] The database is available at www.bvdinfo.com/en-gb/home.
[40] Manufacture of basic metals (24), manufacture of computer, electronic, and optical products (26), manufacture of electrical equipment (27), manufacture of machinery and equipment n.e.c. (28), telecommunications (61), computer programming, consultancy, and related activities (62).
[41] The ORBIS database is maintained by the consultancy Bureau van Dijk.

Table 4. Ten-year (2004–2013) average of companies' median EBITDA margin (%) by sector and country

| Industry (NACE sector) | NACE Code | Italy | France | Germany | United Kingdom |
|---|---|---|---|---|---|
| Manufacture of basic metals | 24 | 7.6 | 15.3 | 7.1 | 35.0 |
| Manuf. of computer, electronic and optical products | 26 | 11.2 | 8.3 | 11.7 | 14.4 |
| Manuf. of electrical equipment | 27 | 10.3 | 16.4 | 11.7 | 11.2 |
| Manuf. of machinery and equipment n.e.c. | 28 | 13.1 | 10.3 | 9.8 | 17.6 |
| Telecommunications | 61 | 40.1 | 13.8 | 11.3 | 10.0 |
| Computer programming, consultancy and related activities | 62 | 11.3 | 15.3 | 8.0 | 8.3 |

Source: Authors' elaborations based on ORBIS database

### 2. Shadow profits of high tech firms

A technology park in country GAMMA is expected to support 45 new enterprises in the red biotechnology field. These enterprises will benefit from the use of testing and prototyping shared laboratories of the infrastructure. These activities will eventually lead to the development of new marketable products. The time horizon of the project is 15 years. According to sector analyses available, the red biotech sector has the following features:

- young, small-cap biotech companies may have low or negative earnings for extended periods because they face high R&D costs throughout the lengthy process of bringing their first product to market;
- the expected annual shadow profit is highly variable. Indeed, some biotech enterprises have no hopes of ever making money, while others which have products already established in the marketplace are quite profitable.

The baseline forecast is that:

- the yearly average amount of revenues per company is zero for the first two years and then real EUR 5 million;
- the yearly salaries for unskilled workers amount to EUR 0.5 million, while those for skilled workers amount to 0.6 million;
- the conversion factor for unskilled labour is estimated at 0.8 due to unemployment in the region, while that for skilled labour is estimated equal to 1;
- the yearly average cost of rents and utilities is EUR 0.3 million and EUR 0.2 million, respectively;
- the yearly average production cost is EUR 1 million;
- the conversion factors for rents, utilities and production costs is assumed equal to one.

The baseline annual shadow profit per each company since year 3 is then:

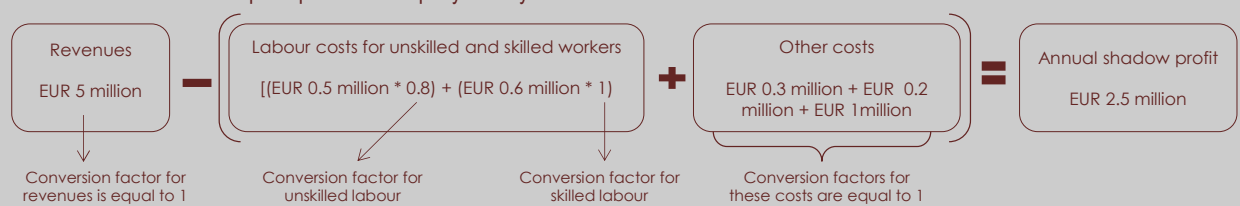

Given the high uncertainty about the shadow profit, the project promoter (based on previous cases) assumes that the shadow profit may take any value between EUR 1.5 million and EUR 3.5 million per year. Thus, a rectangular distribution is hypothesised. This distribution will feed in to the Montecarlo simulation of the ENPV.

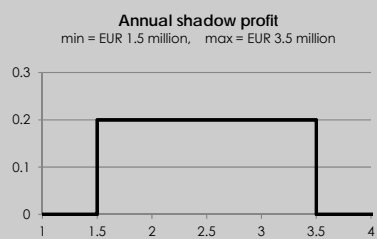

Annual shadow profit
min = EUR 1.5 million, max = EUR 3.5 million

## 5.2 Patents

When a patent is registered, it produces a private return to the inventor and potential knowledge spillover to society. Indeed, a public document is issued containing information on various aspects of the invention, including citations to existing patents. When approved this document grants the inventor an exclusive right for the commercial use of the patented invention for a pre-determined period and serves to delimit the scope of the property right granted to the patent owner. Thus, the cited patents represent the previously existing knowledge on which the citing patent builds, and over which the citing patent cannot have a claim (Jaffe, Trajtenberg, and Henderson, 1993; Deng, 2005).

The fact that patent citations reveal 'prior art' that an inventor has learned makes them potential measures of the knowledge spillovers from past inventions to the current invention. In other words, citations of a patent by many subsequent patents suggest that the patent generated significant technological spillovers because numerous developments build on the knowledge that it embodies. Thus, patent citations have become a broadly used proxy for estimating the social value of patented technologies (see, for instance, Trajtenberg et al., 1997; Caballero and Jaffe, 1993; and Jaffe and Trajtenberg, 1999).

In a CBA framework, both the private returns and the knowledge spillovers brought about by patents granted by a RDI infrastructure represent a benefit that should be considered. Therefore, the *marginal social value of the patent generated by a RDI infrastructure* should be forecasted, provided that double counting from the change in the expected profit from the sale of innovative products is avoided (i.e. when they are appropriated directly by the RDI). Given $i$ as the number of patents over time t, $v_{(pvit,exit)}$ as the patent marginal social value, and $s_t$ as the discount factor, the expected present value of this benefit is expressed as:

$$\mathbb{E}(P) = \sum_{i=1}^{I} \sum_{t=0}^{T} s_t \cdot \mathbb{E}(MSV_{(pvit,exit)}),$$

where the marginal social value ($MSV$) includes both the private value ($pv_{it}$) and the externality ($ex_{it}$), i.e. the knowledge spillover brought about by patents granted by a RDI infrastructure.

*Empirics*

The ex-ante estimation of this benefit involves the following activities.

- The number of patents that will be registered by the RDI project over time is forecasted. This activity involves either using a promoter's track record on patenting or referring to observational data related to similar infrastructures, if available. Alternatively, considering shortcuts may be useful, including the correlation between the existing statistics on the number of patents granted and the number of R&D personnel in a given area/industry/domain.

- The average rate of usage of granted patents ($use$) is forecasted. This activity involves, again, either using a promoter's history on patenting or referring to observational data related to citations of patents issued in the same scientific field or in similar infrastructures, if available. The average rate of patent usage, proxied by the median number of lifetime forward citations[42] per patent, is important to understanding the actual rate of exploitation and, in turn, the knowledge spillovers resulting from patents granted by a RDI infrastructure.

- The average number of references ($ref$), i.e. backward citations[43] to existing patents, is forecasted, which is typically included in patents issued in the relevant technological field.

- The marginal private value of patents is estimated, and double counting given the change in expected profits from the sale of innovations (if they are directly appropriated by the RI) is carefully avoided. In fact, depending on the estimating method used, the value of a patent may or may not already include the market value of the patented invention. In principle, the patent value should be based on the discounted sum of the yearly profits that the patent holder expects to earn because of the patent, net of the equivalent discount stream of profits without the patent (European Commission, 2006).

- The externality of patents in monetary terms is estimated. As mentioned, patent citations mirror the technological importance of a patent for the development of subsequent technologies (Squicciarini *et al.*, 2013). In other words, a citation is a measure of the knowledge spillovers from past inventions to the current invention. However, simply counting citations does not provide information on a patent's value in monetary terms. To attach a monetary value to the stream of citations produced by a patent, the following formula can be used:

$$\mathbb{E}(ex_{it}) = \sum_{i=1}^{I} \sum_{t=0}^{T} use * \frac{\mathbb{E}(pv_{it})}{ref},$$

where $use$ is the average rate of usage of granted patents, $ref$ is the average number of references included in patents issued in the relevant technological field, and $pv_{it}$ is the private value of patents granted in the relevant technological field.

---

[42] To understand the relationship between 'backward citation' (i.e. patents cited by a new patent) and 'forward citation' (i.e. number of times a patent has subsequently been cited), consider the following example. If Patent A (2005) is cited by Patent B (2015) then Patent A is a backward citation of Patent B, whereas Patent B is a forward citation of Patent A. Most search databases allow both backward and forward citation searching.
[43] Ibid.

Figure 2. The social value of patents

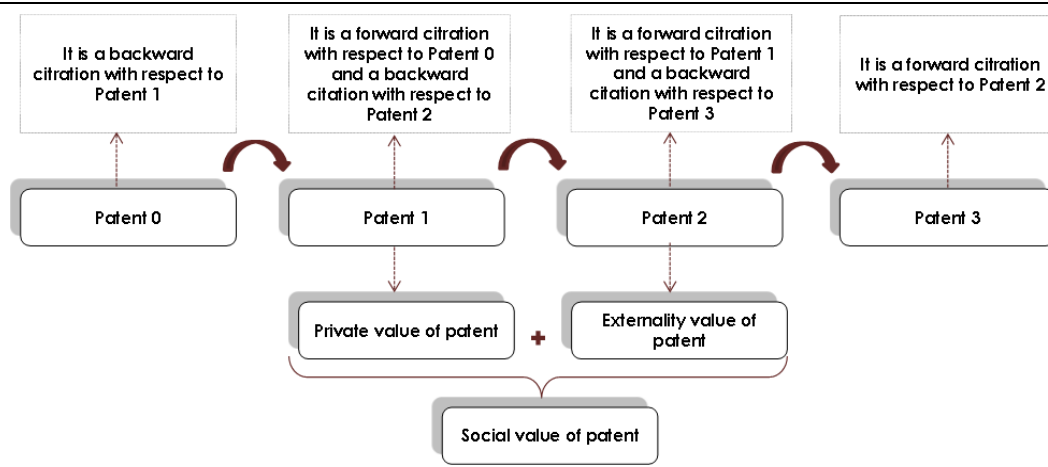

Source: Authors

A range of possible information and statistics (including patents and citations counts) useful for forecasting the number of project patents can be retrieved from several repositories[44]. However, importantly, note that only patents granted by patent offices generate both a private value and knowledge spillover. Instead, patent applications or 'invention disclosures' which are not granted have a private value of zero, while their externalities could be positive. When patents granted statistics are not available, an assumption is made about the number of patents that eventually are registered.

Concerning the estimation of the private value of patents, different empirical approaches can be discussed (see

Measurement of patent private value: overview of different approaches).

Box 9. Measurement of patent private value: overview of different approaches

> The concept of private value takes into account only the value added of the patent for its holder. Thus, private value can be defined as the depreciated sum of the expected cash flows of the owning entity from the patent. The following three main lines of work have been followed by researchers to estimate or infer the private economic value of patents.
> 
> ➡ Estimates based on a patent-holder's behaviour include methods that analyse either the decisions to renew (or not) patents and pay the related fees (see Pakes, 1986; Schankerman and Pakes, 1986; Schankerman, 1998; Lanjouw, 1998; and Bessen, 2006) or the economic terms of actual patent transactions (see Serrano, 2008; Sneed and Johnson, 2009; Leone and Oriani, 2008; and Sakakibara, 2010).
> 
> ➡ Estimates based on inventors' surveys involve directly asking the inventor to provide an estimate of the value of his/her patents on the basis of the price at which he/she would be willing to sell the patent (see Harhoff et al., 1999; Harhoff et al., 2003a; Harhoff et al., 2003b; and Gambardella et al., 2008).
> 
> ➡ Estimates based on external investors' valuations include methods either based on stock market valuations of patent portfolios of publicly listed companies (see Griliches, 1981; Cockburn and Griliches, 1988; Hall et al., 2005; and Hall et al., 2007) or valuations made by venture capital firms of intellectual property-based start-up companies (see Lerner, 1994 and Hsu and Ziedonis; 2008).

At the European level, a reference study was published by the European Commission in 2006[45]. The analysis relies on a questionnaire survey of almost 10,000 inventors in eight European countries[46]. Patents belonging to different technology classes were considered. To obtain a measure of patent value, inventors were asked to provide their best estimate of the value of their patents on the basis of the price at which he/she would be willing to sell his/her patent. This study found that the value of European patents is typically between EUR 100,000 and EUR 300,000, with a small share of patents yielding economic returns higher than EUR 3 million and an even smaller share valued at more than EUR 10 million. Thus, on average a patent in the considered EU-8 countries is worth approximately EUR 3 million. However, because the distribution of patent values is very skewed, the median patent is worth EUR 300,000.

---
[44] For instance, the EPO Worldwide Patent Statistical Database – also known as EPO PATSTAT – the Eurostat statistics available under the 'Science & Technology' statistics; the Trilateral co-operation website, which provides statistics from the EPO, JPO, and USPTO dating back to 1996 in the annual Trilateral Statistical Reports; the reports provide an overview of worldwide patenting activities. The WIPO website provides patent and Patent Cooperation Treaty statistics; the OECD's patent indicators reflect trends in innovative activity across a broad range of OECD and non-OECD countries, with six main sections: EPO, USPTO, and JPO patent families; patenting at the national, regional, and international level; patenting in selected technology areas; patents by institutional sector; international co-operation in patenting; and European and international patent citations.
[45] See European Commission (2006).
[46] The considered countries are Denmark, France, Germany, Hungary, Italy, the Netherlands, Spain, and the United Kingdom

Table 5. Average patent values by country and technological area

| Country | Average patent value (EUR thousands) | Median patent value (EUR thousands) | Technological area | Average patent value (EUR thousands) | Median patent value (EUR thousands) |
|---|---|---|---|---|---|
| Denmark | 2,947 | 300 | Pharmaceuticals, cosmetics | 5,260 | 605 |
| Germany | 2,958 | 305 | Macromolecular chemistry, polymers | 3,980 | 449 |
| Spain | 3,029 | 307 | Space technology weapons | 3,854 | 414 |
| France | 2,922 | 293 | Environmental technology | 3,250 | 354 |
| Hungary | 3,647 | 408 | Biotechnology | 3,134 | 336 |
| Italy | 3,007 | 297 | Semiconductor | 2,555 | 284 |
| The Netherland | 2,788 | 285 | Telecommunications | 2,331 | 247 |
| United Kingdom | 3,355 | 332 | Electrical devices, engineering, energy | 1,938 | 211 |

Source: European Commission (2006).

Because patent values are acknowledged to vary significantly across sectors, technological fields, and geographic areas, considering country/region, sector, and technology-specific statistics when available is useful.

## 3. Estimating the social value of patents

Region DELTA is a hub of specialty and fine chemicals industry. According to regional statistics, one European patent for every 50 researchers was granted every year in the past decade. A consortium of two universities and of five companies has started a feasibility study for a new research infrastructure in molecular chemistry. The infrastructure is envisaged to employ 150 researchers for the entire time horizon. The baseline forecast of the annual average number of patents expected is 3, conservatively based on the past track record in the region.

According to data retrieved from a meta-analysis carried out by the academics of the two universities, the median market value of patents in the field of molecular chemistry is EUR 400,000. Moreover, according to statistics on patent citations retrieved from databases maintained by similar facilities, the median number of backward and forward citations in the molecular chemistry filed is respectively 10 and 15. Assuming that the externalities are linear in the number of citations, the new patent benefits from ten previous discoveries, and will benefit fifteen future ones. Hence, the yearly social value of patents would be EUR 3 million, as follows:

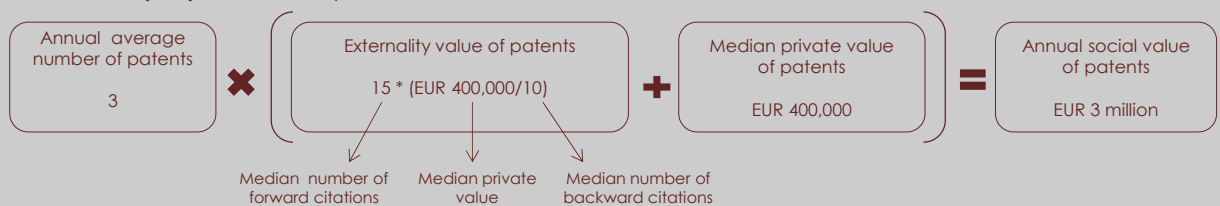

Since the number and the value of patents are highly uncertain ex-ante, probability distributions are considered instead of punctual values. The annual number of patents according to interviewed experts can have a discrete probability distribution taking value of 2,3 or 4, while the value of a patent can have a normal distribution with mean EUR 400,000 and standard deviation of EUR 150,000[47]. Also the ratio between cited and citing patents is uncertain, and a triangular distribution (minimum value = 0.5; modal value = 1.5; and maximum value = 2.5) is assumed. Using a Monte Carlo simulation technique these distribution assumptions can be combined and the conditional expected total benefit of patents estimated.

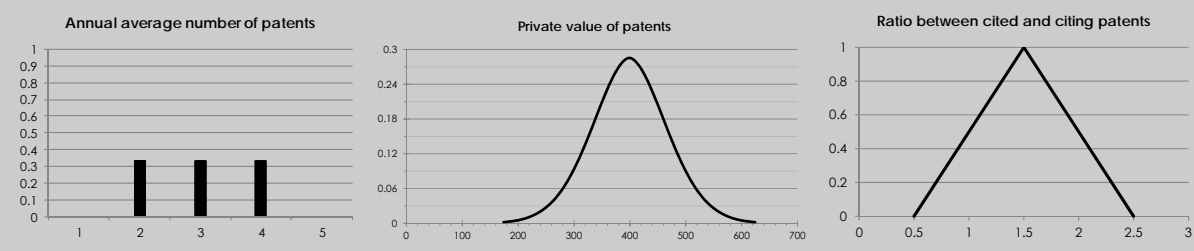

# 5.3 Start-ups and spin-offs

The establishment of start-ups and spin-offs can be one of the intended objectives of innovation infrastructures, as it is for incubator centres. However, this establishment can also be a side effect of fundamental and applied research infrastructures, such as university laboratories. Whatever their origin, the objective of start-ups and spin-offs is ultimately to develop and commercialise new products, services, and technologies.

The benefits produced by a RDI infrastructure to start-ups and spin-offs can be related to either the establishment of new firms or (and) an increase in the survival rate. If the RDI infrastructure contributes to the establishment of start-ups and spin-offs, the economic value of this benefit is valued as the *expected shadow profit* gained by the created business during its overall expected lifetime compared with the without-the-project scenario. Whereas the equity return to investors (e.g. business incubator) and the operating revenues from the sale of consultancy services leading to the establishment of, for example, start-ups, are considered among inflows in the financial analysis, they do not enter in the economic analysis to avoid double counting the considered benefit.

When the RDI infrastructure contributes to increasing the survival rate of start-ups, then the benefit is valued as the incremental expected shadow profit attained by businesses that survive longer than in the without-the-project scenario.

*Empirics*

The ex-ante estimation of this benefit involves:

- Forecasting the *number* of start-ups and/or spin-offs expected to be created by the infrastructure during the entire reference period;

---

[47] It is worth noting that, in general, we suggest to relay on median values instead of average ones when choosing the baseline values in the deterministic model. However, in a probabilistic model, probability distributions should be considered instead of punctual value. Therefore, a probability distribution (characterized by its typical parameters, e.g. mean and standard deviation for normal distribution) should be guessed around this median value.

- Establishing the *expected lifetime and survival rate* of start-ups and spin-offs (when the infrastructure contributes to increasing the life expectancy of start-ups, the expected increase in their survival rate must be estimated); and,
- Estimating the expected *profit* generated by start-up and spin-offs created by the RDI infrastructure.

The median number of start-ups and/or spin-offs created by RDI infrastructures in specific countries and sectors, and their expected lifetimes and survival rates, can be proxied by looking at similar infrastructures in other contexts or retrieved from official statistical database or the literature.[48]

Figure 3. Example of average survival rates in different countries for all sectors of industry, construction, and services except insurance activities of holding companies (*).

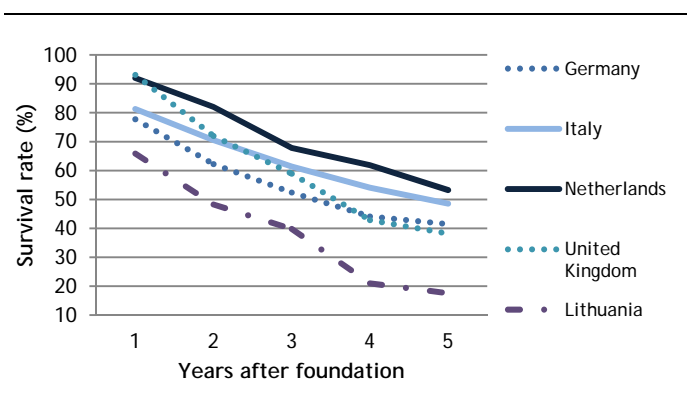

Table 6. Example of survival rates of university spin-offs in various countries and two universities

| Country | Survival rate | Years | Source |
|---|---|---|---|
| Netherlands | 83% | 9 | Shane, 2004 |
| France | 84% | 4 | Mustar, 1997 |
| Sweden | 87% | 34 | Shane, 2004 |
| N. Ireland | 94% | 12 | Shane, 2004 |
| UK – Oxford | 81% | 9 | Lawton Smith and Ho, 2006 |
| ETH – Zurich | 88% | 10 | Oskarsson and Schläpfer, 2008 |

Source: Authors adapted from Oskarsson and Schläpfer (2008)

(*)codes: B-S_X_K642) retrieved from Eurostat business demography statistics (2012)

Concerning the expected profit, the considerations presented in section 4.1 remain valid.

---

[48] Some possible sources of information include Eurostat business demography statistics (2012); Innovation Union Competitiveness report (2011 and 2013); and the European investment Bank (2013).

### 4. The social value of start-ups supported by an incubator

A specialised incubation park in the electronics engineering industry is aimed at supporting the creation of high-tech start-ups. The project time horizon is 15 years, under a regional development policy instrument. According to the initial feasibility study:

- the park is expected to support on average 5 new enterprises per year and on average each firm remains in the park 3 years, meaning:

| Year | 1 | 2 | 3 | 4 | 5 | ... | 14 | 15 |
|---|---|---|---|---|---|---|---|---|
| umulate number of start ups | 5 | 10 | 15 | 15 | 15 | ... | 15 | 15 |

- based on statistics in similar contexts, the average survival rate is 70% after 5 years, 40% after 10 years, 10% after 15 years and 0% after 20 years in both the with- and without-the project scenarios;
- the average shadow profit of the assisted firms for the first three years will be zero; it then increases to EUR 0.5 million per year;

Hence, the curve of the cumulated profit takes the following form.

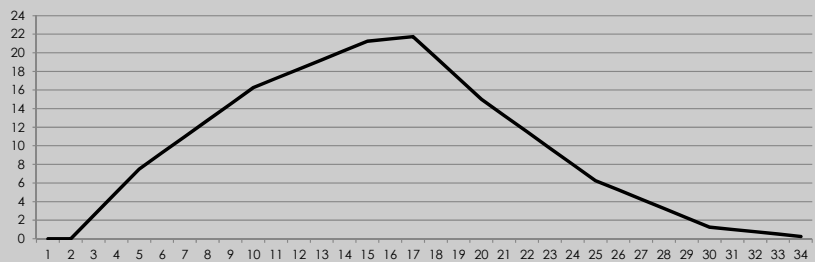

In a deterministic model, the total discounted social benefit is the sum of the shadow profits gained by all the enterprises created thanks to the incubation park during the firms overall expected lifetime provided that the survival rate is 70% after 5 years, 40% after 10 years, 10% after 15 years and 0% after 20 years.

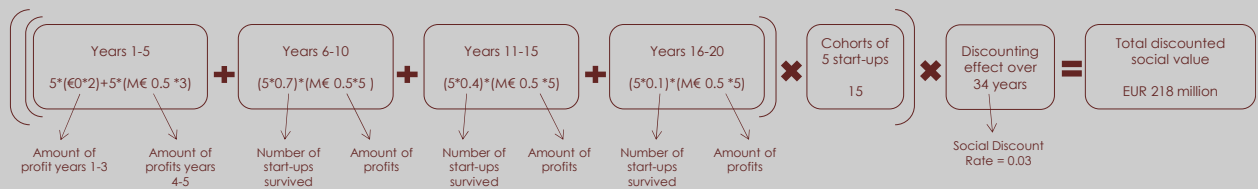

## 5.4 Knowledge spillovers

The RDI infrastructure can produce knowledge spillovers[49] to third parties (businesses, citizens, professionals, public organisations). For instance, open access and open data practices empower everyone to access and re-use, free of charge, results and data from publicly funded research. When the RDI infrastructure produces new knowledge or develops new technologies or products and releases them for free (or at a very low nominal price) and without any form of intellectual property protection, the benefit gained by third parties can be valued using alternative approaches (or a combination of them provided that double counting is carefully avoided) depending on the category of beneficiaries.

- Users are likely to accumulate *incremental shadow profit* from exploiting the knowledge or employing the technology. Given $i$ as the number of entities benefitting from knowledge spillovers over time t, $\mathbb{E}(\Pi_{it})$ as their expected incremental profits directly imputable to the spillover effect, and $s_t$ as the discount factor, the expected present value of technological externalities $\mathbb{E}(F)$ is expressed as:

$$\mathbb{E}(F) = \sum_{i=1}^{I} \sum_{t=0}^{T} s_t \cdot \mathbb{E}(\Pi_{it}).$$

- Users avoid certain *costs* given the exploitation/application of the new knowledge/technology made available for free by the RDI infrastructure. Indeed, in some cases, estimating the costs that no longer need to be sustained for doing something instead of the incremental shadow profit could be more practical. For instance, an innovative combustion technology developed by a research infrastructure offering open access to its research results is considered. The in-novation can be exploited by businesses to improve their own production processes, thereby significantly reducing their energy costs. These avoided costs represent the value of knowledge spillover for businesses.

---

[49] For a theoretical discussion on knowledge spillover, see Mansfield et al. (1977); Griliches (1979); and Hall et al. (2009).

- Users *avoid production costs (or the market cost)* given the transferred knowledge made available by the RDI infrastructure. This phenomenon refers to the costs that no longer need to be sustained to produce the knowledge that has been made available for free (or at a very low price) by the RDI infrastructure (or to purchase the same knowledge on the market). For instance, consider new numerical simulation software developed within a research infrastructure and made available free of charge to other research institutes. These institutes can freely use the software instead of producing it or purchasing it (or similar software) on the market, thereby creating cost savings.

- A group of beneficiaries, such as citizens or professionals, are likely to benefit from the *willingness-to-pay for time saving* given that the technologies or products are released for free[50]. As an example of this case is free online software that makes a type of data storage or transmission for professionals easier and more powerful.

*Empirics*

One way to forecast the possible size of knowledge spillovers of the RDI infrastructure under assessment is to take an already existing similar facility as a benchmark and rely, as far as possible, on the opinion and expectations of experts of the similarity or dissimilarity of technological patterns.

More specifically, depending on the approach followed, the ex-ante estimation of this benefit involves different forecasts. If the incremental shadow profit approach is chosen, the methodology presented in the previous sections applies here as well. Conversely, if the avoided costs approach is preferred, the ex-ante estimation of this benefit requires:

- Forecasting the number of potential beneficiaries affected by the new knowledge or the new/improved technology over time. It should be acknowledged that sometimes innovative products and services are produced for existing but unsatisfied demand (latent demand). In other cases, only potential demand exists. Clearly, the potential demand for a good which does not yet exist should rely on appropriate forecasting methods which may consider among others the potential target users, existing less innovative substitute products or services, similar experiences;

- Estimating the overall cost associated with the production/development of the knowledge/technology and (if relevant) the overall costs avoided given the exploitation/application of the new technology made available for free by the RDI infrastructure (provided that the incremental shadow profit for the same benefit has not been already included). Innovative products (by definition) have no existing market price, however, in some cases it may be possible to determine likely prices by looking at the price of competing, although less innovative, products.

Finally, if the willingness-to-pay approach is considered more suitable, the ex-ante estimation of the benefit requires:

- Forecasting the time saving from the new new/improved technology/products; and,

- Estimating the economic value of time saved; a large body of literature exists on this point.

---

[50] For time savings, see Hensher (1997); Bates and Whelan (2001); Hensher and Goodwin (2004); Antoniou and Matsoukis (2007); and London Economics, (2013).

## 5. Benefits from knowledge spillover to third-parties

After three years since its opening, a university research centre in life sciences (taking 3 years to be constructed) is expected to develop a big data multivariate analysis software called ETA as part of a broader scientific program. The centre will be funded by a government grant provided that the software will be released open source. The potential beneficiaries in the scientific domain are around 10,000 scientists but there are also other 10,000 professional users, most of which private companies in different industries. The socio-economic benefit associated to this software is estimated as the avoided cost for the purchase of an equivalent commercial software. Among the software available on the market, there are "Tool 1", available at an annual licence of EUR 11,000 per computer, and "Tool 2", available at EUR 5,000 per individual and commercial users, but needing adaptation costing 4,000. It is expected that each software is going to become obsolete in 5 years. The discounted total benefit is then:

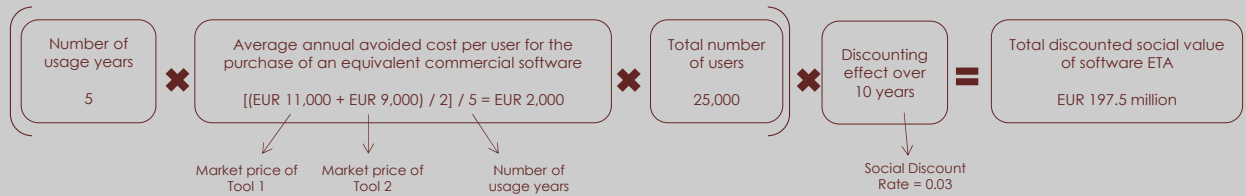

We assume that the displacement effect for the private sector business is negligible (otherwise it should be considered the net effect of positive and negative externalities of the open source release).

Another software ZETA will be developed by the research centre at a total cost of EUR 20 million (calculated as cumulated and discounted time of researchers) after 6 years since its opening. ZETA will be made freely available to the reference scientific community. However, unlike the previous tool, no equivalent commercial software is or will be available on the market. In this case, the benefit associated to the software can be estimated as the avoided cost for all user-entities for developing a new software, equivalent to that made available for free. Based on a benchmarking exercise, the number of institutes, agencies and companies willing to use the software (and with the capacity to develop a similar one in the counterfactual scenario) is 30. The discounted total benefit is then:

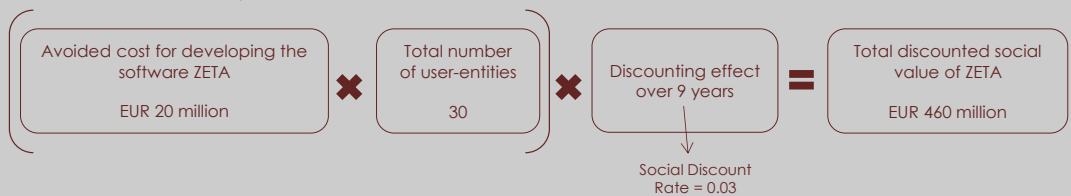

Given the uncertainty associated to forecast, distribution probabilities need to be assigned to each variable. In particular:

- For software ETA, the average annual avoided cost per users for the purchase of an equivalent commercial software is estimated by experts having a triangular probability with mode equal to EUR 2,000 per year, lower bound of EUR 1,000 and upper bound of EUR 3,000.
- For software ZETA the variable 'number of user-entities interested in using the free software' is hypothesised having a uniform probability ranging between a minimum of 50 and a maximum of 120.

## 5.5 Learning-by-doing benefits for the supply chain

High-tech suppliers involved in the design, construction, or operation of infrastructures at the forefront of science or technology can enjoy spillovers from working with/for the RDI infrastructure[51]. Indeed, the firms involved in the supply chain of a RDI infrastructure typically face the challenge of providing non-off-the-shelf industrial solutions to a number of complex technological questions. On the one hand, this situation gives firms the opportunity to collaborate with the scientific and technical staff of the infrastructure and, in turn, to acquire new knowledge and technological skills. On the other hand, the suppliers are incentivised to expand beyond their current state of knowledge. The learning-by-doing benefit of suppliers can yield to different types of developments, ranging from improvements to already existing equipment or manufacturing processes to the invention of new tools that may find applications in other areas of science, services, or industry.

The first attempts to estimate the economic benefit to firms producing equipment for a RDI infrastructure were probably made by the European Organisation for Nuclear Research (CERN)[52]. However, these studies, which typically focus on estimating quantitatively the average 'economic utility'[53] to supplier firms, implicitly assume that the value of learning-by-doing is increased sales and decreased costs. In a CBA framework, the net benefits should be considered and, for this reason, revenues must be considered net of production costs (i.e. profits) and from an incremental perspective.

---

[51] To explore the social returns to R&D, see Bernstein and Nadiri (1991); David et al. (1992); Hertzfeld (1998); and Hall (2009).
[52] See Schmied (1975); Schmied (1982); and Bianchi-Streit et al. (1984). Also in the seventies, first attempts to measure the benefits from the NASA R&D programs were performed (e.g. Midwest Research Institute, 1971). However, these studies usually estimated the productivity changes in the national economy and were not based on production function or cost benefit approaches.
[53] Economic utility was defined as the sum of the increased turnover and cost savings arising directly from the contract, but excluding the value of the contract itself.

The social value of learning-by-doing should be evaluated through the *incremental shadow profit* expected by supplier companies[54]. In principle, this increase in profits should be assessed against a counterfactual group of companies operating in the same sector and sharing other characteristics with the companies that actually worked for the infrastructure. A practical way to value ex ante the incremental increase in profits consists of using a 'benefit transfer' approach[55], exploiting the results of an ex post survey of companies within and outside the supply chain of similar infrastructures. Alternatively, the benefit can be valued as the *cost avoided* given the application of the new knowledge and experience obtained for free as a spillover of the procurement contract (for this concept, refer to the previous section).

*Empirics*

Following the incremental shadow profit approach, the ex-ante estimation of the benefit involves the activities described below.

- The volume of procurement contracts that is likely to generate technological externalities are forecasted. Learning benefits are expected to occur when the procurement contract is for the provision of products that satisfy new technical requirements, usually customised for the infrastructure purpose. Therefore, orders regarding off-the-catalogue products, i.e. items produced for the market and that do not need substantial adaptation for being used, do not entail any spillover effect to suppliers. From an ex ante point of view, determining the technological opportunities opened up by working for the RDI for further profitable investments is a difficult endeavour. Indeed, the potential exploitation opportunities deriving from the development of a new technology might not be evident at the beginning; therefore, immediately identifying all of the technological externalities that might appear in the next years or decades is reasonably impossible. To avoid risking optimism bias, forecasting the possible size of the learning-by-doing benefit by relying, as much as possible, on already existing similar RDI infrastructures as a benchmark and on the opinion and expectations of independent experts is helpful.

- A sales multiplier is estimated to elaborate on the procurement likely to generate learning-by-doing benefits as increased turnover (or decreased costs). For instance, using a multiplier of 3 (as suggested in Bianchi-Streit et al., 1984) indicates that, for every Euro in a procurement contract, a supplier company receive 3 Euros in the form of increased turnover or cost savings. The following table presents the results of different studies aimed at estimating the 'economic utility' ratio in the field of RDI.

Table 7. Economic utility' ratios in the literature

| Average values | Organisation | Method | Source |
|---|---|---|---|
| 3 | CERN | Survey of firms | Schmied (1975); |
| 1.2 | CERN | Survey | Schmied (1982); |
| 3 | CERN | Survey | Bianchi-Streit et al. (1984) |
| 3 | ESA | Survey of firms | Brendle et al. (1980) and Bach et al. (1988) |
| 1.5-1.6 | ESA | Survey | Schmied (1982); |
| 4.5 | ESA | Survey | Danish Agency for Science (2008) |
| 2.1 | NASA Space Programmes | Input-Output model | Bezdek and Wendling (1992) |
| 2-2.7 | INFN | Input-Output model | Salina (2006) |
| 3.03 | John Innes Centre | Input-Output model | DTZ (2009) |

Source: authors based on cited sources.

---

[54] According to Florio (2014), we maintain that the change in sales does not need to be considered, but instead the change in *net* output (i.e. profit) at shadow prices.
[55] The benefit transfer approach refers to the process of extrapolating the results of existing primary studies (i.e. surveys or other ad-hoc analyses) and transferring them to different populations and contexts. In other words, when a parameter has been previously estimated for a similar project in a different context (e.g. different country, different region), it can be used in another analysis after proper adjustment to take into account technical, socio-economic, geographic, and temporal specificities of the project under evaluation. On the benefit transfer method, see for instance Pearce et al. (2006), the Asian Development Bank (2013), and Florio (2014).

- A profitability measure (e.g. the EBITDA margin) is estimated to multiply the turnover previously calculated. This step is key because it allows for the consideration of increased profits rather than simply increased sales or decreased costs as a benefit for supplier firms.

### 6. Social value of learning-by-doing

Thirty supplier firms are involved in the provision of customised high-tech items for the construction of a research space satellite for an inter-governmental body. These firms are potentially beneficiaries of technological and knowledge spillovers because the items will be co-designed with the client. Based on forecasted production cost and according to a benchmark analysis to similar infrastructure elsewhere, the following hypotheses holds:

- each firm will be involved in the procurement for 1 year and, on average, the co-design of high-tech items will take 500 working hours. Specifically, the breakdown of suppliers' involvement over the construction period (lasting 4 years) is: five firms in the first year, ten firms respectively the second and the third years, and five firms in the fourth year.
- each firm's volume of procurement potentially associated with learning-by-doing benefit has a uniform probability distribution ranging from EUR 2 million to EUR 4 million (the baseline value is EUR 3 million);
- the sales multiplier has a uniform probability distribution ranging from 1 to 3 (the baseline value is 2);
- the incremental expected profit registered on average by the supplier firms can be approximated by a triangular probability distribution ranging from 1% to 10% with a modal value of 7%. This PDF is the result of a profitability forecasting analysis of suppliers' sub-sectors (spacecraft component; propulsion; lander, rover and probe; micro and nanospacecraft applications) based on data of the last ten years, future market trends and interviews to companies' managers.

In a deterministic model, the total discounted learning-by-doing benefit to suppliers is then:

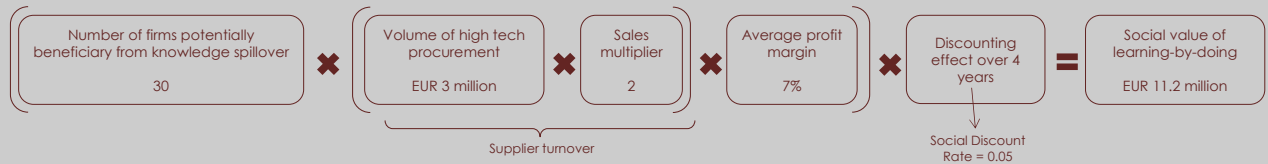

In a probabilistic model, the computation of the benefit's expected present value requires a Monte Carlo simulation conditional to the probability distribution functions of critical variables such as the profit margin and the sales multiplier.

## 5.6 Human capital formation

Typically, an RDI infrastructure employs four broad groups of staff: a) scientific personnel, b) technical personnel (technicians and engineers), c) administrative and support personnel, and d) PhD students, postdoctoral researchers, and visiting young academics and other short-term users.

For students, postdoctoral researchers, and visiting young scientists who enjoy the possibility of spending time working within a major RDI infrastructure, the main expected benefit is a 'premium' on their future salaries. This premium results from the acquisition of human capital, i.e. new capacity and skills, from experience with the project[56].

The mentioned 'premium' is the *incremental lifelong salary* earned by students and young scientists over their entire careers compared with the without-the-project scenario. Conceptually, two slightly different effects contribute to the formation of this 'premium' salary. On the one hand, the premium reflects the marginal salary increase gained by a former student who spent time at the RDI project relative to the salary that would have been earned anyway, i.e. without the experience offered by the infrastructure. On the other hand, the increase is the result of the fact that people having spent a training period at the infrastructure tend to increase their chances of being hired in labour markets that offer higher average wages.

Figure 4. Example of effects determining the salary premium

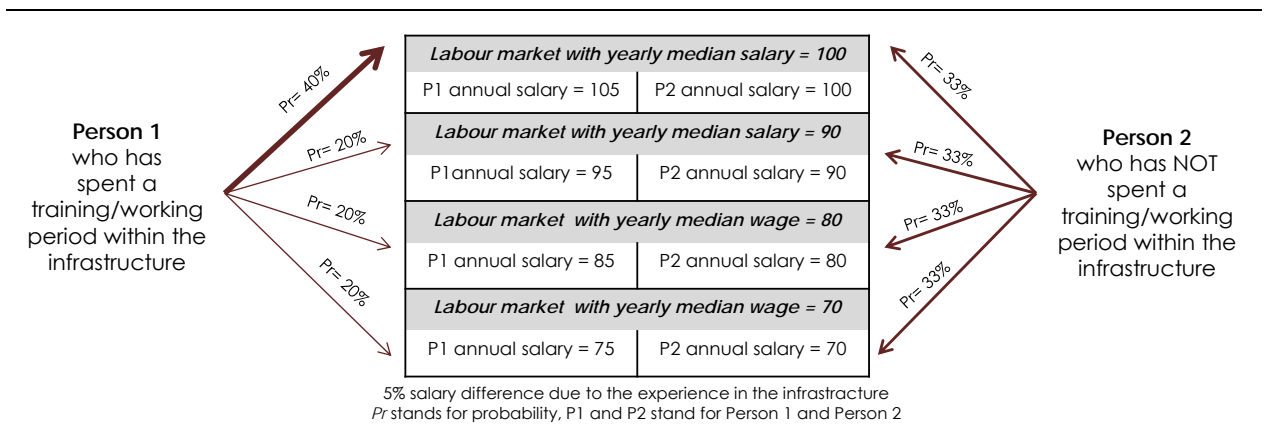

Note: Pr = probability, P1 = Person 1, P2 = Person 2. Source: Authors

The expected present value of human capital accumulation benefits, $\mathbb{E}(H)$, can be defined as the sum of the expected increasing earnings, $\mathbb{E}(E_{it})$, gained by RDI students and young scientists and commonly indexed by $i$, from the moment (at time $\varphi$) they leave the RDI infrastructure.

$$\mathbb{E}(H) = \sum_{i=1}^{I}\sum_{t=\varphi}^{T} s_t \cdot \mathbb{E}(E_{it}).$$

*Empirics*

In principle, assessing the effect produced by the RDI infrastructure on students and young scientists requires a quasi-experiment perspective. Such a perspective implies tracking the careers of cohorts of students in the long run and matching data on the careers of young people with experience in the RDI infrastructure with those who lack such experience. Alternatively, the effect could be estimated using an econometric model, e.g. based on Mincer's human capital earnings function (1974)[57].

---

[56] Similarly, some capacities and skills can also be acquired by scientists, engineers, and technical staff working at the RDI infrastructure. Once they leave the facility, the increase in earnings they receive compared with what they would have received without their experience at the RDI infrastructure is a 'premium' similar to that enjoyed by young researchers. The evaluation approach presented for young scientists holds for former employees as well.

[57] Jacob Mincer (1974) was the first to derive an empirical formulation of earnings over the lifecycle. In the 1974 formulation, Mincer modelled the natural logarithm of earnings as a function of years of schooling and years of labour market experience. The most broadly used version of Mincer's human capital earnings function is: $\log y = \log y_0 + rS + \beta_1 X + \beta_2 X^2$, where y is earnings, $y_0$ is the level of earnings of an individual with no education and no experience, S is years of schooling, and X is the years of potential labour market experience.

Box 10. The marginal return to human capital

The estimation of the return to human capital, for both the individual (or private return) and society as a whole, has been the focus of considerable debate in the economics literature. In particular, the private return is defined as the extra salary earned as a result of an increase in human capital, typically proxied by years of schooling. The benchmark model for an empirical estimation of the returns to education is the relationship derived by Mincer (1974), which includes on-the-job training and experience beyond schooling. However, the literature on the impact of education on earnings reveals a broad range of empirical approaches (one factor vs multiple factors model; homogeneous vs heterogeneous returns model; OLS regression; the instrumental variable method; the control function method; the method of matching; and the discount method) that have been adopted to estimate the return and an equally broad range of estimates (see Psacharopoulos, 1994; Psacharopoulos and Patrinos, 2004; Psacharopoulos, 1995; and Heckman et al., 2005 for a review).

As an example, Table 8 presents the results of a small sample of empirical studies, which have attempted to yield comparable results by using cross-country data sources.

Table 8. Average return to education

| Country | Harmon et al. (2003) Returns to education, 1995[58] | Blöndal et al. (2002) Private internal rates of return to tertiary education, 1999-2000[59] | Boarini and Strauss (2007) Private internal rates of return to tertiary education, 2001[60] |
|---|---|---|---|
| Austria | 6.8 | - | 6.4 |
| Denmark | 5.6 | 11.3 | 9.1 |
| Finland | 8.7 | - | 7.8 |
| France | 7.8 | 14.8 | 9.0 |
| Germany | 8.8 | 8.7 | 6.3 |
| Ireland | 11.3 | - | 13.1 |
| Italy | 6.9 | 6.5 | 5.1 |
| Netherland | 5.7 | 12.3 | 6.2 |
| Portugal | 9.7 | - | 12.2 |
| Spain | 7.8 | - | 5.7 |
| UK | 10.4 | 16.1 | 12.0 |

Source: Authors adapted from different sources.

However, from an ex-ante perspective, the estimation of a future premium on salary may require benefit transfer approaches from other contexts, interviews, and expert opinions by specialists such as recruiters in the labour market of interest. Moreover, the marginal increase in earnings ascribable to the RDI infrastructure needs to be carefully tested in the risk analysis.

A benchmark with a similar infrastructure is also needed to forecast the number of students and young scientists spending time within the infrastructure and then entering different labour markets. According to the RAMIRI online handbook[61], the personnel needed during the design, construction, operation, and – finally – decommissioning or upgrade/reorientation of an RDI infrastructure have a very different composition of skills, attitudes, age, and mobility (see Staff evolution during the lifecycle of a facility).

The personnel statistics of similar infrastructures are the main source of information for forecasting the stream of flows of different incoming students, i.e. undergraduates, PhD students, fellows, and postdoctoral researchers. Instead, the median gross annual salary associated with the different labour markets that the former infrastructure's students enter after their study career can be derived from national or European statistics (e.g. the European Community Statistics on Income and Living Conditions – EU-SILC).

---

[58] In this study the rate of return to education is based on multivariate (OLS) analysis from the International Social Survey Programme (ISSP). In particular, in this table we report OLS estimates using potential experience (age minus education leaving age).
[59] In this study the private internal rate of return is defined as the discount rate that equalises the real costs of education during the period of study to the real gains from education thereafter. For more details on methodology adopted see Blöndal et al. (2002):19.
[60] In this study private return are calculated follows the approach developed in De la Fuente and Jimeno (2005), combining the discount method and the estimation of Mincerian wage premia and other labour market premia from micro-level data. For more details on methodology adopted see Boarini and Strauss (2007):8 et seq..
[61] See www.ramiri-blog.eu.

Figure 5. Staff evolution during the lifecycle of a facility

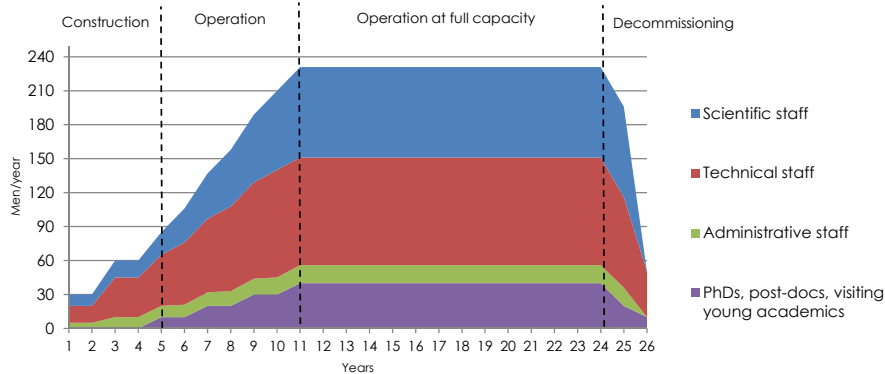

Source: Authors based on RAMIRI project handbook.

In practice, the following operational steps can be applied to estimate the human capital benefit:

- Forecast the number of incoming young researchers by category (e.g. Master students, PhD students, fellows, post-doctoral researchers);
- Assume the possible professional sectors in which students leaving the infrastructure are expected to find a job, such as in other research centres, in the academia, and in different industry sectors;
- Assume the probability distribution of different categories of students who find a job in the previously identified professional sectors;
- Estimate the median gross annual salary for each of the identified professional sectors at different career levels (entry level, mid-career, experienced, late career);
- Use an appropriate (e.g. logarithmic) function to estimate the continuous salary curve for each professional sector; and,
- Estimate the 'premium' salary associated with having spent a training period at the considered RDI infrastructure, i.e. the incremental earnings compared with the average salary curve previously described.

### 7. Social value of human capital formation

A public sector supported research laboratory in the green biotechnology field is one of the top European institutes in terms of reputation. A project of enlarging it has been proposed, and it is based inter alia on the following assumptions:

- the project time horizon is 15 years ;
- it will host additional 15 PhD students every year for a training period of two years;
- and will host 10 Post-docs every year for a contract period of three years;
- after their training period, students and post-docs are expected to immediately enter the labour markets. In particular, students are supposed to enter four possible professional sectors with the following probabilities:

| Professional sector | PhD students | Post-doc students |
|---|---|---|
| Academia | 20% | 40% |
| Other research centre in biotechnology | 30% | 30% |
| Biotechnology industry | 30% | 20% |
| Other industry (including of chemical, medical and pharmaceutical industries) | 20% | 10% |

- the salary curve associated with the four possible future professional careers are those presented in the graph below;

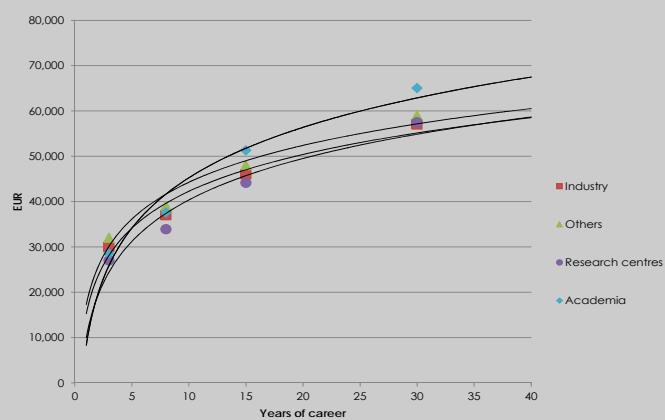

- based on statistical information, a salary premium of 5 % over the total future salary is expected for students having spent a training period at the considered laboratory as compared to their peers who have not enjoyed the same experience;
- a work career of 40 years.

The expected total discounted human capital benefit, estimated as the present value of the total annual gross incremental salary gained by all students trained during the project time horizon over their entire work career, is EUR 15.5 million. The following formula applies:

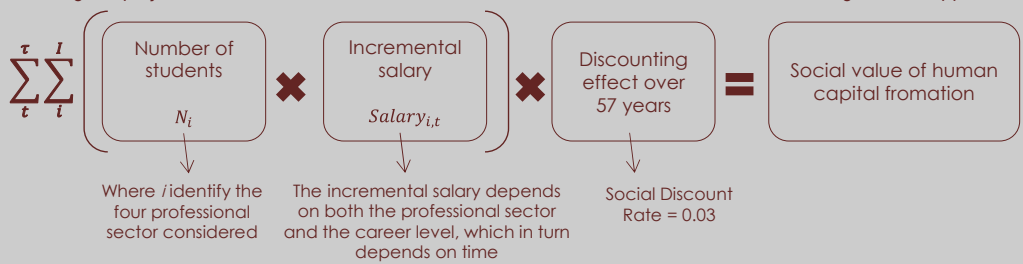

$$\sum_{t}^{\tau} \sum_{i}^{I} \left( \underbrace{N_i}_{\substack{\text{Number of students}}} \times \underbrace{Salary_{i,t}}_{\substack{\text{Incremental salary}}} \times \underbrace{\text{Discounting effect over 57 years}}_{} \right) = \text{Social value of human capital fromation}$$

Where $i$ identify the four professional sector considered. The incremental salary depends on both the professional sector and the career level, which in turn depends on time. Social Discount Rate = 0.03.

## 5.7 Knowledge outputs and their impact

For scientists and researchers, particularly academics, one of the main benefits to working at a RDI infrastructure is the opportunity to access and process new experimental data, to contribute to the creation of new knowledge, and – ultimately – to produce scientific output that may take the form of technical reports, proceedings, preprints or working papers, articles in scientific journals, and research monographs.

The peculiarity of the RDI infrastructure is that the demand for the knowledge creation function of a RDI project is driven by scientists and researchers in a given field(s) who are often simultaneously users and producers of knowledge. This fact implies that when scientists and researchers spend time on a project, they have an opportunity cost from not working on an alternative project. If this opportunity cost is assumed equal to the average scientist's hourly compensation, a reasonable proxy of the value of scientific output is its marginal scientific cost. This marginal scientific cost represents the time spent by scientists to conduct research and produce knowledge outputs valued at appropriate shadow wages. Hence, the social benefit related to the production of scientific publications can be valued using their *marginal production cost*.

Clearly, not all scientific output has the same value for the relevant community. For this reason, weighing the influence of a paper by multiplying the value of the scientific publications *by a multiplier of impact* is advisable and entails the social value attributed to the degree of influence of that piece of knowledge on the scientific community. In other words, the multiplier captures the additional value attributable to citations that the outputs receive. However, it does not include the indirect social benefits of knowledge that are either accounted for under other items or completely uncertain and set to zero.

Given $\mathbb{E}(Y_t)$ as the expected social cost of producing knowledge outputs at time $t$, $s_t$ as the discount factor (shadow cost of production), and $\mathbb{E}(m)$ as the expected multiplier of impact, the expected present value of this benefit is expressed as:

$$\mathbb{E}(O) = \sum_{t=0}^{\mathcal{T}} (s_t \cdot \mathbb{E}(Y_t) \cdot \mathbb{E}(m)).$$

*Empirics*

The ex-ante estimation of this benefit involves two operations. The first operation is the estimation of the social value of producing new knowledge as embodied in technical reports, proceedings, preprints or working papers, articles in scientific journals, and research monographs. The second operation is the estimation of the multiplier of impact, i.e. a synthetic multiplicative factor capturing the social value attributed to the degree of influence of that piece of knowledge on the scientific community. These two operations can be synthesised using an appropriate function of time. For example, among different forms, a double-exponential model is described by the formula (see Bacchiocchi and Montobbio, 2009):

$$O = Y_t \, \alpha \, \exp(-\beta_1(\mathcal{T} - t)) \, (1 - \exp(-\beta_2(\mathcal{T} - t))),$$

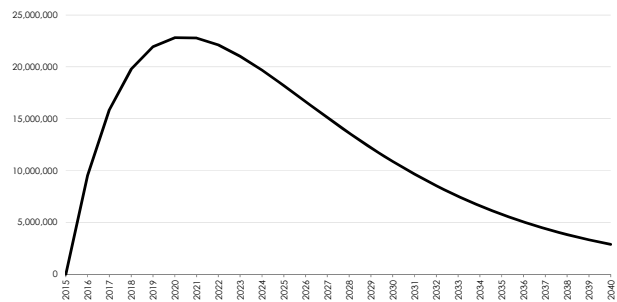

where the expected value of knowledge outputs produced by a RDI infrastructure depends on:

$Y_t$, the social cost of producing knowledge output;

$\mathcal{T}$, the total number of years for which the number of papers is estimated;

$t$, the number of remaining years from a given year to the end of the simulation period;

$\alpha$, a combination of a set of multiplicative parameters representing the characteristics of the scientific community (number of authors, summary measure of their productivity, and others); and,

$\beta_1$ and $\beta_2$, two parameters that determine the shape of the curve that changes according to the papers weights, distinguishing between excellent and mediocre papers.

To fit the equation parameters, bibliometric[62] techniques analysing the patterns of the scientific literature generated over time around a similar RDI infrastructure or its experiments can be conveniently exploited to associate a measure of scientific output to the RDI infrastructure[63]. However, note that although the use of bibliometric techniques is a well-established approach to provide a quantitative characterisation of scientific activity, relying on publication records available in online repositories may lead to bias when the dominant mode of production in a specific scientific domain is not the journal article. The limited coverage of particular scientific fields by reference databases is a well-known issue in some specific disciplines, such as social sciences and humanities (see Hicks, 2004 and Nederhof, 2006), law, and computer science in which peer-reviewed conferences are a major form of communication. This issue must be carefully taken into account when dealing with the evaluation of the considered benefit and, if necessary, bibliometric analysis needs to be complemented by a detailed analysis of unpublished scientific outputs. Also, attention should be paid to not double counting articles. Actually, in some scientific field such as theoretical physics, exper-imental physics, materials, engineering the same article is often produced in several different versions (same contents, different titles, and different publication channels).

---

[62] Biblometrics is the discipline dealing with citation data and the statistics derived from them.
[63] These techniques are discussed in Carrazza *et al.* (2013).

Box 11. The diffusion of knowledge outputs

According to the literature, curves describing the dynamics of knowledge diffusion over time can be proxied using citation curves. Citation patterns can vary significantly depending on the document studied. Some articles may never be cited, whereas others receive citations in the years immediately after publication and before becoming obsolete. Additionally, some articles may remain rarely cited in the years following their publication, but then become recognised (Andrés, 2009). A 'typical citation curve' describes the history of an article that receives a few citations in the first subsequent years after publication, then peaks, but subsequently is gradually less cited (Sun et al., 2015; Larivière et al., 2008). In some cases, lognormal functions best fit typical citation curves (Egghe and Rao, 1992). Most studies on obsolescence find that the use of the literature declines exponentially with age, and parameterise this phenomenon with a single number, often called the 'half-life' (Burton and Kebler, 1960). However, some argue that an exponential increase in citations is sometimes recognisable over a long period, thereby leading to an exponential function (Li and Ye, 2014).

An operational shortcut to estimate the social value of knowledge outputs consists of a series of steps, as concisely presented in *Figure 6*. In what follows, the data and computations required for each step are discussed.

Figure 6. Social value of scientific output

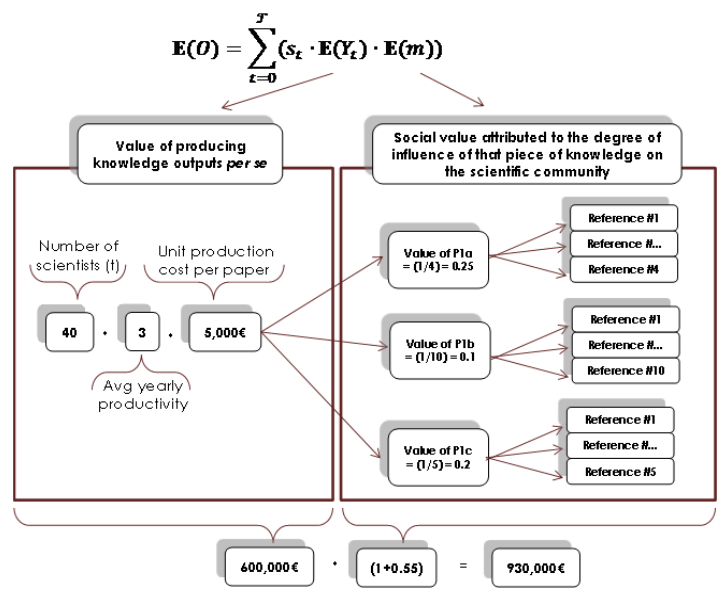

Note: (*) The benefit is valued through the time that scientists took to read and understand someone else's paper and to decide whether or not to cite it, which is assumed to be one hour. (**) The number of papers, P1, citing parent one is forecasted by reviewing the median number of citations of scientists involved in the infrastructure.
Source: Authors.

First, forecasting the knowledge production capacity of the infrastructure is needed and involves:

- Forecasting the number of scientists working at the infrastructure over the time horizon; and

- Forecasting scientists' yearly average productivity, i.e. the average number of knowledge outputs per author per year. Clearly, predicting the number of knowledge outputs produced within the RDI infrastructure can be influenced by the standards of the personnel expected to be recruited and by the scientific field. Hence, the scientists' yearly average productivity and the average number of authors per paper may largely differ from one discipline to another due to different practices in use. In case of multiple authorships, the main difficulty when estimating average productivity[64] is the definition of the individual contribution[65] to an article.

The second step consists of estimating the value of a scientific output in money terms. In applied welfare economics using marginal costs to estimate the output value is well established, and even in the international accounting convention on GDP public sector costs are used to estimate the contribution of government to the product of a country. The unit production cost per knowledge output may be estimated using the ratio of the gross salary of the author over the number of scientific outputs produced per year. Clearly, only the salary amount for the time dedicated to research within the infrastructure should be considered.

---

[64] For a review of the literature on scientists' research productivity, see IVA (2012).
[65] Attention should be paid to two opposed situations. Experiment collaboration articles are typically signed by the so called "authorlists", i.e. persons who did not directly contribute to the contents of the article. Conversely, in other cases articles are only signed by the persons who wrote the article, although many more persons contributed to the work. In both cases, the authors' productivity can be distorted if these aspect are not appropriately reflected in the calculation.

Data on scientists' salaries according to different scientific fields are found in various country-specific or worldwide databases. Table 9 provides illustrative benchmarks for different scientific fields and countries.

Table 9. Examples of scientists' annual salary

| Country | Scientific field and experience degree | Benchmark values (EUR) | Reference year | Source |
| --- | --- | --- | --- | --- |
| Austria | All fields – senior researcher | Median 66,038 | 2010 | Ates and Brechelmacher (2013) |
| Finland | All fields – senior researcher | Median 48,387 | 2008 | Ates and Brechelmacher (2013) |
| France | All field | Average 49,332 | 2011 | Altbach et al. (2012) |
| Germany | All fields - Entry-Level Research Scientist | Median 48,677 | 2015 | PayScale (2015) |
| Poland | All fields - senior at university | Median 32,078 | 2010 | Ates and Brechelmacher (2013) |
| United States | Biotechnology | Median 64,932 | 2015 | PayScale (2015) |
| United States | Material science | Median 74,744 | 2015 | PayScale (2015) |
| United States | Clinical research | Median 59,504 | 2015 | PayScale (2015) |

Source: Authors based on cited sources.

The third step consists of forecasting the median number of citations per scientific output.[66] The median value is considered a more accurate indicators instead of the average number of citations. Also, an analysis of the median individual h-index of scientists involved in the infrastructure, which captures the *n* number of articles that have received at least *n* citations, could be useful (see H-index vs individual H-index).[67]

As revealed by Table 10, the citation statistics show high variability in different scientific domains[68]. This variability results from different factors. For instance, the chance of being cited is related to the number of publications (and the number of scientists) in the field (Moed et al., 1985); thus, small fields attract fewer citations than more general fields (King, 1987). Therefore, bibliometric comparisons should be conducted only within a field unless a normalising factor is applied.

Box 12. H-index vs individual H-index

> The h-index was proposed by Hirsch (2005) with the aim to provide a robust single-number metric of an academic's impact by combining quality with quantity. A scientist has an h-index equal to n if n of his or her papers have at least n citations each and the other papers have no more than n citations each. Hence, a scientist with an h-index of 20 has produced 20 articles with at least 20 citations each.
> A number of studies addressed the attempts to correct the h-index for the number of co-authors (see, for instance, Batista et al. (2006) or the alternative provided by the Publish or Perish software program), the scientific field (see, for instance, Iglesias and Pecharromán, 2007); Radicchi et al., 2008); and Malesios and Psarakis, 2012), and the recentness (see, for instance, Sidiropoulos et al., 2006).
> In particular, if the h-index is corrected for the number of co-authors, the resulting metric is called the individual h-index. According to the literature, the two indexes shows differences between disciplines. As an example, Table 10 presents the analysis provided in Harzing (2010).
>
> Table 10. Metrics comparisons across disciplines
>
> | Scientific field | Average H-index | Average number of authors | Individual H-index |
> | --- | --- | --- | --- |
> | Cell biology | 24 | 3.90 | 15 |
> | Computer science | 34 | 2.57 | 22 |
> | Mathematics | 15 | 2.95 | 8 |

---

[66] It should be acknowledged that estimation of scientific outputs based on citations is somehow problematic in some fields such as theoretical and experimental physics, material and engineering for two main reasons: 1) the science does not work with referencing other papers frequently due to text size limitations; and 2) the same article is often produced in several different versions (same contents, different titles, different publication channels).

[67] Moreover, the h-index is considered a conservative measure that avoids incurring an overestimation of benefits. Indeed, the proposed framework assumes that an article has been accessed and read before being cited, but this may not be true. In some cases, citation data may overstate the extent to which the scientific literature has been consulted. In fact, an author may cite from the abstract of an article or simply copy a reference from another paper (Simkin and Roychowdhury, 2007; Broadus, 1983).

[68] Citations' skewness was identified early on by Price (1965).

| | | | |
|---|---|---|---|
| Pharmacology | 39 | 3.08 | 23 |
| Physics | 30 | 2.66 | 18 |

Source: Authors based on analyses presented in Harzing (2010).

The fourth step consists of assigning a monetary value to citations by deriving the value from the time scientists need to download, read someone else's paper, and decide to cite it[69]. When estimating the value of an article (P1) that cites a scientific output (P0) produced by a scientist within the considered RDI infrastructure, that other articles beyond P0 have contributed to the production of P1 should be considered. This consideration is reflected in the number of references included in P1. Thus, the value of P1 attributable to P0, i.e. the time needed to read and cite P0, must be divided by the number of references contained in P1.

### 8. Social value of publications

A new integrated network of ten marine biological stations in different coastal locations will be constructed and provided with state-of-art equipment. On average ten scientists will be employed in each facility for the entire operational period (16 years). Given the field, and the track record of existing similar infrastructures in other countries:

- the expected average yearly productivity of scientists within the infrastructure is 3 articles;
- the average time devoted to research is 60% (the remaining time is devoted to teaching and administrative issues);
- the average gross yearly salary of scientists using the infrastructure is EUR 40,000;

The marginal cost of an article produced by scientists working in the RDI infrastructure is then:

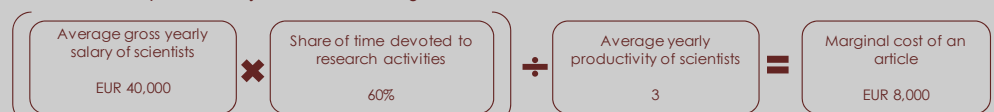

Average gross yearly salary of scientists EUR 40,000 × Share of time devoted to research activities 60% ÷ Average yearly productivity of scientists 3 = Marginal cost of an article EUR 8,000

Moreover, according to bibliometric analysis and field experts' opinions, the following information holds:

- the median number of citations of scientists expected to work in the infrastructure is 10;
- the average number of references in the scientific field is 40 per paper;
- the time needed to evaluate someone else's paper and decide to cite it is 1 hour;

As a result, the total non-discounted social value of publications is:

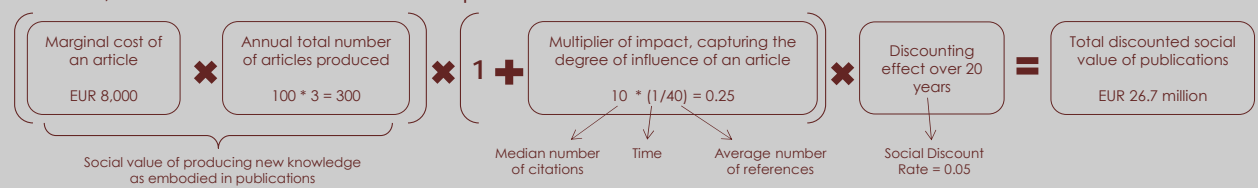

Marginal cost of an article EUR 8,000 × Annual total number of articles produced 100 * 3 = 300 × (1 + Multiplier of impact, capturing the degree of influence of an article 10 * (1/40) = 0.25) × Discounting effect over 20 years = Total discounted social value of publications EUR 26.7 million

Social value of producing new knowledge as embodied in publications

Median number of citations | Time | Average number of references

Social Discount Rate = 0.05

## 5.8 Provision of services

Some RDI infrastructures provide services to external users, including industries, governmental bodies, and other research teams. These users may pay a fee for accessing and using the infrastructure's equipment and/or specific services offered by the facility for research or technological development and testing purposes.

Services provided by RDI infrastructures may include, among others, machine time, computing resources, software, data, communication, sample preparation, access to archives or collections, education and training, expert support, and analytical services. Some examples are provided in Table 11.

Table 11. Examples of services offered to external users by RDI infrastructures

| Infrastructure | What does it offer? |
|---|---|
| NASA Glenn Research Center (https://facilities.grc.nasa.gov/using.html) | It provides ground test facilities for/needing: Acoustics; Engine Components; Full-Scale Engine Testing; Flight research; Icing research; Microgravity research; Space power and propulsions; Wind tunnels. In addition, a range of test consultation services are offered. |
| European Synchrotron Radiation Facility (http://www.esrf.eu/Industry/applications) | It provides synchrotron light and techniques (e.g. of imaging, microtomography, topography, microscopy RX and FTIR microscopy ) which have many industrial applications. For instance, pharmaceutical and biotech companies use synchrotron techniques to help develop new products at all stages of research, from drug design and formulation to pre-clinical phases. Automotive industries use synchrotron techniques to obtain more efficient catalytic exhaust converters. The synchrotron techniques are also commonly used for the study of inclusions in surfaces or identification of a defect on silicon wafers used to produce semiconductors. |
| High Magnetic Field Laboratory in | Access to the facility is given to all researchers which have their research proposal for access granted by an |

---

[69] For simplicity, the time needed to download, read someone else's paper, and decide to cite it can be set equal to one hour.

| The Netherlands (http://www.ru.nl/hfml/facility/access_to_the) | external review committee. Access involves the use of the installation, the use of all available auxiliary equipment, and (if necessary) support of the local staff. |
|---|---|
| Laserlab-Europe (http://www.laserlab-europe.net/transnational-access) | The 20 laboratories under the Consortium offer access to their facilities for European research teams. Access is provided: to world-class laser research facilities, to a large variety of inter-disciplinary research, including life sciences, free of charge, including travel and accommodation. Access is provided on the basis of scientific excellence of the proposal, reviewed by an external and independent Selection Panel. Priority is given to new users. A typical access project has a duration of 2 to 6 weeks. |

Source: Authors based on information retrieved from facilities' websites.

Typically, customers contact the facility manager and, after reviewing the requests, the infrastructure manager provides external users with a cost estimate for a time shift on the machine or the service required. In some cases, the cost estimate by the RDI infrastructure reflects market prices. In other cases – typically when the external users are researchers – the fees charged only cover the costs incurred by the facility to make the service available.

The preferred way to value this benefit is by either using the *long-run marginal cost* of the services provided *or* estimating external users' *WTP for the service*. Alternatively, when market prices are available and are supposed to be non-distorted, i.e. they reflect economic prices, the *nominal (market) prices* can be used.

### Empirics

The ex-ante evaluation of this benefit involves the following activities.

- The range of services to be provided by the infrastructure or the amount of time dedicated to commercial uses is forecasted. Benchmarks with similar infrastructures are helpful. As an example, the share of time dedicated to commercial purposes in five European synchrotron radiation facilities is provided in Marks (2007).

- The number of potential external users that may be interested in exploiting the infrastructure's equipment or services is forecasted. Typically, during the design phase of a RDI infrastructure, the promoters investigate external users' interests through surveys. The data collected can be exploited for the forecast.

- The economic value of the services offered is estimated. The long-run marginal cost of the services can be proxied by the costs incurred by the infrastructure to make the services available. Alternatively, external users' WTP for the services can be estimated using contingent valuation methods. When market prices exist for similar services offered by a similar infrastructure, they could be exploited, provided they are not distorted.

### 9. Benefit from services provided to third parties

A new laser facility is planned and will be operational for 15 years. Among others, the laboratory will offer free access to a complete line of ultrafast lasers and experimental set-ups for high-intensity laser-matter interaction. According to preliminary forecast performed by the project promoters, the following holds:

➡ based on estimated economic production cost, the beam time costs is EUR 2,000/hour;
➡ the total hours of beam activity is 3,780 per year;
➡ the yearly share of beam time dedicated to third-parties access is 10%.

In a deterministic model, the total discounted benefit for the laser provision to third-parties, valued at the marginal cost, is then:

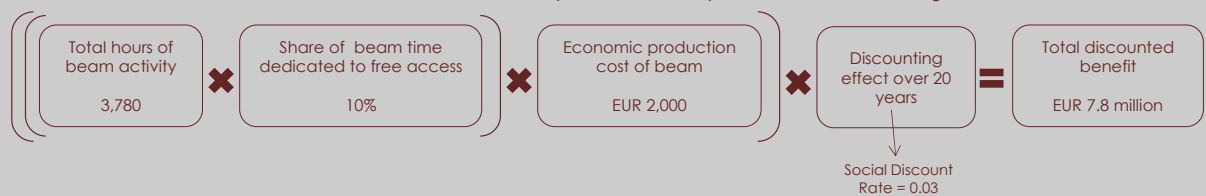

Total hours of beam activity 3,780 × Share of beam time dedicated to free access 10% × Economic production cost of beam EUR 2,000 × Discounting effect over 20 years (Social Discount Rate = 0.03) = Total discounted benefit EUR 7.8 million

# 5.9 Social benefits of RDI services for target groups of consumers

Infrastructures for applied research and development are expected to use new knowledge to deliver innovative services and products addressing specific societal needs, e.g. tackling climate change, finding new ways to ensure energy security and efficiency, reducing environmental pollution, mitigating the risk of natural disasters, improving health conditions. Therefore, benefits arise to users who are better off by the delivery of the innovative service or product.

This category of benefits refers only to applications that fall within the infrastructure mission since its funding decision. Situations in which the practical use of a good can be, in principle, expected but is still unknown (in the technical sense, that a probability distribution function is unknown) and cannot be considered under this category of benefits because forecasting and estimating ex ante are not possible. In the latter case, the benefit value is determined by what is generally called 'quasi-option value'. Further elaboration on this issue is provided in the next section on 'non-use benefits'.

The methods to quantify and value the set of benefits derived from the practical application of a research effort depend on the type of innovative service or products made available by the infrastructure. However, these methods are generally based on the willingness to pay or avoided cost approaches and may refer to the more traditional CBA approaches developed for specific sectors. Table 12 presents - as illustrative examples - the evaluation methods referring to typical benefits in the environmental, energy, and health sectors[70].

It is worth noting that the most challenging aspect in the estimation of such benefits is not the translation in monetary value, which usually draws upon well-known and established methodologies and techniques, but the quantification of the amount of benefits. Indeed, given the innovative nature of the supplied service/product, the reach, magnitude and extent of the actual materialisation of benefits is highly uncertain. Hence, what is challenging is forecasting the probabilities of success and the different levels of effectiveness associated to the innovation implemented. In this context, different scenarios – at least one pessimistic and one carefully optimistic – should be forecasted, and each of them needs to be carefully tested through a risk assessment.

---

[70] This list is not intended to be exhaustive.

Table 12.Examples of social benefits derived from the practical application of a research

| Benefit | Evaluation methods | References |
|---|---|---|
| Reduction of GHG and air pollutant emissions | To estimate the externality of greenhouse gas (GHG) and pollutant emissions[71], the usual approach consists of *quantifying the emissions avoided because of the project* (measured in kg per tonne of waste) *and valuing them with a unit economic cost* (measured in Euro per kg emission). However, when a new eco-friendly technology is developed and sold to enterprises, for example, the selling price could already incorporate the environmental benefit. In such a case, the externality should not be estimated to avoid double counting. | <ul><li>IMPACT study (European Commission, 2008), which lists unit cost values for the main relevant air pollutants (in Euros per ton) on the basis of HEATCO[72] and CAFE CBA[73] reports;</li><li>NEEDS Integrated Project[74], which provides unit damage costs for air pollutants from emerging electricity generation technologies. NEEDS also provides reliable cost factors for ecosystem and biodiversity damage from air pollution;</li><li>Extern-E study[75] provides the unit values of air pollutants produced by energy infrastructures in EU member states.</li><li>Teichmann, D. and Schempp, C. (2013). 'Calculation of GHG Emissions of Waste Management Projects'. JASPERS Staff Working Papers.</li></ul> |
| Improved energy efficiency | The improved energy efficiency benefit is valued through the *decrease in energy costs*, whether incurred by the energy producer, distributor, or final user. The cost reduction is not expressed at market prices, but by considering *the opportunity cost* (shadow price) *of the avoided energy sources*, which should be calculated as the long-run marginal cost of production and (if relevant) transportation.<br>Note that producing electricity from a renewable source could be, at least initially, more expensive than from other sources. In fact, emerging renewable technologies are typically not competitive with fossil fuel alternatives (HM Treasury, 2006). Thus, the project would produce a cost and not a benefit. However, this cost would be (partly or fully) compensated from higher benefits from reduced GHG and pollutant emissions.[76] | <ul><li>European Commission (2014) chapter 5 on Energy sector.</li><li>EIB (2013) contains a chapter on Energy Efficiency and District Heating.</li><li>ENTSOE (2015) Guideline for Cost Benefit Analysis of Grid Development Projects.</li><li>World Health Organisation (2006). Guidelines for conducting cost-benefit analysis of household energy and health interventions, by Hutton G. and Rehfuess E., WHO Publication</li><li>Clinch, J.P. and Healy, J. D. (2001). 'Cost-Benefit Analysis of Domestic Energy Efficiency', *Energy Policy*, 29(2): 113-124.</li></ul> |
| Reduction in vulnerability and exposure to natural hazards | When a RDI infrastructure project is aimed at developing tools and disaster management systems to facilitate disaster resilience and risk prevention and management for natural risks, a benefit from *avoided damage to capital and natural stocks* is expected[77].<br>The cost of the avoided damages is estimated using information and data contained in risk and hazard maps and modelling. A shortcut consists of adopting the market insurance premiums available for different typologies of risks to proxy the value of the avoided damage to the capital stock. Instead, for assets for which an insurance market does not exist, averaged calculations on the basis of the avoided costs of the public administration for civil protection activities, compensation | <ul><li>The Word Bank (2003) Building Safer Cities The Future of Disaster Risk contains chapter 3 on Natural Disaster Risk and Cost-Benefit Analysis.</li><li>Guha-Sapir, D. and Santos, I. (2013) *The Economic Impacts of Natural Disasters*. New York, NY: Oxford University Press.</li><li>MMC (2005). Natural Hazard Mitigation Saves: An Independent Study to Assess the Future Savings from Mitigation Activities. Volume 2-Study Documentation. Washington DC: Multi-hazard Mitigation Council.</li><li>Benson C. and Twigg, J. (2004). 'Measuring mitigation: Methodologies for assessing natural</li></ul> |

---

[71] Examples of pollutants are nitrogen oxides (NOx), sulphur dioxide ($SO_2$), particulate matter (PM10 and PM2.5), non-methane volatile organic compounds (NMVOC) as a precursor of ozone ($O_3$), and ammonia ($NH_3$).
[72] European Commission (2004).
[73] Clean Air for Europe (CAFE) Programme, at: http://ec.europa.eu/environment/archives/cafe/activities/pdf/cafe_cba_externalities.pdf.
[74] New Energy Externalities Development for Sustainability www.needs-project.org.
[75] www.externe.info.
[76] This do not hold true for project within the EU ETS - Emissions Trading System.
[77] The estimation of this benefit involves two main sources of difficulties. First, predicting when an actual disaster will occur and at what intensity is not always possible. Second, the effectiveness of the investments is estimated through vulnerability assessments that include a degree of uncertainty. Therefore, the avoided damages are probabilistic, at best.

|  | paid to citizens, relocation of buildings, and other activities should be carried out and added to the economic analysis (European Commission, 2014). Alternatively, the people's WTP for decreasing the vulnerability and exposure to a natural hazard could be estimated.<br><br>Finally, when the project addresses natural assets, additional effects should be evaluated in terms of increased use or non-use values. Regarding use values, the typical additional effects to be considered are increased recreational value (typically valued through the Travel Cost Method) and preservation of productive land (typically valued through its opportunity cost). Regarding non-use value, the preservation of a natural asset in good condition must be estimated by eliciting its existence value (typically through contingent valuation or benefit transfer)[78]. | hazard risks and the net benefits of mitigation - A scoping study'. Geneva: International Federation of Red Cross and Red Crescent societies / ProVention Consortium.<br>• Kunreuther, H. and Michel-Kerjan, E. (2014). 'Economics of natural catastrophe risk insurance', in *Handbook of the Economics of Risk and Uncertainty*, Volume 1, MJ Machina and WKViscusi (Eds), Elsevier. |
| Improved health conditions | Changes in human mortality and morbidity rates can be triggered by a RDI infrastructure[79] with different aims, such as improving the health conditions of the people affected by a certain disease by producing a new drug or an innovative treatment technology; improving the health safety of people (or a group of people), such as through food and transport; remediating a polluted environment, e.g. a radioactive dump or a site contaminated by chemical waste; and mitigating the risk of natural disasters.<br><br>In such cases, the project's marginal benefit is the reduction in mortality or morbidity rates, or improved health conditions. Following the literature, these reductions can be valued using the Value of Statistical Life (VOSL), defined as the value that society deems economically efficient to spend on avoiding the death of an undefined individual. The Quality Adjusted Life Year (QALY) may also be used, which measures the value of a change in both life expectancy and quality of life. The preferred approach to value changes in health outcomes is to calculate the willingness-to-pay of people affected by the project. This calculation can be done using the stated preference methods (surveys) or revealed preference methods (hedonic wage method). However, in practice, the human capital approach (for mortality) or the cost of illness approach (for morbidity) is more frequently used. Each method has benefits and drawbacks. | • On Value of Statistical Life, see for instance: Landefeld and Seskin (1982); Viscusi and Aldy (2003); Abelson (2008 and 2010), Sund (2010); and Viscusi (2014).<br>• On Value of a Life Year, see for instance: Johannesson and Johansson (1996), Chilton et al. (2004) and Desaigues et al. (2011).<br>• For a major meta-analysis of Value of Statistical Life estimates, derived from surveys that asked people around the world about their willingness to pay for small reduction in mortality risks, see OECD (2012) or Lindhjem et al. (2011).<br>• For a meta-analysis of Value of Statistical Life estimates, derived from revealed preference studies, see Mrozek and Taylor (2001).<br>• On human capital approach see for instance Landefeld and Seskin (1982) and Brent (2003, chapter 11.<br>• On the cost of illness approach see for instance Rice (1967) and Rice at al. (1985) and Byford et al. (2000).<br>• World Health Organisation (2006), Guidelines for conducting cost-benefit analysis of household energy and health interventions, by Hutton G. and Rehfuess E., WHO Publication. |

Source: Authors.

---

[78] Some examples of existence values retrieved from the literature are provided in Table 13.
[79] Or project is carried out within the RDI infrastructure.

## 10. Benefit of CO2 reduction

A research-based biotechnological company in collaboration with a university laboratory wants to develop an innovative enzyme to enhance energy efficiency in detergent industries. In particular, the enzyme-driven industrial processes is expected to reduce the carbon dioxide emissions from 300 kg to a range of 40-80 kg every tonne of product produced (baseline value = 50 kg/tonne).

Assume there are 20 cosmetic plants interested in using the new enzyme (which will become obsolete in 10 years) and that each of them produce 100 tons of products per year. If the shadow price of CO2 is EUR 35/tonne, the total benefit for reduction in CO2 emission in a deterministic model is then:

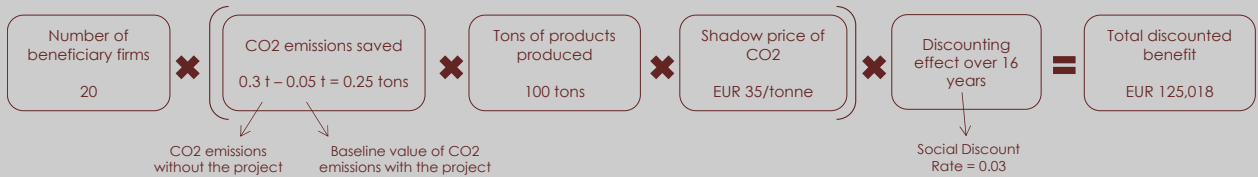

In a pessimistic scenario, the carbon dioxide emissions are 80 kg/tonne of product produced, thus the total discounted benefit is EUR 110,016. Instead, in an optimistic scenario, the emissions are 40 kg/tonne and the total discounted benefit is EUR 130.019. Taking 40 and 80 as the lower and upper bounds of a triangular probability with modal value equal to 50 kg/tonne, the expected benefit of CO2 reduction can be calculated through a Monte Carlo simulation.

## 11. Social value of energy saving new technology

A laboratory meant to develop a new technology which allows reduced energy-consumption costs to keep the temperature inside high temperatures process used in the primary metals industry at the same level as in the without-the-project scenario. It is assumed that an energy bill of EUR 100,000 is annually paid by the manufacturing plants potentially interested in the new technology which corresponds to a steady temperature of 2500 K. After the new technology's implementation, the energy efficiency of the manufacturing plants increases and this is reflected in a decrease of annual energy costs (to EUR 75,000) that is required to maintain the same required temperature.

In the economic analysis, the opportunity cost of energy should be considered, by applying a conversion factor to the cost saving. Based on border prices approach, the conversion factor is estimated to be 1.1.

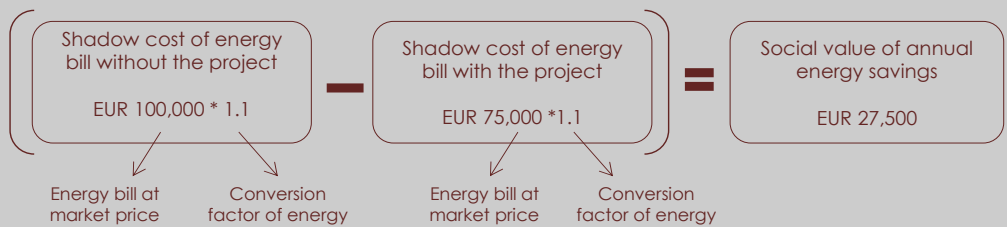

> ## 12. Benefit from increased life expectancy
>
> An applied research infrastructure specialised in hadrontherapy, an advanced oncological treatment showing clinical advantages as compared to traditional radiotherapy and exploiting accelerator technology, is planned. The facility will provide health treatments to patients affected by some other fatal types of solid cancer, for whom gains in terms of longer or better lives are expected as compared to a counterfactual situation where they are treated with conventional therapies. Health improvements related to patients affected by chordomas and chondrosarcomas of the skull base are considered. According to forecasts made by project promoters, the following holds:
>
> - during routine operation nearly 300 patients affected by chordomas and chondrosarcomas of the skull base will be treated every year. In particular, 60 patients per year are expected under six classes of age with median age corresponding to 10, 30, 50, 70 and 90 years.
> - no alternative treatments are available for patients affected by chordomas and chondrosarcomas of the skull base;
> - the marginal percentage of patients who fully recover compared to the counterfactual situation is 70%. This means that 70% of total treated patients thanks to hadrontherapy gains the same life expectancy of the average healthy population (80 years);
> - following the human capital approach, the VOLY values for each of the six classes of age identified has been calculated. Namely:
>
> | Age class   | VOLY   |
> |-------------|--------|
> | 10 (1-20)   | 29,000 |
> | 30 (21-40)  | 28,000 |
> | 50 (41-60)  | 27,000 |
> | 70 (61-80)  | 26,000 |
> | 90 (81-100) | 25,000 |
>
> In a deterministic model, the total discounted benefit for each age class is calculated in the following way:
>
> 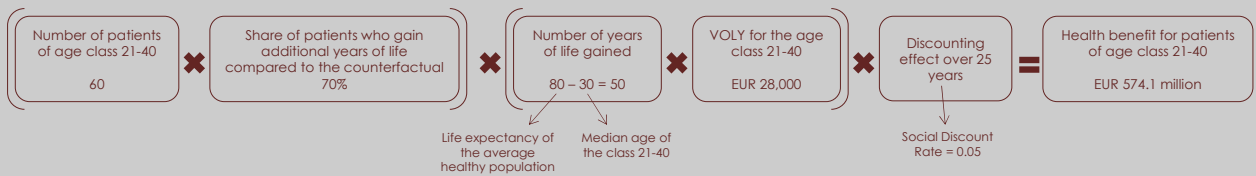
>
> Number of patients of age class 21-40 (60) × Share of patients who gain additional years of life compared to the counterfactual (70%) × Number of years of life gained (80 – 30 = 50, Life expectancy of the average healthy population – Median age of the class 21-40) × VOLY for the age class 21-40 (EUR 28,000) × Discounting effect over 25 years (Social Discount Rate = 0.05) = Health benefit for patients of age class 21-40 (EUR 574.1 million)
>
> Aggregating the discounted health benefits related to all the age classes, the total discounted benefit stemming from the infrastructure is calculated.

## 5.10 Recreational benefits for the general public

Some RDI infrastructures, in particular but not only large-scale ones, have an outreach strategy to attract the interest of the general public, such as through the organisation of permanent or temporary exhibitions, open days, or guided tours – typically for free or at a modest price. For instance, large research infrastructures follow such a strategy, including the Daresbury Synchrotron Radiation Source in Cheshire (UK), the Kennedy Space Center (KSC) Visitor Complex in Florida, the CERN in Geneva, the Max Planck Institute for Plasma Physics in Greifswald (DE), the High Magnetic Field Laboratory in Nijmegen (NL), and the European Southern Observatory in the Atacama Desert of northern Chile.

The ultimate beneficiaries of outreach activities are visitors to the infrastructure. In line with standard CBA approaches in the cultural economics sector[80], the expected marginal social value of this benefit is the expected visitors' implicit willingness-to-pay for a visit. Hence, the following formula applies:

$$\mathbb{E}(R) = \sum_{i=1}^{I} \sum_{t=1}^{T} s_t \cdot \mathbb{E}(W_{it})$$

where $W$ is the user-benefit of general public individuals ($i = 1, ... I$) for visiting the RDI infrastructure.

In addition to in-person visits, participating in activities on social media, in television audiences, and through websites are further indicators of the size of the cultural impact produced by the infrastructure. When relevant, the cultural benefits enjoyed by virtual visitors should be considered as well.

---

[80] See, for instance, Clawson and Knetsch (1966); Caulkins et al. (1986); Cuccia and Signorello (2000); and Bedate et al. (2004).

*Empirics*

The ex-ante estimation of the benefit enjoyed by personal onsite visits involves:

- Forecasting the number of visitors over the time horizon; and,

- Estimating the willingness-to-pay for a visit. As with recreational sites, the standard way to estimate the WTP is using the travel cost method (see box below) or the benefit transfer approach[81]. In the economic analysis, that the WTP replaces any possible revenues from visitors included in the financial analysis is worth noting.

Box 13. The travel cost method

The travel cost method (TCM) was suggested by Hotelling (1947) and developed by Clawson and Knetsch (1966) to assess the value of environmental resources and recreational sites[82]. TCM has also gained popularity in cultural economics, particularly regarding cultural heritage[83]. The method attempts to place a value on a non-market good by drawing inferences from expenditures incurred to 'consume' it, including the cost of the trip (e.g. fuel, train, or airplane ticket), the opportunity cost of time spent travelling, entry fees, on-site expenditures, and accommodation costs.

In particular, two types of TCMs exist: the 'individual demand approach' and the more common 'zone of origin approach' (Anex, 1995). The latter is the simplest and least expensive approach and is applied by collecting information on the number of visits to the site from different distances. This information is used to construct the demand function and to estimate the economic benefits for the recreational services of the site. Instead, the individual demand approach uses survey data from individual visitors in the statistical analysis rather than data from each zone.

Although widely adopted, the TCM is affected by a major limitation that should be carefully addressed. The limitation is related to the apportionment issue arising whenever it is reasonable to assume that a trip is made for different reasons (a multi-purpose trip) and not to visit a specific RDI infrastructure. Actually, disentangling the willingness to pay of visitors for a given infrastructure when more than one attraction is located in the same site or in the same area could be arduous.

The ex-ante estimation of the benefits enjoyed by virtual visitors involves the following activities.

- The number of virtual visitors is forecasted over the time horizon. First, the possible communication mediums people can use to virtually approach the infrastructure, i.e. social media, website, television, and radio, are established. Second, virtual visitors per type of mean is forecasted through proper techniques commonly used by marketing specialists, such as by using the number of 'tweets' or followers in Twitter, posts or pages in Facebook, subscribers of the YouTube dedicated channel or number of views of a video, estimated number of people watching an event on television, number of blog conversations, volume of web traffic, registrations on the official website, and so on.

- The willingness-to-pay for a virtual visit is estimated[84]. A broadly used method to attach a monetary value to non-market goods is contingent valuation. Contingent valuation consists of asking people to state the maximum amount they would be willing to pay to obtain a good or would accept as compensation to give away a good, contingent on a given scenario. However, empirical studies show that when consumers are accustomed to receiving an online service or content for free, their willingness to pay is very low or nil (see, for instance, Chyi, 2005). Difficulties in obtaining values of willingness to pay through contingent valuation have also been experienced in the cultural sector (Snowball, 2008). In this context, the choice experiment or conjoint analysis methods are considered more useful than traditional contingent valuations. Still based on stated preferences, these techniques imply asking a sample population to choose or rank different combinations of attributes of the same good (e.g. a museum, an archaeological site), for which the price is included as an attribute. This method enables the uncovering of preferences in terms of willingness to pay for each attribute and the entire set to be more effective. The same techniques could be usefully exploited to attempt to value public interest for the RDI infrastructure.

---

[81] On WTP of recreational sites/activities, see Sorg and Loomis (1984); Pearce (1993); Loomis and Walsh (1997); and Mendes and Proença (2005).
[82] For a review, see, for instance, Garrod and Willis (2001); Hanley and Barbier (2009); and Tietenberg and Lewis (2008).
[83] See, for instance, Alberini and Longo (2005); Bedate et al. (2004); Poor and Smith (2004); and Ruijgrok (2006).
[84] On WTP for social media, see Westland (2010); Han and Windsor (2011); and Vock et al. (2013).

| 13. Cultural effects of on-site visits |
|---|

A research facility hosting a particle accelerator is expected to organise 40 guided tours per years for general public. The tours are free of charge. The project promoters have forecasted that the average number of visitors per tour is 15 persons. No revenue from the visitors is recorded in the financial analysis, but the visitors' willingness-to-pay need to be accounted in the economic analysis. According to the travel cost method to estimate the WTP for a visit, the following information need to be collected:

- Breakdown of visitors by origin. According to a sample of experts, 80% of visitors are expected to come from an area with in a radius distance of 200 km and the remaining 20% from a longer distance.
- Breakdown of visitors by transport mode used. Assume that all the visitors coming from an area with radius distance of 200km are expected to travel by car, while the other visitors are expected to travel by bus (40%) or plane (60%).
- Estimate the cost of travel by transport mode. It includes the cost of fuel, tolls and other operating costs if travelling by car. Otherwise it includes the cost of ticket.
- Estimate the value of time spent in travelling. HEATCO reference values for leisure trips can be used.
- Estimate the cost of meals and the possible cost of accommodation for the share of visitors coming from larger distance.

Summing up:

|  | Transport mode | % of total visitors | Cost of travel (A/R) | Time (hour) | Value of time | Cost of meals and accommodation |
|---|---|---|---|---|---|---|
| <200 Km | Car | 80% | EUR 140 | 5 | 21 | EUR 15 |
| >200 Km | Bus | 8% | EUR 90 | 9 | 15 | EUR 20 |
|  | Plane | 12% | EUR 180 | 6 | 31 | EUR 100 |

The average WTP for different classes of visitors, i.e. coming from different origins and by different transport modes, is then multiplied to the share of expected number of visitors per year in order to obtain a valuation of the economic benefit. In a deterministic model, the total discounted value of visits is obtained in the following way:

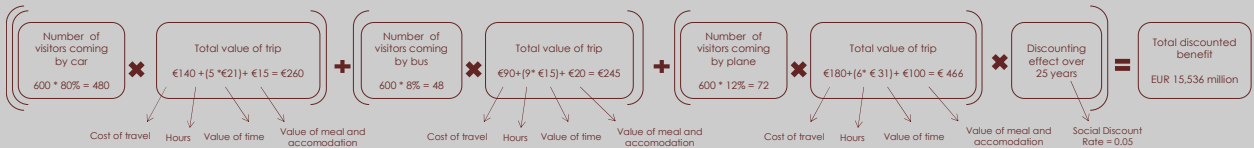

## 5.11 Non-use benefits: new knowledge as a public good

In most cases, for applied research and innovation infrastructures, the estimation of use-benefits ($B_u$) should probably be sufficient to justify the worthiness of the infrastructure in CBA terms, i.e. the (ENPV)>0. However, for a basic research infrastructure, an additional impact on social welfare may be related to its discovery potential. The unpredictable and maverick nature of discovery makes estimations of the results possible only in probabilistic terms and only to some extent. Basic research experiments are associated with a broad set of all possible outcomes, usually called the sample space, and each outcome has a probability of occurring (see Figure 7)[85].

---

[85] In this context, it is worth mentioning that when research hypotheses are not confirmed or designed discoveries do not occur, these results are, nevertheless, valuable and lead to at least the production of knowledge outputs, such as publications.

Figure 7. Illustrative example showing probabilities (Pr) associated with different research outcomes

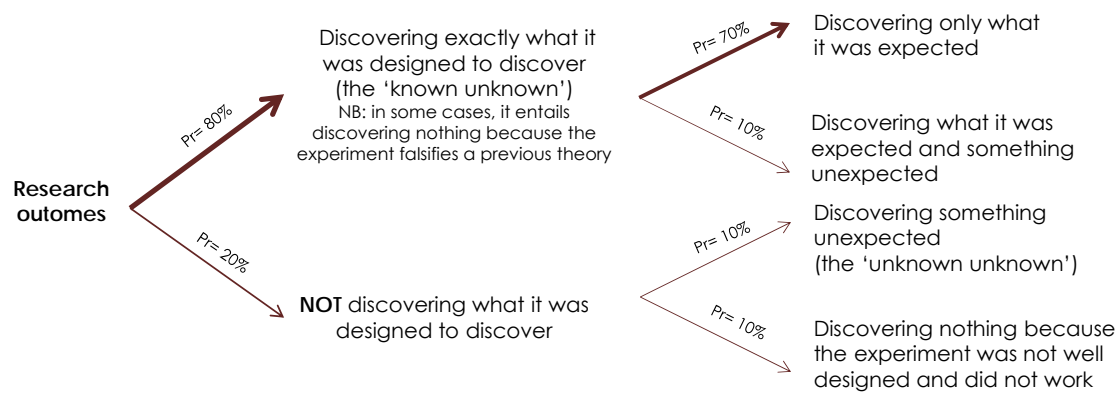

Note: Pr = probability. Source: Authors

Although the value of publications stemming from such discoveries crudely reflects, in statistical terms, the social benefit to scientists of advancing knowledge within their community, the discovery itself could have an intrinsic social value and could bring a number of further improvements to human wellbeing that have not been accounted for until now. These additional benefits are defined as non-use benefits and are captured by a residual term ($B_n$) in the present framework. This approach and terminology is borrowed by environmental economics, for which any good or natural resource can be assigned a total economic value[86] that, in turn, can be decomposed into the following two general classes.

- The *use value* refers to the direct or indirect benefits arising from the *actual use* of an asset or its potential or *option use*[87], which indicates the value attached to the future, based on known opportunities.

- The *non-use value* denotes the social value for simply preserving a natural resource compared with not preserving it[88]. Non-use value includes a *bequest value* that arises from the desire to preserve certain resources for the benefit of future generations[89] and an *existence value* related to knowing that a good simply exists even if it has no actual or planned used for anyone and is independent of any altruistic motives[90]. In addition, situations could also exist in which the practical use of a good can be expected in principle but is still unknown. In these cases, its value is determined by what is generally called '*quasi-option value*'.

---

[86] See Daily (1997); OECD (1999); Turner (1999); and Pearce *et al.* (2006).
[87] The concept of option use value was first introduced by Weisbrod (1964).
[88] The concept of non-use value was first noted by Krutilla (1967).
[89] On 'bequest value', see, for instance, Walsh et al. (1984) and Schuster *et al.* (2005).
[90] On 'existence value', see, for instance, Boyle and Bishop (1987) and Blomquist and Whitehead (1995).

Box 14. 'Option value' vs 'quasi-option value' in the literature

> The concept of option value originated in Weisbrod et al. (1964), where he responded to Friedman's (1962) advocacy of a policy of closing down a national park if the commercial value of lumber or minerals exceeded the willingness to pay for the recreational use. Weisbrod et al. (1964) argued that the WTP by recreation users of the park understates its value to society because many individuals expect they may visit the park and would be willing to pay for an option that guarantees their future access. Therefore, option value becomes significant under conditions of uncertainty regarding future demand and/or supply, but when a potential use is identified.
>
> In OECD (2006), the option value is formulated as the difference between the option price, i.e. the maximum WTP expressed now for something uncertain in the future, and the expected value of a consumer's surplus. In the literature, the concept of 'option value' is sometimes considered equal to the 'quasi-option value', which is confusing (Reiling and Anderson, 1980). Although the two values are interlinked and can coincide under certain conditions, they are built on slightly different assumptions, in particular for what concerns the irreversibility condition of the investment decision.
>
> The concept of 'quasi-option value' was introduced by Arrow and Fisher (1974) when studying how the uncertain effects of certain economic activities could be irreversibly detrimental to future environmental preservation. This concept describes the impact of a development intervention in one period on the expected costs and benefits in the next, i.e. the expected net benefits in future periods that are conditional on the realised benefits in the present period. Elaborating on this concept, Conrad (1980) highlighted the notion of quasi-option value as being equivalent to the expected value of information. The value of lost and new options allowed by an investment project implemented today is an expected value based on what one might learn. The same interpretation is found in Atkinson et al. (2006: 21), who defined the quasi-option value as the 'difference between the net benefits of making an optimal decision and one that is not optimal because it ignores the gains that may be made by delaying a decision and learning during the period of delay'. In the context of CBA of RDI, the same definition may apply to the unknown losses that may occur by delaying the same decision. Using the concepts of 'quasi-option value' originally conceived for environmental goods and natural resources to also value other categories of goods is not an entirely new concept. Arrow and Fisher (1974: 319) conceded that the quasi-option value is a general notion that may be applied outside environmental economics because it is linked to uncertainty, information, and irreversibility issues that affect decision making in general.

Retaining this terminology, the term $(B_n)$, i.e. the value of unknown non-use value effects, is determined through the sum of two different components: the quasi-option value (QOV) and the existence value (EXV), while if there is an option value for some specific benefit, this should be included as a use value.

More specifically, a RDI infrastructure may create a quasi-option value in the sense that it could generate discoveries that produce positive impacts that, however, cannot be estimated when the funding decision is made[91]. The irreversibility aspect is that once something is known, destroying such knowledge and going back to the previous state of the world is virtually impossible. Hence, the welfare effect in principle is the opposite of the Arrow-Fisher effect, which claims that quasi-option value leads to similar risk aversion results. In the RDI context, as information is created, the effect is similar to risk seeking, i.e. it increases the value of a project because it has the chance to discover something.

Figure 8. Preferences to risk

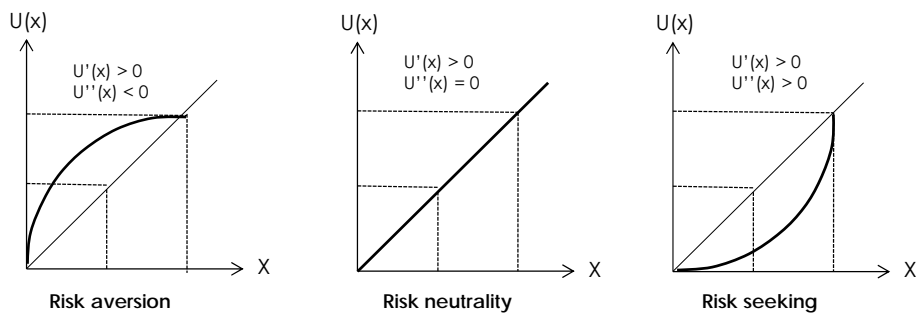

Note: First and second derivatives of a utility function related to good x determine the shape of utility and express preference for risk-taking.
Source: Authors.

In general, we expect that QOV is to be considered a benefit. However, in certain cases, the social preference may be not to know something, as reflected in legislation that forbids certain types of research on humans, for either ethical reasons or as a precautionary principle. Given the fact that, in general, the PDF of QOV is unknown, we suggest conservatively setting this component equal to zero. This assumption may be considered excessively prudent but also is consistent with the first principles of CBA. For example, building a new highway may have far-reaching broader effects in the future because the highway may create new opportunities to connect distant people. From this fact, new cultural and economic circumstances may arise. However, our present knowledge of this concept is too uncertain and usually it is not included in the CBA of a highway, i.e. it is implicitly set to zero, even if transport infrastructures in the long run will literally change the world over many dimensions.

---

[91] QOV usually remains completely unknown for a long time, even ex post.

In contrast, an *existence value* can be attributed to the discoveries of an RDI infrastructure, reflecting a social preference for pure knowledge *per se*. In other terms, EXV refers to the intrinsic value of knowing the object of the discovery, regardless of the fact that it might find some use sooner or later. Contrary to the QOV which is unknown, preferences for the existence value of discoveries in principle can be detected ex ante. In practice, when $ENPV_u < 0$, the net present value of the non-use benefits, proxied by the EXV of discoveries, should be greater than the net costs for the infrastructure to be deemed socially beneficial[92].

Box 15. 'Existence value' vs 'quasi-option value' in our framework

> Quasi-option value and existence value are two distinct concepts with the following main differences.
> - In principle, the quasi-option value could be either positive or negative, producing either an increase or a decrease in social welfare. However, we assume that advancing knowledge has at least a zero value and, in general, a positive one, unless for extreme situations in which such knowledge has potentially detrimental uses for society. In fact, the irreversible effect is new knowledge itself. Setting this component to zero represents taking a neutral attitude about unknown uses of new knowledge. Instead, existence value can always be regarded as intrinsically positive or at least nil: people can be expected to be better off or completely indifferent to a discovery, just for the pure value of such knowledge.
> - The two concepts have an empirical dimension. The quasi-option value for the unknown effects of a discovery is usually completely uncertain ex ante and, thus, no preferences can be imputed to it as long as the effects remain unknown. However, ex post, the value will be revealed by new information. Instead, people could have some ex-ante preferences about knowing that something is discovered; they are unlikely to have preferences if they do not know or understand the issue at stake. However, if they obtain 'some' information, then assuming that preferences will arise that allow them to choose between two states of the world is reasonable, i.e. one state in which the scientific discovery occurs and one state in which it does not. Indifference is also possible. In the latter case, the existence value is also zero.

*Empirics*

The standard method for estimating non-use values for which no observable price system exists is to recur to stated preference techniques[93]. In particular, the use of a contingent valuation methodology has found widespread application in environmental economics for estimating the economic value of species and natural resources[94]. In the same vein, one could attempt to grasp the WTP of taxpayers having preferences for having a discovery, regardless of its actual or potential use.

There are two possible objections to contingent valuation[95]. The first objection concerns the cost required to perform a proper contingent valuation exercise. Although this consideration is important, the typical cost per capita of a well-designed contingent valuation is often a modest fraction of the overall cost of the infrastructure in the first place, particularly for the large ones.

A second objection is that asking individuals their WTP for the mere existence of any good may not be easy and may bias the results in a number of individual, cultural, and socio-economic circumstances (Carson and Groves, 2007; Carson, 2012). To address these issues, the evaluator should take into account a number of recommendations developed since the early nineties, particularly those followed by a panel of distinguished economists for the US National Oceanographic and Atmosphere Administration (NOAA, 1993), including indications about the modalities and structure of the interviews. In
NOAA panel guidelines adapted to the RDI infrastructure evaluation context, we slightly adapt the NOAA panel guidelines to our context.

---

[92] This statement holds true under the reasonable assumption that the EXV is always positive.
[93] Stated preference techniques involve soliciting responses to hypothetical questions regarding the value that people place on goods. Thus, they are based on answers given by a representative sample of the population of interest to derive a respondent's willingness to pay for a good. Within the class of stated preference methods, two main alternative groups of techniques exist: choice modelling and contingent valuation. The latter seeks measures of willingness to pay through direct questions such as, 'What are you willing to pay?' and 'Are you willing to pay €X'? The former seeks to secure rankings and ratings of alternatives from which WTP can be inferred (Bateman et al., 2002).
[94] Empirical studies that used contingent valuation include, for example, Vasely (2007); Togridou et al. (2006); Sattout et al. (2007); Carson (1998); Walsh et al. (1984); and Greenley et al. (1981).
[95] The literature on contingent valuation has debated numerous issues. Reviews of these debates can be found in Carson et al. (2001), Portney (1994), and Mitchell and Carson (1989).

Box 16. NOAA panel guidelines adapted to the RDI infrastructure evaluation context

- The target population should be identified, and could be the entire population of the country (or region) in which the infrastructure is located or a defined group of people in a reference geographical area (world, nation, region). Because new knowledge can be a global public good, in some cases all of humankind is the potential beneficiary, but only tax-payers of some countries would in fact support the project.
- The sample type and size should be identified, which should be the closest practicable approximation to the target population and might consist of, for example, all taxpayers in a region, all students from certain universities, and all subscribers to certain magazines. Appropriate sampling is essential. This selection process involves the use of a randomised procedure.
- Careful pretesting should be done because the interviewers may contribute to 'social desirability bias' in the event that face-to-face interviews are used to elicit preferences. To avoid the bias that certain things may be broadly viewed as something positive, pretesting for the interview effect should be done. Pretesting is also essential to verify whether respondents understand and accept the context description and the questions.
- The purpose of the questionnaire should be stated and an accurate description of the RDI infrastructure, its mission, and research potential should be provided, which ensures that respondents understand the context, are motivated to cooperate, and are able to participate in an informed manner. The use of pictures could help.
- The payment vehicle, i.e. the manner in which the respondent is (hypothetically) expected to pay for sustaining the infrastructure's research activity/mission should be described.
- The elicitation format, for which open-ended, bidding game, payment card, and single-bounded or double-bounded dichotomous choices are the most broadly used formats, should be carefully selected. Dichotomous (i.e. 'referendum-like') valuation questions that allow for uncertainty by including a 'don't know' option are recommended by the NOAA panel.
- Follow-up questions should be inserted, which are essential to understanding the motives behind the answers to WTP elicitation questions.
- Questions allowing for cross-tabulations should be inserted, including a variety of other questions that assist in interpreting the responses to the primary valuation questions. For example, income, education level, prior knowledge of the infrastructure, prior interest in the issue, and attitude towards RDI are items that would be helpful in interpreting the responses.
- Both sample non-response and item non-response should be minimised. A reasonable response rate should be combined with a high but not forbidding standard of information.
- The conservative approach should be preferred: when the analysis of the responses is ambiguous, the option that tends to underestimate WTP is preferred. Similarly, the reliability of the estimate should be increased by eliminating outlier answers that can implausibly bias the estimated values.

Alternatively, to overcome the difficulty of explicitly stating a willingness to pay, in some circumstances, valuation methods based on the revealed preference can be conveniently employed. These methods assume that the existence value can be determined through the observation of economic behaviour in a related market, such as voluntary contributions to organisations devoted to the preservation of a public good. For instance, in the RDI context in some countries, several scientific institutions are supported by taxpayers who can name a charity or a research body to which a percentage of their taxable income is donated (Florio and Sirtori, 2014). Additionally, universities regularly receive donations for research from firms and individuals.

A third approach, not necessarily an alternative to the previous approaches, is to recur to benefit transfers. In this case, a meta-analysis of contingent valuation studies on the existence value of goods produced by other projects is used to establish a benchmark median value or a range of values. If the project is well within the range or in the median to lower bound of the range, the guess is made that the project is as beneficial as other goods for which empirical analysis of an existence value is available (Florio and Sirtori, 2014).

Box 17. The Environmental Valuation Reference Inventory

The Environmental Valuation Reference Inventory (EVRI) is a searchable storehouse of empirical studies on the economic value of environmental benefits and human health effects. The EVRI was developed in the 1990s by Environment Canada ( a governmental body) in collaboration with a number of international experts and organisations as a tool to help policy analysts use the benefits transfer approach to estimate economic values for changes in environmental goods and services or human health. Currently, the database makes available more than 4,000 valuation study records contained in more than 30 fields. The main categories in an EVRI record include study reference, study area and population, environmental focus, study methods, a table of estimated values, and an abstract (McComb, 2006). In particular, the EVRI's Searching Module helps the user identify studies with the potential for transfer, whereas the Screening Module helps the user assess the suitability of the studies identified in the search according to criteria outlined in the benefits transfer literature. The EVRI can be visited at www.evri.ca.

As a final remark, attention should be given to the fact that some research projects are much more appealing or in vogue than others, even if less promising in terms of the probability of achieving the designed results. This interest is reflected in the estimate of their existence value regardless of the approach followed. For instance, fighting climate change, finding a remedy for Alzheimer's, and discovering exoplanets are appealing issues that could be attached to higher value relative to other research quests that address less known issues, have less visibility, and have a weaker impact on the general public. This possible bias needs to be taken into account when estimating the existence value of discovery, and the information setting of the contingent valuation should be carefully designed.[96] These examples also suggest that, in fact, WTP as elicited by a CV approach will usually be a blend of perceived QOV and of pure EXV. This fact *per se* is not disturbing, provided that the design and interpretation of the survey results are careful. While there is wide experience of these methods in other fields, experimenting them in the RDI domain will need several adaptations and a learning by doing process. Against a relative modest cost, the advantage of this approach is to elicit tax-payers preferences in quantitate terms, and the correlations with individual and social features, which is *per se* important information in a science policy perspective.

---

[96] The information and the questions provided in the questionnaire should not have emotionally charged effect.

## 14. Existence value

A new terrestrial exoplanet research institute equipped with a Large Binocular Telescope Interferometer is planned in country LAMBDA. It will be entirely supported by the national government budget. It has a mission to explore terrestrial bodies and seek out the existence of potential life beyond the confines of our own solar system. The net present use-value of this government-owned research infrastructure is negative: EUR –40 million discounted.

There are however a number of people in the general public interested in discovery of exoplanets, as suggested by website and other media data. In order to elicit people willingness to pay for the potential discoveries the project promoters design and carry out a contingent valuation survey. Specifically, the survey was performed in NOAA-panel referendum format on a random of 600 taxpayers of country LAMBDA (counting a population of 5 million taxpayers) drawn using a simple random selection method in such a way as to be representative of the population. Thanks to the face-to-face interaction, the survey showed a quite high response rate of 80%.

First, respondents have been inquired about their personal and household incomes with the aim to measure their financial means and possible contribution to support the institute. Then, their previous knowledge and interest for the topic of research infrastructure has been investigated before going to the specific case of exoplanet research institute on which a summary information, including visual material, is given. Finally, the willing to contribute to support the potential discoveries of the institute is elicited.

In line with NOAA recommendations, the "No-answer" option was allowed and followed-up by questions with the aim to non-directly explain the choice. The answers to these follow-up questions suggest individuals' rough indifference towards the survey matter. Thus, the "No-answers" were not considered in the calculation of the average WTP. The survey results take the following form:

| Classes (EUR per year) | Average EUR per year | Relative frequency |
|---|---|---|
| 0 | 0 | 10% |
| 0.1 - 0.5 | 0.3 | 10% |
| 0.6 – 1.0 | 0.8 | 45% |
| 1.1 -1.5 | 1.3 | 20% |
| 1.6 -2.0 | 1.8 | 15% |

As an example, summaries of cross-tabulations of willingness to pays and on how important is for the respondent is investing in R&D infrastructure is presented in the following table.

| Response category | % with WTP 0 | % with WTP 0.3 | % with WTP 0.8 | % with WTP 1.3 | % with WTP 1.8 |
|---|---|---|---|---|---|
| Not important at all | 85% | 60% | 40% | 30% | 15% |
| Slightly important | 5% | 10% | 10% | 15% | 10% |
| Moderately important | 5% | 10% | 20% | 30% | 15% |
| Very important | 0% | 10% | 15% | 15% | 30% |
| Extremely important | 0% | 5% | 12% | 8% | 25% |
| Don't know | 5% | 5% | 3% | 2% | 5% |

Statistical inference from the sample to the population of LAMBDA's taxpayers is done. Hence, considering a time horizon of 20 years, the total discounted existence value for the discoveries of the exoplanetary research institute is then:

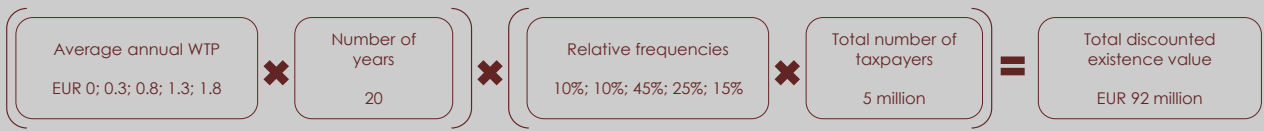

Average annual WTP: EUR 0; 0.3; 0.8; 1.3; 1.8 ✕ Number of years: 20 ✕ Relative frequencies: 10%; 10%; 45%; 25%; 15% ✕ Total number of taxpayers: 5 million = Total discounted existence value: EUR 92 million

Table 13. Examples of existence value retrieved from the literature

| Category | Good | Country | Average WTP (EUR per person, per year) | Source |
|---|---|---|---|---|
| Cultural goods | British Library | United Kingdom | 7.62 | British Library (2004) |
| | Royal Theatre of Copenhagen | Denmark | 27.69 | Hansen (1997) |
| | The Arts, Kentucky | USA | 3.95 | Thompson et al. (2002) |
| Environment - Habitats | Wilderness areas, Portugal | Portugal | 16.64 | Nunes (1999) |
| | Ecological agricultural fields | The Netherlands | 9.87 | Wiestra (1996) |
| | Desert protection in California | USA | 45.31 | Richer (1995) |
| | Protection of Peat Meadow Land | The Netherlands | 25.36 | Brouwer (1995) |
| | Protection of the Kakadu Conservation Zone and national park | Australia | 29.25 | Carson et al. (1994) |
| | Grand Canyon | USA | 33.32 | Pearce (1993) |
| | Colorado Wilderness | USA | 18.40 | Pearce (1993) |
| | Increased forest biodiversity conservation | Finland | 81.12 | Lehtonen at al. (2003) |
| | Preserve the Rakaia river in its existing state | Australia | 9.21 | Sharp and Kerr (2005) |
| Environment - Single species | Bald Eagle | USA | 15.55 | Pearce (1993) |
| | Whooping Crane | USA | 1.50 | Pearce (1993 |
| | Grizzly bear | USA | 23.19 | Pearce (1993) |
| | Coyote | USA | 6.59 | Stevens et al. (1991) |
| | Salmon | USA | 9.49 | Stevens et al. (1991) |
| | Gray Whale | USA | 19.75 | Loomis and Larson (1994) |
| | Wolf in Sweden | Sweden | 55.80 | Boman and Bostedt (1995) |
| Environment - Various species | Endangered species in West Germany | Germany | 66.64 | Hampicke et al. (1991) |
| | Preservation of 300 endangered species in Sweden | Sweden | 125.23 | Johansson (1989) |
| | All endangered species in Victoria | Australia | 46.58 | Jakobsson and Dragun (1996) |

Source: Authors based on cited sources.

# 6. The net benefit test

## 6.1 Estimating the probability distribution of the economic net present value

Once the inflows and outflows (financial and economic) associated with the infrastructure have been identified and their baseline values have been valued in monetary terms, the usual financial and economic performance indicators of the project are computed.

- Similar to financial performance (see section 2.4, 'Financial profitability and sustainability'), the investment's economic performance can be calculated using the following indicators:

- the economic net present value (ENPV), expressed in monetary terms, which is defined as the difference between the discounted total social benefits ($B_u + B_n$) and costs;

- the economic internal rate of return (EIRR), which is the specific social discount rate value that produces an ENPV equal to zero; and,

- the benefit-cost (B/C) ratio, i.e. the ratio between the discounted benefits and costs.

In particular, a project with positive performance, i.e. a project that shows a positive return to society, is associated with the following results:

- The ENPV is higher than zero – the higher the ENPV the larger the social benefits achieved, net of costs and negative externalities.

- The EIRR is higher than the adopted social discount rate.

- The B/C has a value higher than one.[97]

However, from an ex-ante perspective, the probability of an error related to each forecast and estimate included in the analysis should be considered to be high. To address this issue, a full-fledged quantitative risk assessment is required[98], meaning that costs and benefits become part of a probabilistic (stochastic) model instead of a deterministic one. The baseline ENPV is calculated as the discounted sum of a set of 'most likely' (or 'best guess') values, subjectively assumed by the project promoter in accordance with the empirical evidence and given his/her knowledge about the specificities of the infrastructure. Instead, the probabilistic model requires assigning each critical variable a specific probability distribution. As a result, the probability distribution of the outcome of interest, i.e. ENPV or EIRR, is considered to assess project performance instead of punctual performance indicators based on baseline values.

The steps necessary to perform a risk assessment and, in turn, to estimate the probability distribution of the ENPV are discussed in section 5.4, 'Uncertainty of the social impacts of research: the role of risk analysis'. In section 5.5, 'How to present the CBA results of RDI infrastructures', the distribution functions of the CBA result indicators and the simple statistics resulting from the risk analysis are presented.

## 6.2 Time horizon

Some of the benefits produced by the RDI infrastructures last and even materialise beyond the operational phase of the project (i.e. after it useful technical or economic life) and, in some cases, are permanent. In principle, this phenomenon implies setting a time horizon for the analysis that is longer than the infrastructure's life cycle, or even one that is infinite.

For instance, when a major telescope detects a new phenomenon in the sky, an accelerator observes a new type of particle, or a therapy to cure a cancer pathology is discovered, such knowledge exists forever and is transmitted generation after generation to future researchers without any clear endpoint. Thus, the accumulation of knowledge has a longer – possibly almost infinite – time horizon than the period of operation of the infrastructure[99]. However, an infinite time horizon leads to a paradoxical result because any large investment cost spread throughout a finite range of years is less than the sum of any small benefit spread over an infinite time horizon, whatever the non-strictly positive social discount rate.

---

[97] A B/C ratio that exceeds one indicates that the project has a positive ENPV. Indeed, the B/C ratio uses the same information as the ENPV but presents it in a slightly different manner. However, the B/C ratio can be misleading when ranking projects, which are mutually exclusive alternatives because they do not account for the scale of the project (Boadway, 2006). Conversely, the NPV is also an appropriate criterion for choosing from a group of comparable projects. In this case, the objective is to select the project that maximises the NPV (de Rus, 2010).
[98] The need to adopt formal risk-based approaches to address an uncertain future is widely recognised by the literature (Pouliquen, 1970; Reutlinger, 1970; Clarke and Low, 1993; and Savvides, 1994).

Another reason for the adoption of a finite reference period is related to the obsolescence process of both the equipment and the value of knowledge over time. For example, this obsolescence is observable in the trend of citations of both scientific publications and patents. In retrospect, indeed, we know that past discoveries and inventions have lost some of their scientific/technological value and have been surpassed by current knowledge and technology.

As a result, for the RDI infrastructure to be able to demonstrate a positive economic return, assuming a long but finite time horizon seems reasonable, such as one that ranges from 20 to 30 years depending on the nature of the project. Generally, the end year can be set by considering the useful life of the main equipment. Specifically, the following criterion applies: when extraordinary maintenance of the main equipment or machine became so frequent and expensive that replacing them with new ones is more convenient, the equipment/machine has arrived at the end of its life. Additionally, considerations in terms of advancements in the scientific field and subsequent possible obsolescence of the technology used should be considered.

To take into account the fact that some assets and benefits from the end of the period of analysis still have an expected value, a residual value should be computed. Similar to the residual value of a fixed asset whose economic life is not yet completely exhausted – which reflects the remaining service potential[100] – the project effects that last beyond the reference time horizon should be included in the last year of the analysis as a residual value.

Box 18. Residual value

> The residual value of a project reflects the remaining value of the investment at the end of its project lifetime. Since an investment involves many fixed assets, the residual value should be calculated through its asset components, i.e. by calculating a residual value for each infrastructure item and then summing them (Jones et al. 2013). The following different calculation methods exist.
> - The European Commission (2014) recommends calculating residual value as the residual market value of fixed capital as if the capital was sold at the end of the time horizon, in other words, the discounted value of all net future revenues.
> - In practice, the residual value of fixed assets is frequently calculated as the present value of expected net cash flows during the years of the economic life outside the reference period, i.e. the considered time horizon, if the economic life exceeds the project lifetime period.
> - Another method consists of estimating the amount that an entity would currently obtain from the disposal of the assets net the estimated cost of disposal.
> - A simple and commonly used method is the straight-line depreciation method, in which the residual value is equal to the non-depreciated amount of the asset, and the concept of remaining service life (RSL) is exploited. The formula is: residual value = [(RSL/total service life) * initial capital cost].
>
> In the RDI context, in which some benefits may continue after the infrastructure's decommissioning (such as technological spillovers on firms and human capital effects on former students and young researchers), the calculation of the residual value should not only rely on the remaining value of the fixed capital but also on the discounted value of the benefit that exceeds the project's time horizon. Indeed, even if such effects occur beyond the CBA's time horizon, they have been generated as a result of the infrastructure.

# 6.3 The social discount rate

The long time span of effects influences the decision over the most appropriate social discount rate[101] to adopt to discount economic flows to reflect the social view on how benefits and costs are to be valued against present ones. In most CBA practices, a constant social discount rate is used, which implies an exponential discounting process of the project's flows. In other words, benefits occurring far in the future are discounted more than the costs of investments that, instead, typically occur in the initial years of the time horizon. This greater discounting could lead to negative project performance indicators and to the decision of not implementing the project, disregarding the concept that the same project may bring significant benefits to the welfare of future generations.

---

[99] Possible similarities can be found in environmental economics and, in particular, the economics of climate change, in which a few hundred years are frequently considered to evaluate the impact of a policy intervention. For example, the Stern Review (HM Treasury, 2006) sets a time horizon of 200 years. Another 'extreme' example is suggested by Boardman et al. (2006). The latter, to stress that the benefits of some projects may continue to flow from many years even if the project is finished from an engineering or administrative perspective, mentions the case of roads originally laid down by the Romans more than eighteen centuries ago and that are still the basis of contemporary motorways.

[100] The residual value should not be taken into account unless the asset is actually liquidated in the last year of the analysis.

[101] The social discount rate reflects the inter-temporal opportunity cost of capital for the entire society and is used in the economic analysis to discount flows. Different approaches exist in the literature to estimate the social discount rate. The most popular approaches are the social rate of return on private investments and the social rate of time preferences.

One possible way to address the issue raised by a constant rate is to adopt a sufficiently low discount rate. One example is in the Stern Review on the Economics of Climate Change (HM Treasury, 2006). See The social discount rate in the Stern Review of Climate Change.

Box 19. The social discount rate in the Stern Review of Climate Change

> The Stern Review adopts the Social Rate of Time Preference (SRTP) approach and assumes a near-zero pure time preference (equal to 0.1%), representing the possibility of extinction of the human race. The elasticity of the marginal utility of consumption is set to 1 and the expected rate of future per capita growth is assumed equal to 1.3% per annum, in accordance with historical data. When these values are entered into the SRTP formula[102], they result in a rate of 1.4% in real terms. However, the social discount rate estimated by the Stern Review has been criticised for being too low. In fact, it does not consider the impatience component, which implies that a greater weight be attached to the present generation. Moreover, to be consistent with the saving rates, the elasticity of the marginal utility of consumption should be higher (Nordhaus, 2007; Weitzman, 2007; Dasgupta, 2007 and 2008).

Another possibility suggested in the literature is to use a declining discount rate, following a hyperbolic discounting function (Laibson, 1977) and characterised by relatively high discount rates in the short term and relatively low rates over long horizons. Figure 9 exhibits some declining social discount rate patterns proposed in the economic literature.

Figure 9. Some declining social discount rates

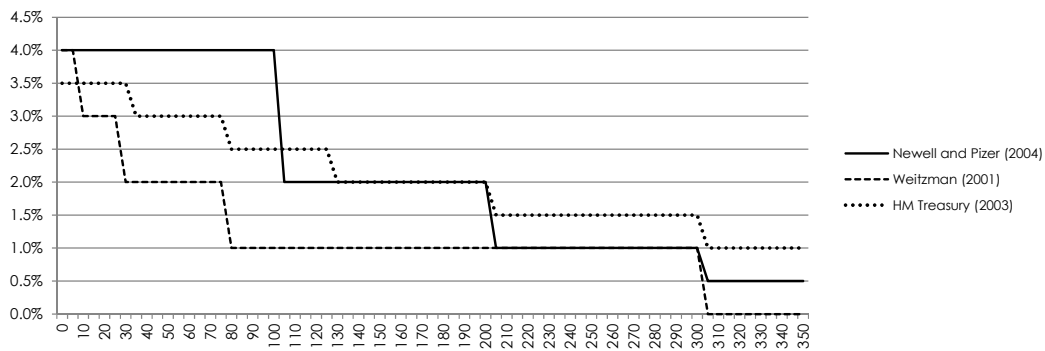

Source: Authors based on cited sources.

With respect to empirical estimation, significant variations in the social discount rate adopted by different governments exist depending on the estimation method used and the specific underlying parameters (see Table 14). However, when assessing the RDI infrastructure, a further aspect that should be considered when choosing the most suitable social discount rate is that, in most instances, the rate cannot be country specific because the project benefits spread relatively quickly and globally – for instance, consider the technological spillovers and knowledge outputs.

---

[102] The SRTP formula is based on a formula obtained from the Ramsey growth model. SRTP = ρ + ε* g, where ρ is the pure time preference, ε is the elasticity of the marginal utility of consumption, and g is the expected growth rate of per capita consumption.

Table 14. Social discount rate adopted in selected countries

| Country | Social discount rate | Estimation approach | Source |
|---|---|---|---|
| Australia | 8%, with sensitivity test over the range 3-10% | Social rate of return on private investments | Harrison (2010) |
| Canada | 8%, with sensitivity test over the range 3-10% | Social rate of return on private investments | Treasury Board Secretary Canada (2007) |
| China | For short and medium term projects: 8%. For long term projects: 8% | Weighted average | Zhuang et al. (2007) |
| European Union | 3% for cohesion countries and 5% for other member states | Social Rate of Time Preference | European Commission (2014) |
| France | 4% (declining after 30 years) | Social Rate of Time Preference | Quinet (2007) |
| Germany | 3% | Social Rate of Time Preference | Florio (2006) |
| Italy | 5% | Social Rate of Time Preference | Florio (2006) |
| Netherland | 4% | Social rate of return on private investments | Florio (2006) |
| Slovak republic | 5% | Social Rate of Time Preference | OECD (2007) |
| UK | 3.5% (declining after 30 years) | Social Rate of Time Preference | HM Treasury (2003) |
| USA (office of management and budget) | 7% | Social rate of return on private investments | Zhuang et al. (2007) |

Source: Authors adapted from Florio (2014).

## 6.4 Uncertainty of the social impact of research: the role of risk analysis

To account for the uncertainty and risk that characterises the future, the risk assessment is generally required as the last step of the project's ex-ante appraisal (see *inter alia* Zerbe and Dively, 1994; de Rus, 2010; Florio, 2014).

In general, deviations from predictions can be related to a number of different causes. A broad distinction exists between endogenous (e.g. errors incurred ex ante or during project implementation) and exogenous (e.g. changes to the project context caused by an unpredictable event) sources of risks (Florio, 2014). Whereas the latter is hardly predictable and, thus, outside the control of the project manager and evaluator, the former can be identified and minimised ex ante. According to Flyvbjerg *et al*. (2003), three broad categories of explanations exist for endogenous errors: technical, psychological, and political-economic[103]. Also available is the intrinsic uncertainty associated with the forecast of future trends. All of these factors can potentially affect the appraisal exercise to provide the decision maker with a more or less biased outlook of the expected project outcomes (Flyvbjerg, 2013). Even if some of the mentioned causes of deviation from predictions, such as accidents, are unpredictable, the sources of uncertainty can at least be identified to enable the parties to be ready to react and able to manage them if they appear. Instead, when a probability distribution function reflecting the degree of risk can be assigned to an event, doing so can be embedded in a risk assessment.

However, for the RDI infrastructure, the uncertainty of the CBA results can be even greater than that for other traditional infrastructures. This result is affected by the following range of factors.

- Frequently, RDI infrastructures are designed to be unique and breakthrough compared with the past; therefore, no references are available or suitable enough to forecast future trends.

---
[103] Technical explanations refer to errors and pitfalls in forecasting techniques; psychological explanations refer to planning fallacy and optimism bias (i.e. the tendency ex ante to overestimate benefits and underestimate costs and timing for a project); political-economic explanations refer to the fact that project promoters may deliberately overestimate benefits and underestimate costs when forecasting project outcomes.

- The experiments and tests are subject to some probability of success or failure (see Figure 6). Actually, research, development, and innovation involve the generation of new knowledge, and the probability of success depends on many factors, not all of which are under the direct control of RDI projects' promoters or are easy to be forecasted.

- Experiments and tests could generate discoveries that produce positive impacts that cannot be estimated when the funding decision is made.

- Only some experiments and tests are planned from the beginning and a new set of services and possible uses of the RDI infrastructure could arise during the RDI life cycle. This potentiality makes a RDI infrastructure intrinsically different from a standard infrastructure that, at least to a certain extent, is designed from the beginning to deliver a relatively well-defined set of services.

The two latter points describe inherently unpredictable effects. They have a quasi-option value (see *Boxes 15 and 16*); however, because forecasting the probability of occurrence and magnitude of this quasi-option value is usually impossible, the convenient working hypothesis is simply to assume that their values are non-negative and skip them. Conversely, the uncertainty characterising the first situation can be defined as measurable because it can be embedded into a stochastic model through probability distributions and, thus, tested using a risk assessment.

The set of procedures for overall risk assessment is traditionally split into three steps.

The first step is a sensitivity analysis. The impact of each variable entering the analysis on the predefined outcome measure (such as ENPV or the EIRR) is assessed by changing each 'best guess' value in absolute terms or by arbitrary percentages, one by one. Having then set the criterion for deciding whether the variation in the output is sufficiently large, the most critical variables for the CBA can be identified. The European Commission (2014) suggested focusing on a neighbourhood of 1% around the 'best guess'. In other words, a 1% change in the value of the independent inputs of the CBA is assessed, and the variables leading to a greater than 1% change in the outcome measure are considered critical. However, a good practice is to consider a range of percentage variations around the best estimate on a continuum scale ranging from, e.g. –10% to +10% (Florio, 2014). This practice allows detecting the non-linear and non-symmetric effects of the variables on the project outcome (see *Figure 10*). A specific category of the sensitivity test is the scenario analysis (see

- Scenario analysis on this matter).

Figure 10. Sensitivity analysis plot

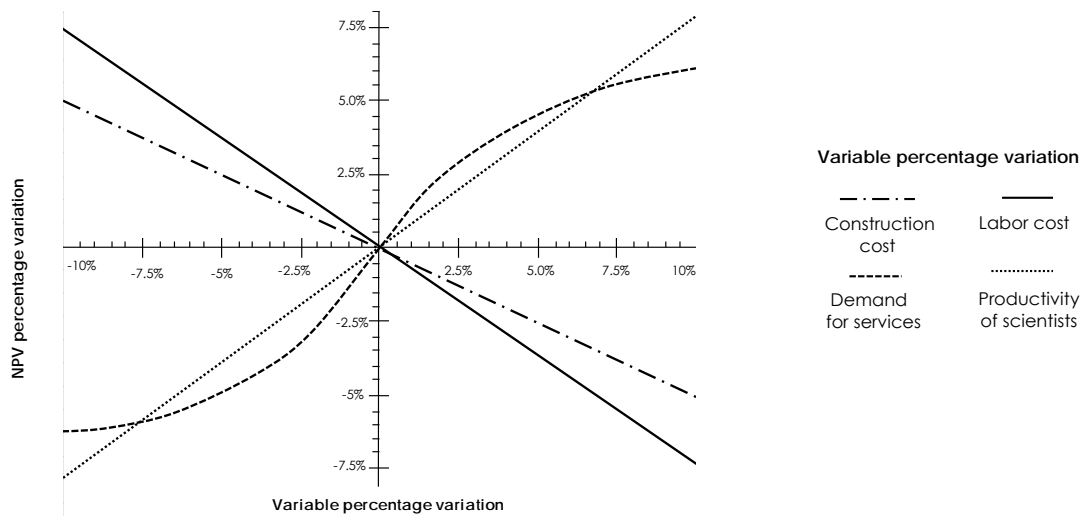

Source: Authors

Box 20. Scenario analysis

> Scenario analysis, also called 'what-if' analysis (Vose, 2008), is a specific form of sensitivity test. Whereas the influence of each independent variable on project performance is tested separately under standard sensitivity analysis, scenario analysis studies the combined impact of sets of values assumed by the critical variables[104]. In particular, combinations of 'optimistic' and 'pessimistic' values of a group of variables could be useful to build extreme scenarios and calculate the extreme limit of the project performance indicators. To define the optimistic and pessimistic scenarios, the extreme values defined by each critical variable's distributional probability should be used.

- Second, a range of variations and a specific probability distribution function are assigned to each identified critical variable. Probability distributions are highly dependent on the specific type of project under evaluation and may be determined from various sources of information, including experimental data, distributions found in the literature and adopted in projects similar to the one under assessment, and time-series or other types of historical data (Vose, 2008). When insufficient data exist to construct probability distributions, the range and likelihood of possible values rest on project promoter and evaluator judgments. Some exemplificative probability distributions are provided in Figure 11.

Figure 11. Example of probability distributions

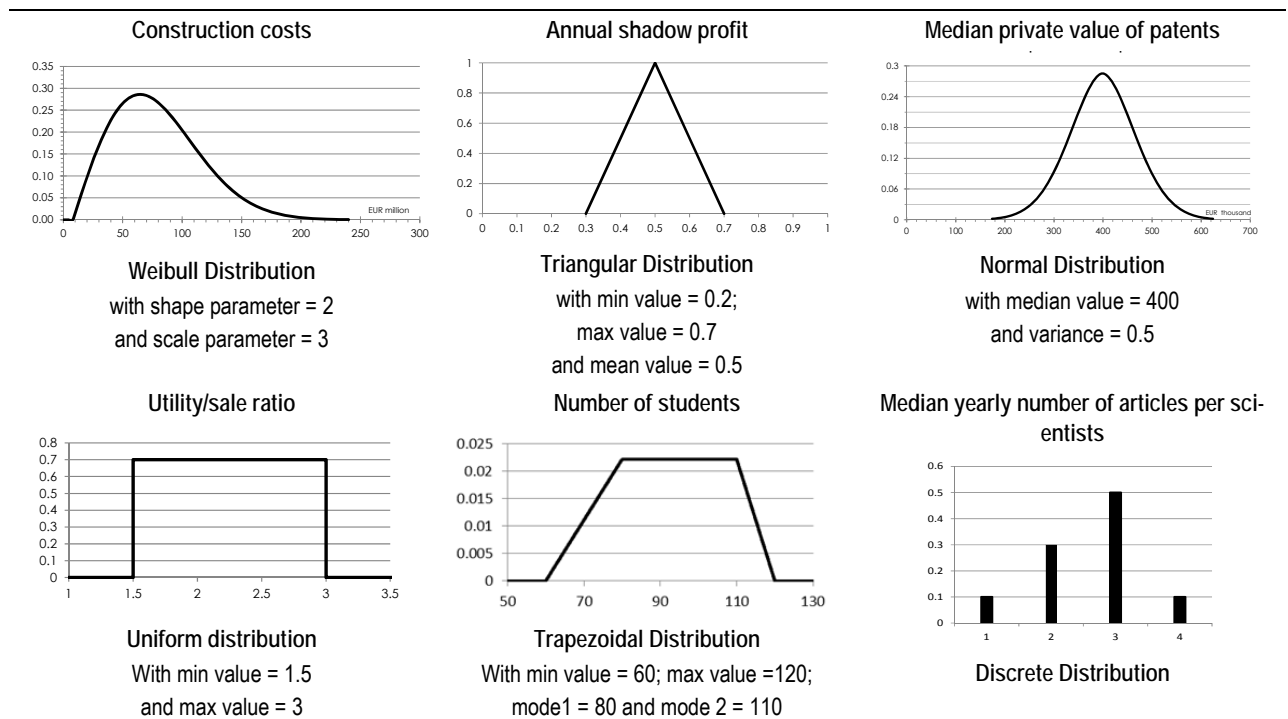

Source: Authors

- Third, the project's riskiness is assessed using the Monte Carlo simulation method[105], which allows estimations of the integral corresponding to the probability distribution function of the project performance indicator of interest (e.g. ENPV). By extracting one value of each critical variable from the respective cumulative distribution function and plugging it into the CBA model, the associated NPV is computed. This process, if repeated over a large number of iterations, leads to the probability distribution of the project ENPV. In other words, through the law of large numbers, which implies the convergence of the ENPV empirical distribution to its 'true' counterparts, the CBA result can be considered in probabilistic terms and the minimum, maximum, mean, and standard deviation values of the NPV can be computed.

Table 15 contains a list of typical risks affecting RDI projects and the relative variables that are likely to be critical and should be tested in the sensitivity analysis. As a final remark, it is worth mentioning that, in addition to the riskiness attached to variables entering the social CBA on either the cost or the benefit side, the results of the CBA model for RDI infrastructures could be strongly influenced by two important parameters: the time horizon of the analysis and the social discount rate. Our indication is to assess the var-iation in the CBA results also subject to different assumptions on the length of the time horizon and of the chosen discount rate and discounting function (exponential, hyperbolic, or others).

---

[104] Defined as critical through the standard sensitivity analysis.
[105] For a review, see, for instance, Robert and Casella (2004) and Balcombe and Smith (2011).

Table 15. Typical risks and critical variables in RDI infrastructure projects

| Stage | Risk | Critical variables |
|---|---|---|
| Design and construction | • Inadequate site selection<br>• Inadequate design cost estimates<br>• Project cost overruns<br>• Delays in completing the project design/in construction/during installation of equipment<br>• Accidents | • Number of years necessary for the construction of the infrastructure<br>• Investment costs |
| Operation | • Delays in making the equipment fully and reliably running<br>• Unexpected environmental externalities/accidents<br>• Insufficient production of research results<br>• Lack of academic staff/researchers<br>• Demand of students/young researchers different than predicted<br>• Demand of industrial users different than predicted<br>• Interest in general public different than predicted | • Expected incremental shadow profit;<br>• Number of patents expected to be registered over the project time horizon;<br>• Economic value of patents;<br>• Survival rate of spin-offs/start-ups;<br>• Number of stakeholders benefitting from technological spillovers;<br>• Expected incremental profit earned by suppliers;<br>• Number of scientific publications expected to be produced over the project time horizon;<br>• Estimate of the unit economic value of scientific publications;<br>• Average number of citations received by scientific publications;<br>• Number of young researchers and students benefiting from human capital development;<br>• Expected incremental salary obtained by students as a result of human capital development over their professional career;<br>• Size of targeted population;<br>• Avoided cost or WTP for reduced environmental or health risk;<br>• Forecast of the success rate of the project;<br>• Estimated WTP of visitors;<br>• Value of environmental impacts. |
| Financial | • Inadequate estimate of operating costs<br>• Inadequate estimate of financial revenues<br>• Insufficient success in obtaining research funding<br>• Loss of existing clients/users due to competition from another RDI infrastructure | • Operating costs<br>• Licence revenues gained from patents' commercialisation<br>• Revenues from services sold to third parties<br>• Revenues from target population using the research outputs<br>• Revenues from outreach activities |

Source: Authors adapted from European Commission (2014).

## 6.5 How to present the CBA results of RDI infrastructures

As noted, project performance may be assessed in probabilistic terms using a Monte Carlo simulation that approximates the probability distribution functions of the ENPV and the EIRR and their cumulative distribution functions, whose shapes are reported in Box 21. Whereas the probability distribution function summarises the likelihood of occurrence of all outcome values randomly extracted during the Monte Carlo simulation, the cumulative distribution function returns the probability that the outcome is equal to or smaller than any given value in the range of the variation in the considered performance indicator[106]. Thus, the latter can be directly exploited to observe the cumulated probability that corresponds to some feasibility threshold. Namely:

- When considering the ENPV, the interest is on the probability (Pr) that the ENPV is equal to or smaller than zero. Hence, if Pr {ENPV ≤ 0} ≈ 0, then the project can be judged as socially desirable.

- When considering the EIRR, the interest is on the probability that the EIRR on the probability is smaller than the adopted social discount rate (r). Hence, if Pr {EIRR ≤ r} ≈ 0, then the project can be judged as socially desirable.

Additional relevant information on the project performance is provided by various summary statistics of the estimated PDF of the outcome values, i.e. the ENPV and the EIRR. In particular:

- The range of variations consists of the window of values – from minimum to maximum – within which the ENPV and the EIRR vary. The range of variations provides a picture of the project's variability. In general, a project with a narrower range of variability in its performance indicators is preferable, *ceteris paribus*.

- The mean value is the estimate of the expected value of the ENPV and the EIRR, and is interpreted as the outcome expected to occur over a large number of potential project realisations. Thus, the expected mean values provide an immediately readable synthesis of the indicators of the most likely discounted social value of the project.[107]

- The standard deviation consists of the variation around the mean values of the ENPV and the EIRR. In general, no rule exists for interpreting the standard deviation as 'high' or 'low' in absolute terms. However, this synthesis indicator can provide useful information if compared with those of similar type projects.[108]

All of these elements form the CBA results of the RDI infrastructures and are used to judge the investment's social worthiness and riskiness, along with other criteria.

---

[106] To further explore this topic, see chapter 9 in De Rus (2010) and chapter 8 in Florio (2014).
[107] The ENPV calculated in a deterministic model, i.e. as the outcomes of the best guess value of variables, do not exactly coincide with the expected value of the ENPV, i.e. the mean ENPV resulting from a Monte Carlo simulation process.
[108] A useful synthetic ratio is the coefficient of variation, i.e. the ratio of standard deviation to the mean, which is a dimensionless number and thus comparable between projects.

Box 21. Probability distributions of the ENPV and synthesis statistics

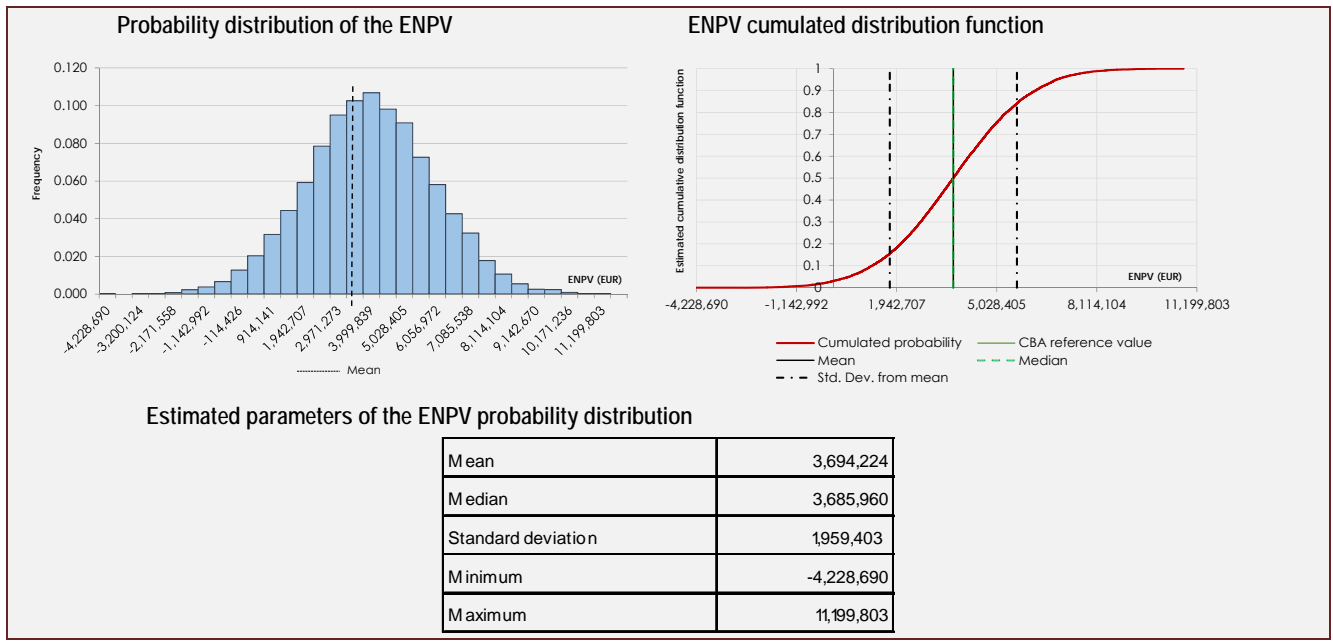

Finally, for communication purposes, presenting CBA results in a disaggregated manner could be useful, such as by breaking down each discounted benefit, by distinguishing between the expected value of use and non-use benefits, or according to a group of beneficiaries (firms, students, the general public). The use of pie charts could assist in presenting the various breakdown of benefits arising from the infrastructure in a reader-friendly manner.

Figure 12. Example of distributions of the discounted benefits for two hypothetical research infrastructures

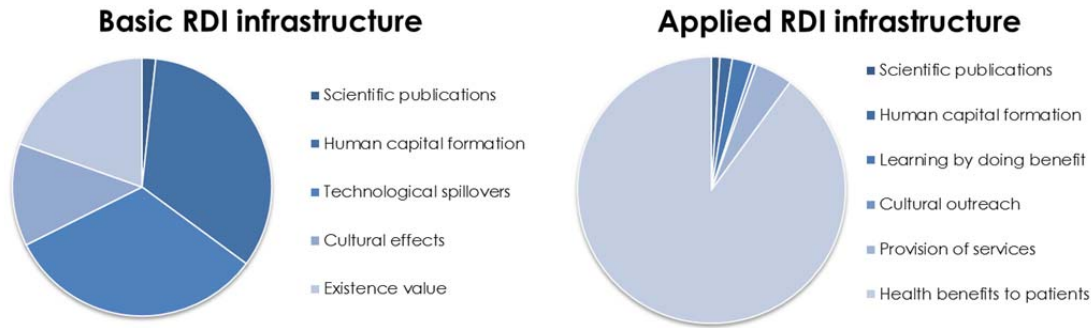

Source: Authors based on analysis of case studies.

# 6.6 Wider economic impacts

RDI infrastructure investments are often financed within broader territorial development strategies (regional or national). As a matter of fact, RDI facilities may become a crucial element in the development path of a territory, as witnessed by experiences of high-tech clusters developed also thanks to the relationship with research facilities located in the same region[109]. In such cases, relevant economic impacts in the form of employment or technological spillover effect on existing or newly created SMEs in the region are included in the CBA as described in the previous sections. Additional wider benefits, for example in terms of contribution to regional GDP, are not accounted in quantitative terms within the CBA model as their inclusion within the socio-economic assessment is still subject to a lot of discussion.[110]

Typically, these impacts include agglomeration economies, multiplier effects, labour supply impacts, impact on competition, or changes in the value of land or houses. In a RDI context, there may be demonstration effects on the general population, particularly the young, about the role of science and technology. For instance, the proximity of universities to a RDI project may convince a larger share of students to achieve a degree in sciences and this, in turn, could be somewhat correlated to the long-term regional growth rate. Also, an RDI facility which attracts high quality personnel, possibly from other regions or abroad, may contribute to opening the cultural horizon of the local society. This in turn can contribute to an increase in local social capital and in some particularly beneficial cases, even to the improvement of the overall quality of institutions.

---

[109] See for example OECD (2009).
[110] The discussion on wider economic impacts is particularly developed in the transport sector. See inter alia Venables (2007), Betancor et al. (2013).

All these impacts are appealing. Nevertheless, there are two main problems regarding their inclusion within the CBA (Jorge-Calderón, 2014): firstly, many of these effects are already accounted for within the direct effects heading. In this respect, it is worth noting that wider economic impacts in terms of labour are already accounted for by means of shadow wages (see section 3.3). Secondly, in many cases these effects do not incorporate a counterfactual comparison, and hence would not be measuring the incremental relevant benefits. Therefore, such type of impacts within a CBA framework shall not be included in the quantitative analysis, but better described in qualitative terms.

The territorial dimension of an RDI infrastructure is an important element for the decision taker, which must be carefully considered in the option analysis[111]. In fact, the availability of relevant know-how and human capital on the territory may have important effects in terms of costs on the RDI construction and operation.

---

[111] The selection of infrastructure project should always result from a throughout analysis of the alternative project options. The criteria to be considered when assessing each option are discussed in in the new Guide for the CBA of investment (European Commission, 2014). They concern, for example, the technological solutions employed and the choice between building a brand new facility rather than improving an already existing one. The localisation of the project can also be a relevant selection criterion. Actually, the territorial dimension of human, technological, physical and production capital endowment can be decisive to determine the project localisation, which in turn affects the potential for the project to promote regional development.

# 7. Concluding remarks

This paper contributes to the literature on the ex-ante appraisal of large-scale, capital intensive RDI infrastructures. Specifically, it presents the main findings and lessons learned from a three-year research on how to advance social CBA methods for RDI facilities. The approach proposed is intended to complement and not substitute for other evaluation methods, including peer review assessments or road mapping.

Our approach can be summarised as follows. RDI facilities are, after all, infrastructures. As such, the CBA principles and rules for the appraisal of infrastructure projects should be exploited and adapted as best as possible. In line with this goal, our model recurs to: 1) shadow prices to capture social benefits beyond the market or financial value; 2) a counterfactual scenario to ensure that all costs and benefits are estimated in incremental terms; 3) discounting to convert any past and future value in their present equivalent; and 4) a consistent identification of social benefits with respect to CBA theory.

At the same time, given some features of RDI infrastructures, such as the intangible nature of certain benefits and the uncertainty associated with achieving the results, a tailored approach to these specificities has been devised. There are three novelties in our approach. First, we identify the core types of beneficiaries in terms of the standard economic agents: firms, consumers, employees, taxpayers, and we present the specificities of the RDI arena in this perspective (such as the coincidence of producers and consumers for certain types of research).

A second novelty of our approach consists of breaking down the intertemporal benefits into two broad classes: use and non-use benefits. The former refers to benefits held by different categories of infrastructure users, such as scientists, firms, students, and general public visitors. The latter benefit denotes the social value for the discovery potential of the RDI infrastructure regardless of its actual or future use. In an heuristic perspective this break-down has the advantage of avoiding the confusion that has often delayed until know the application of CBA to RDI infrastructure projects.

A third significant feature of our approach is the stochastic nature of the CBA model. Indeed, all critical variables entering the model are expressed in terms of expected values instead of punctual values. This approach circumvents another issue that has until now hindered the application of CBA to the RDI infrastructure, i.e. the unpredictability of the social value of knowledge.

Hence, the approach presented in this paper is consistent with the main principles of CBA, but also novel and heuristic. Although many of the ideas that we use are firmly rooted in CBA tradition in different fields, their application to the RDI context is in its infancy. Therefore, we hope that this discussion paper encourages further applications and testing to fine tune and expand the current methodologies and techniques. In particular, forecasting technological spillovers and human capital effects would need econometric 'treatment' techniques; the estimation of the social value of discoveries using contingent valuation methods needs to be tested by relying on both existing and future facilities; forecasting media impact of outreach is another interesting topic to be studied in greater detail. Further research is also necessary to investigate issues that go beyond what we have been able to address until now. Such possible issues include for example assessments of RDI projects' broader economic impacts (including in the territorial perspective), the possible use of real option analysis to value the flexibility and risk of RDI infrastructures[112], advances in risk analysis to consider correlations of stochastic variables, use of Bayesian networks in the study of social preferences. Exploring the use of CBA in such a new field of application will reveal other challenging issues.

---

[112] As suggested by the EIB (2013). See footnote 17.

# List of references

Abelson, P. (2008). 'Establishing a Monetary Value for Lives Saved: Issues and Controversies', WP 2008-02 in cost-benefit analysis, Office of Best Practice Regulation, Department of Finance and Deregulation, Sydney University.

Abelson, P. (2010). 'The Value of Life and Health for Public Policy', Macquarie University, Available at: http://www.appliedeconomics.com/au/pubs/papers/pa03_health.htm

Adler, H.A. (1987). Economic Appraisal of Transport Projects, The World Bank Economic Development Institute, Washington DC.

Alberini, A. and Longo, A. (2005). 'The Value of Cultural Heritage Sites in Armenia: Evidence from a Travel Cost Method Study', Working Papers 2005.112, Fondazione Eni Enrico Mattei.

Altbach, P.G., Reisberg, L., Yudkevich, M., Androushchak, G., and Pacheco, I.F (2012). *Paying the Professoriate: A Global Comparison of Compensation and Contracts*, Routledge.

Andres, A. (2009). Measuring Academic Research: How to Undertake a Bibliometric Study. Chandos Publishing.

Anex, R. (1995). 'A travel-cost method of evaluating household hazardous waste disposal services', *Journal of Environmental Management*, 45: 189-198.

Antoniou, C. and Matsoukis, E. (2007). 'A Methodology for the Estimation of Value-of-Time using State-of-the-Art Econometric Models', *Journal of Public Transportation*, 10(3): 1-19

Arrow, K.J. and Fisher, A.C. (1974). 'Environmental Preservation, Uncertainty, and Irreversibility', *The Quarterly Journal of Economics*, 88 (2): 312-19.

Asian Development Bank (2013). Cost-Benefit Analysis for Development: A Practical Guide. Mandaluyong City, Philippines: Asian Development Bank, 2013.

Ates, G. and Brechelmacher A. (2013). 'Academic Career Paths', in Teichler, U. and Höhle, E.A., Eds. *The work Situation of the Academic Profession in Europe. Findings of a Survey in Twelve European Countries*, Dortrecht: Springer.

Atkinson, G., Mourato S., Pearce D.W. (2006), Cost Benefit Analysis and the Environment. Recent developments, Paris: OECD Publishing.

Bacchiocchi, E. and Montobbio, F. (2009). 'Knowledge diffusion from university and public research. A comparison between US, Japan and Europe using patent citations', *The Journal of Technology Transfer*, Springer, vol. 34(2): 169-181, April.

Bach L., Lambert G., Ret S., Shachar J., with the collaboration of Risser R., Zuscovitch E., under the direction of Cohendet P., Ledoux M-J. (1988). *Study of the economic effects of European space expenditure*, ESA Contract No. 7062/87/F/RD/(SC).

Balcombe, K.G. and Smith, L.E.D. (1999). 'Refining the use of Monte Carlo techniques for risk analysis in project planning', *Journal of Development Studies*, Taylor & Francis Journals, 36(2): 113-135.

Bateman, I.J., Carson, R.T.; Day, B., Hanemann, M., Hanley, N., Hett, T., Lee, M.J., Loomes, G., Mourato, S., Ozdemiroglu, E., Pearce, D.W., Sugden, R. and Swanson, J. (2002). Economic valuation with Stated Preference Techniques: A Manual. Cheltenham/Northampton: Edward Elgar.

Bates, J. and Whelan, G.A. (2001). *Size and Sign of Time Savings*, Working Paper 561, Institute Transport Studies, University of Leeds, Leeds, UK.

Batista, P.D.; Campiteli, M.G.; Konouchi, O.; Martinez, A.S. (2006). Is it possible to compare researchers with different scientific interests? *Scientometrics*, vol. 68, no. 1, pp. 179-189.

Bedate, A., Herrero, L. C. and Sanz, I. (2004). 'Economic Valuation of the Cultural Heritage: Application to Four Case Studies in Spain'. *Journal of Cultural Heritage*, 5: 101-111.

Belli, P., Anderson, J.R., Barnum, H.N., Dixon, J.A., Tan, J. (2001). Economic Analysis of Investment Operations – Analytical Tools and Practical Applications. World Bank Institute (WBI)

Benson C. and Twigg, J. (2004). 'Measuring mitigation: Methodologies for assessing natural hazard risks and the net benefits of mitigation - A scoping study'. Geneva: International Federation of Red Cross and Red Crescent societies / ProVention Consortium.

Bernstein, J.I. and Nadiri, M.I. (1991). *Product demand, cost of production, spillovers, and the social rate of return to R&D*. Technical report, National Bureau of Economic Research.

Bessen, J.E. (2006). 'Estimates of Firms' Patent Rents from Firm Market Value'. Boston, MA: Boston University School of Law Working Paper Series 06-14.

Betancor, O., Hernández, A. and Socorro, M.P. (2013). 'Revision of infrastructure project assessment practice in Europe regarding impacts on competitiveness'. Deliverable 2.2 of the I-C-EU Project. European Commission.

Bezdek, R.H. and Wendling, R.M. (1992). 'Sharing out NASA's spoils'. *Nature*, 355(6356):105–106.

Bianchi-Streit, M., Blackburne, N., Budde, R., Reitz, H., Sagnell, B., Schmied, H., and Schorr, B. (1984). 'Economic utility resulting from CERN contracts'. CERN Research Communications, Geneva.

Blomquist, G. C., and Whitehead, J. C. (1995). Existence value, contingent valuation, and natural resource damage assessment. *Growth and Change* 26: 573–89.

Boadway, R. (2006). 'Principles of cost-benefit analysis', *Public Policy Review*, Vol.2, No.1.

Boardman, A.E., Greenberg, D.H., Vining, A.R. and Weimer D.L. (2006). *Cost Benefit Analysis – Concepts and Practice*, Pearson Education, 3rd edition.

Boarini, R. and H. Strauss (2007). 'The Private Internal Rates of Return to Tertiary Education: New Estimates for 21 OECD Countries', OECD Economics Department Working Papers, No. 591.

Boman, M. and Bostedt, G. (1995). 'Valuing the wolf in Sweden', report No.110, Swedish University of agricultural Sciences, Department of Forest Economics, Umea.

Boyle, K.J. and Bishop, R.C. (1987) Valuing Wildlife in Benefit-Cost Analyses: A Case Study Involving Endangered Species. Water Resources Research 23(5), 943-950;

Brendle P., Cohendet P., Heraud JA., Larue de Tournemine R., Schmied H. (1980). *Les effets économiques induits de l' ESA*. ESA Contracts Report Vol. 3

Brent, R.J. (2003) Chapter 11: Cost–Benefit Analysis and the Human Capital Approach in *Cost–Benefit Analysis and Health Care Evaluations,* Elgar.

British Library (2004). 'Measuring our value: results of an independent economic impact study commissioned by the British Library to measure the Library's direct and indirect value to the UK economy'.

Brouwer, R. (1995), 'The Measurement of the Non-Marketable Benefits of Agricultural Wildlife Management: The Case of Dutch Peat Meadow Land'. Wageningen Economic Paper, 1995-1, Wageningen Agricultural University.

Blöndal, S., Field, S. and Girouard, N. (2002). "Investment in Human Capital through Upper-Secondary and Tertiary Education", *OECD Economics Studies*, Vol. 34 (1), pp. 41-90.

Brealey, R.A. and Myers, S.C. (2000). *Principles of Corporate Finance*, 6th edition, Irwin/McGraw-Hill.

Brigham, E. F., Gapenski, L. C. and Daves, P. R. (1999). *Intermediate Financial Management*, 6th edition, The Dryden Press.

Broadus R.N. (1983). An investigation of the validity of bibliographic citations. *J Am Soc Inf Sci*. 1983 Mar; 34(2):132–5.

Burton, R.E. and Kebler, R.W. (1960). 'The Half-Life of some Scientific and Technical Literature', *American Documentation*, 11(1): 18-22.

Byford, S., Torgerson, D. J., and Raftery, J. (2000). Cost of illness studies. *British Medical Journal*, 320(7245): 1335.

Caballero R.J. and Jaffe A.B. (1993). 'How High Are the Giant Shoulders: An Empirical Assessment of Knowledge Spillover and Creative Destruction in a model of Economic Growth'. In O. Blanchard and S. Fisher, eds., *NBER Macroeconomics Annual*, v. 8. MIT Press.

Carrazza, S., Ferrara A. and Salini S. (2014). 'Research infrastructures in the LHC era: a scientometric approach', draft paper produced in the frame of the research project "Cost/Benefit Analysis in the Research, Development and Innovation Sector", part of the EIBURS programme.

Carson, R.T. (1998). Valuation of tropical rainforests: philosophical and practical issues in the use of contingent valuation. *Ecological Economics* 24(1998): 15-29.

Carson, R.T. (2012). 'Contingent Valuation: A Practical Alternative when Prices Aren't Available', *Journal of Economic Perspectives*, 26(4), 27-42.

Carson, R.T. and Groves, T. (2007). 'Incentive and informational properties of preference questions', *Environmental and Resource Economics*, 37(1): 181-210.

Carson, R.T., Flores, N.E., and Meade, N.F. (2001). Contingent Valuation: Controversies and Evidence. *Environmental and Resource Economics* 19 (2): 173–210

Carson, R.T., Wilks, L., Imber, D. (1994). 'Valuing the preservation of Australia's Kakadu conservation zone'. *Oxford Economic paper* 46: 727-749

Caulkins, P.P, Bishop, R.C., Bouwes, N.W. (1986). The travel cost model for lake recreation: a comparison of two methods for incorporating site quality and substitution effects, *American Journal of Agricultural Economics* 68 (2): 291–297.

Chilton, S. et al. (2004), Estimating a Value of a Life Year Gained from Air Pollution Reduction: A Comparison of Approaches, Paper prepared for the EAERE Annual Conference, Thessaloniki, 27-30 June, 2007.

Chyi, H.I. (2005). 'Willingness to Pay for Online News: An Empirical Study on the Viability of the Subscription Model', *Journal of Media Economics*, 18(2): 131-142.

Clarke, R. and Low, A. (1993). Risk analysis in project planning: a simple spreadsheet application using Monte-Carlo techniques, *Project Appraisal*, 8(3): 141-146.

Clawson, M. and Knetsch, J.L. (1966). *Economics of outdoor recreation*, Johns Hopkins Press.

Clinch, J.P. and Healy, J. D. (2001). 'Cost-Benefit Analysis of Domestic Energy Efficiency', *Energy Policy*, 29(2): 113-124.

Cockburn, I., Griliches, Z. (1988), 'Industry Effects and Appropriability Measures in the Stock Market's Valuation of R&D and Patents, *American Economic Review*, 78(2): 419-23.

Conrad, J.M. (1980). 'Quasi-Option Value and the Expected Value of Information', *The Quarterly Journal of Economics*, Vol. 94 (4): 813-820.

Courtney, H., Kirkland, J. and Viguerie, P. (1997). Strategy under uncertainty. *Harvard Business Review*, November-December, 67-79.

Cuccia, T. and Signorello, G.A. (2000). 'A contingent valuation study of willingness to pay for visiting a city of art: the case study of Noto', Proceedings of the 11th International Conference on Cultural Economics, Minneapolis, 28–31 May, 2000.

Czech Ministry of Education, Youth and Sport and JASPERS (2009). Background methodology for preparing feasibility and cost-benefit Analysis of R&D infrastructure projects in Czech Republic, supported by the Cohesion Fund and the European Regional Development Fund in 2007-2013.

Daily, G.C. (1997). Nature's Services. Societal Dependence on Natural Ecosystems, Washignton DC: Island Press.

Danish Agency for Science (2008). 'Evaluation of Danish industrial activities in the European space agency (ESA). Assessment of the economic impacts of the Danish ESA membership'. Technical report, Danish Agency for Science.

Dasgupta, P. (2007). 'Commentary: The Stern Review's Economics of Climate Change', *National Institute Economic Review*, 199(4): 4-7.

Dasgupta, P. (2008). 'Discounting climate change', *Journal of Risk and Uncertainty*, 37(2): 141-169.

David, P. A., Mowery, D., and Steinmueller, W. E. (1992). Analysing the economic payoffs from basic research. *Economics of innovation and New Technology*, 2(1):73–90.

Day, B. (2001), The Valuation of Non-Market Goods 2, Imperial College London, mimeo

De la Fuente, A., and Jimeno, J.F. (2005). "The Private and Fiscal returns to Schooling and the Effect of Public Policies on Private Incentives to Invest in Education: A General Framework and Some Results for the EU", CESifo Working Paper No. 1392.

Del Bo, C., Fiorio C. and Florio M. (2011). 'Shadow Wages for the EU Regions', *Fiscal Studies*, 32 (1): 109-143.

Deng, Y. (2005). The Value of Knowledge Spillovers, University of Illinois at Urbana-Champaign's Academy for Entrepreneurial Leadership Historical Research Reference in Entrepreneurship. Available at SSRN: http://ssrn.com/abstract=1513816

de Rus, G. (2010). *Introduction to Cost–Benefit Analysis*, Books, Edward Elgar, number 13756, March.

Desaigues, B. et al. (2011). Economic valuation of air pollution mortality: A 9-country contingent valuation survey of value of a life year (VOLY). *Ecological Indicators*, 11(2011): 902–910.

DTZ (2009). 'Economic impact of the John Innes Centre'. Technical report, DTZ, Edinburgh.

Dupuit, A.J. (1844). 'On the measurement of the utility of public works'. *International Economic Papers* 2, 1952.

Dupuit, A.J. (1853). 'On utility and its measure – on public utility'. *Journal des Économistes* 36: 1-27.

Economics and Development Resource Center (1997). 'Guidelines for the economic analysis of projects'.

Egghe, L., and Rao, I. K. R. (1992). Citation age data and the obsolescence function: Fits and explanations. *Information and Processing Management*, 28(2), 201-217.

EIROforum (2015). 'Long-term sustainability of Research Infrastructures', EIROforum discussion paper.

ENTSOE (2015). Guideline for Cost Benefit Analysis of Grid Development Projects. FINAL- Approved by the European Commission.

ESF (2013). Research infrastructures in the European Research Area – A report by the ESF Member Organisation Forum on Research Infrastructures.

ESFRI (2008). European Roadmap for Research Infrastructures – Roadmap 2008.

ESFRI (2010). Strategy Report on Research Infrastructures – Roadmap 2010.

European Commission (2004). HEATCO: Developing Harmonised European Approaches for Transport Costing and Project Assessment, Deliverable 5, Brussels.

European Commission (2006). Study on Evaluating the Knowledge Economy. What are Patents Actually Worth? The value of patents for todays' economy and society, Final Report Tender No MARKT/2004/09/E, Lot 2, 23 July 2006.

European Commission (2008), IMPACT: Internalisation Measures and Polices for All external Cost of Transport, Handbook on estimation of external costs in the transport sector, Version 1.1, Brussels.

European Commission (2010). Communication from the Commission Europe 2020 - A strategy for smart, sustainable and inclusive growth, COM(2010) 2020.

European Commission (2011). *Innovation Union Competitiveness report 2011*. Analysis. Part III Towards an innovative Europe – contributing to the Innovation Union, Brussels.

European Commission (2013). Innovation Union Competitiveness report 2013. Brussels.

European Commission (2014). Guide to Cost-benefit Analysis of investment projects. Economic appraisal tool for Cohesion Policy 2014-2020, Directorate-General for Regional and Urban Policy.

European Investment Bank (2013). The Economic Appraisal of Investment Projects at the EIB.

European Space Agency (2012). Design of a Methodology to Evaluate the Direct and Indirect Economic and Social Benefits of Public Investments in Space

Feller, I. (2013). Peer review and expert panels as techniques for evaluating the quality of academic research, in Link, A.N., Vonortas, N.S. (eds) *Handbook on the theory and practice of program evaluation*, Edward Elgar.

Florio, M. (2006). 'Cost-Benefit Analysis and the European Union Cohesion Fund: On the Social Cost of Capital and Labour', *Regional Studies*, 40(2): 211-224.

Florio (2014) Applied Welfare Economics: Cost-Benefit Analysis of Projects and Policies, London: Routledge.

Florio M. and Sirtori E. (2014), 'The Evaluation of Research Infrastructures: a Cost-Benefit Analysis Framework', paper produced in the frame of the research project "Cost/Benefit Analysis in the Research, Development and Innovation Sector", part of the EIB University Sponsorship programme (EIBURS).

Florio, M., Forte, S. and Sirtori, E. (2015). 'Cost-Benefit Analysis of the Large Hadron Collider to 2025 and beyond', *arXiv*:1507.05638 [physics.soc-ph].

Flyvbjerg, B. (2013). 'Quality control and due diligence in project management: getting decisions right by taking the outside view'. *International Journal of Project Management*, 31: 760–774.

Flyvbjerg, B., Bruzelius, N., and Rothengatter, W. (2003). *Megaprojects and Risk: An Anatomy of Ambition*, Cambridge University Press, Cambridge.

Gambardella, A., Harhoff, D. and Verspagen, B. (2008). 'The value of European patents'. *European Management Review* 5: 69-84.

Garrod, G. and Willis, K. (1991). *Economic Valuation of the environment: methods and case studies*, Massachusetts: Edgar Elgar.

Greenley, D.A., Walsh, R.G., Young, R.A., 1981. Option value: Empirical evidence from a case study of recreation and water quality. *Quarterly Journal of Economics* 96, 657-673.

Griliches, Z. (1979). Issues in assessing the contribution of research and development to productivity growth. *The Bell Journal of Economics*, pages 92–116.

Griliches, Z. (1981). 'Market Value, R&D, and Patents'. *Economics Letters*, 7: 183–187.

Guha-Sapir, D. and Santos, I. (2013) *The Economic Impacts of Natural Disasters*. New York, NY: Oxford University Press.

Halevi, g. (2013) 'Citation characteristics in the Arts & Humanities', MLS PhD

Hall, B.H., Jaffe, A., Trajtenberg, M. (2005), "Market value and patent citations", RAND Journal of Economics, Vol. 36, No. 1, Spring 2005 pp. 16–38.

Hall, B.H., Thoma, G. and Torrisi, S. (2007). 'The Market Value of Patents and R&D: Evidence from European Firms'. NBER Working Paper 13426, Cambridge, MA.

Hall, B. H., Mairesse, J., and Mohnen, P. (2009). Measuring the returns to R&D. *Handbook of the Economics of Innovation*, 2:1033–1082.

Hampicke, U., Tape, K., Kiemstedt, K., Horlitz, H., Walters, U., Timp, D. (1991). The economic importance of preserving species and biotopes in the federal republic of Germany, results cited in Romer, A.U., Pommerehne, W.W. (1992). Germany and Switzerland. In Navrudu (Ed.). Pricing the European Enviroment, Scandinavia University Press.

Han, B. and Windsor, J. (2011). 'User's willingness to pay on social network sites', Journal of Computer Information Systems, 51 (4): 31.

Hanley, N. and Barbier, E.B. (2009) *Pricing Nature: Cost-benefit Analysis and Environmental Policy*, Edward Elgar, Cheltenham.

Hansen, T.B. (1997), 'The Willingness-to-Pay for the Royal Theatre in Copenhagen as a Public Good', *Journal of Cultural Economics*, 21: 1-28.

Harhoff, D., Narin, F., Scherer, F. M., Vopel, K. (1999), "Citation Frequency and the Value of Patented Inventions", Review of Economics and Statistics, 81, No. 3, pp. 511-515.

Harhoff, D., Scherer, F. M., Vopel, K. (2003a), "Citations, Family Size, Opposition and the Value of Patent Rights", Research Policy, 32, pp. 1343-1363.

Harhoff, D., Scherer, F. M. and Vopel, K. (2003b),"Exploring the Tail of Patented Invention Value" Distribution", in O. Granstrand (ed.), Economics, Law and Intellectual Property: Seeking strategies for research and teaching in a developing field, Kluwer Academic Publishers, Boston

Harmon, C., Oosterbeek, H. and Walker, I. (2003). "The Returns to Education: Microeconomics," *Journal of Economic Surveys*, Wiley Blackwell, vol. 17(2), pages 115-156, 04.

Harrison, M. (2010). 'Valuing the Future: the social discount rate in cost-benefit analysis', Visiting Researcher Paper, Productivity Commission, Canberra. Online.

Harzing, A-W. (2010). Citation analysis across disciplines: The impact of different data sources and citation metrics. Available at: http://www.harzing.com/data_metrics_comparison.htm.

Heckman, J.J., Lochner, L.J. and Todd, P.E. (2005), 'Earnings Functions, Rates of Return, and Treatment Effects: The Mincer Equation and Beyond', NBER Working Paper No. 11544.

Hensher, D.A. (1977). *Value of Business Travel Time*, Oxford: Pergamon Press.

Hensher, D.A., and Goodwin, P. (2004). 'Implementation values of travel time savings: the extended set of considerations in a toll road context', *Transport Policy*, 11(2): 171-181.

Hertzfeld, H.R. (1998). Measuring the returns to NASA life sciences research and development. In AIP conference proceedings, pages 810–815. IOP Institute of Physics Publishing LTD.

Hicks, D. (2004). The four literatures of social science. In: Glaenzel, W., Moed H.F., Schmoch, U. (eds) Handbook of quantitative science and technology research: the use of publication and patent statistics in studies of S&T systems. Kluwer Academic Publishers, Dordrecht, pp. 473–496.

Hirsch, J.E. (2005) An index to quantify an individual's scientific research output, arXiv:physics/0508025

HM Treasury (2003). The Green Book – Appraisal and Evaluation in Central Government, Treasury Guidance.

HM Treasury (2006), Stern Review: The Economics of Climate Change, London.

Honohan, P. (1998). 'Key Issues of Cost-Benefit methodology for Irish Industrial Policy', General Research Series 172, The Economic and Social Research Institute, Dublin.

Hotelling, H. (1947). 'Letter to the National Park Service'. Reprinted in: *An economic study of the monetary evaluation of recreation in the national parks.* (1949). Washington, D.C.: U.S. National Park Service.

Horlings, E. and Versleijen, A. (2008) Groot in 2008. Momentopname van Grootschalige Onderzoeksfaciliteiten in de Nederlandse Wetenschap.

Hsu, D.H. and Ziedonis, R.H. (2008). 'Patents as Quality Signals for Entrepreneurial Ventures'. Academy of Management Best Paper Proceedings.

Iglesias, J.E, Pecharromán, C. (2006) Scaling the h-Index for Different Scientific ISI Fields, arXiv:physics/0607224

IVA (Kungl. Ingenjörsvetenskapsakademien) (2012). Review of literature on scientists' research productivity.

Jaffe, A.B., Trajtenberg, M. and Henderson, R. (1993). "Geographic Localization of Knowledge Spillovers as Evidenced by Patent Citations," *The Quarterly Journal of Economics*, MIT Press, vol. 108(3), pages 577-98, August.

Jägle, A.J. (1999). Shareholder value, real options, and innovation in technology-intensive companies, *R&D Management,* 29(3): 271-287.

Jakobsson, K.M. and Dragun, K. (1996). *Contingent Valuation and Endangered Species: Methodological Issues and Application*. Cheltenham, UK and Broodfield US: Edward Eldgar..

JASPERS (2013). 'Project Preparation and CBA of RDI Infrastructure Project', Staff Working Papers, JASPERS Knowledge Economy and Energy Division.

Jenkins, G-P., Kuo, C-Y, Harberger, A.C. (2011). *Cost-benefit analysis for investment decisions.*

Johanssson, P.O. (1989). 'Valuing public goods in a risk world: an experiment'. In: Folmer, H., Iereland, E. (Eds.). *Evaluation Methods and Policy Making in Environmental WI Economics*, Amsterdam, North-Holland.

Johannesson, M. and Johansson, P.O. (1996), 'To Be or Not To Be, That Is The Question: An Empirical Study of the WTP for an Increased Life Expectancy at an Advanced Age', *Journal of Risk and Uncertainty*, 13:163-174.

Jones, H., Domingos, T., Moura, F. and Sussman, J. (2013). 'Transport infrastructure evaluation using cost-benefit analysis: improvements to valuing the asset through residual value - a case study', for the 93rd Annual Meeting of the Transportation Research Board, January 2014.

Jorge-Calderón, D. (2014). *Aviation investment: economic appraisal for airports, air traffic management, airlines and aeronautics*. Ashgate Publishing Limited.

JRC-IPTS (2002) 'RTD Evaluation Toolbox - Assessing the Socio-Economic Impact of RTD-Policies', IPTS Technical report Series.

King, J. (1987). A review of bibliometric and other science indicators and their role in research evaluation. *J Inf Sci*, 13: 261–276.

Krutilla, J. (1967). 'Conservation Reconsidered', *American Economic Review*, September, pp. 777-786.

Kunreuther, H. and Michel-Kerjan, E. (2014). 'Economics of natural catastrophe risk insurance', in *Handbook of the Economics of Risk and Uncertainty*, Volume 1, MJ Machina and WKViscusi (Eds), Elsevier.

Landefeld, S. and Seskin, E. (1982). 'The economic Value of Life: Linking Theory to Practice', *AJPH* Vol. 72, No. 6.

Laibson, D. (1997). 'Golden Eggs and Hyperbolic Discounting', *Quarterly Journal of Economics*, 112(2): 443-477.

Lanjouw, J.O. (1998). 'Patent protection in the shadow of infringement: Simulation estimations of patent value', *Review of Economic Studies*, 65: 671-712.

Larivière, V., Archambault, É. and Gingras, Y. (2008). Long-term variations in the aging of scientific literature: from exponential growth to steady-state science (1900-2004). *Journal of the American Society for Information Science and Technology*, 59, 288-296.

Lehtonen, E., Kuuluvainen J., Pout E., Rekola M. and Li C-Z. (2003). 'Non-market benefits of forest conservation in southern Finland', *Environmental Science & Policy*, 6: 195-204.

Leone M.I. and Oriani, R. (2008). 'Explaining the Remuneration Structure of Patent Licenses'. In: 2008 Annual Meeting of the Academy of Management, 8-13 Agosto, Month 1.

Lerner, J. (1994), "The Importance of Patent Scope: An Empirical Analysis", RAND Journal of Economics, Vol. 25, No. 2, pp. 319-333.

Li, J., and Ye, F. Y. (2014). A Probe into the Citation Patterns of High-quality and High-impact Publications. *Malaysian Journal of Library and Information Science*, 19(2), 31-47.

Lindhjem, H., Navrud, S., Braathen, NA and Biausque, V. (2011). Valuing mortality risk reductions from environmental, transport, and health policies: A global meta-analysis of stated preference studies. *Risk Analysis,* 31: 1381-1407.

Little, I.M.D. and Mirrlees, J.A. (1974). *Project appraisal and planning for developing countries*, London: Heinemann Educational Books.

Londero, E.H. (2003). *Shadow Prices for Project Appraisal*, Books, Edward Elgar, number 3064, March.

London Economics, (2013), Guidance Manual for Cost Benefit Analysis (CBAs) Appraisal in Malta.

Loomis, J.B., Larson, D.M., (1994). Total Economic Value of Increasing Gray Whale Populations: result from a contingent valuation survey of visitors and household. *Marine Resource Economics* 9: 275-286

Loomis, J.B. and Walsh, R.G. (1997) *Recreation Economic Decisions: Comparing Benefits and Costs*, 2nd edition, Pennsylvania: Venture Publishing.

Luehrman, T.A. (1998). Strategy as a Portfolio of Real Options, *Harvard Business Review*, September-Octeber, 1998, p. 89-99.

Malesios, C.C. and Psarakis, S. (2012). 'Comparison of the h-index for different fields of research using bootstrap methodology', *Qual Quant* 48: 521–545.

Manigart, S., Wright, M., Robbie, K., Desbrières P. and De Waele, K. (1997) Venture capitalists' appraisal of investment projects : An empirical European study. *Entrepreneurship Theory and Practice*, 21(4):29-43.

Manigart, S., K. De Waele, M. Wright, K. Robbie, P. Desbrières, H. Sapienza and Beekman, A. (2000). Venture capital, investment appraisal and accounting information: a comparative study of the US, UK, France, Belgium and Holland. *European Financial Management*, 6, 380-404.

Mansfield, E., Rapoport, J., Romeo, A., Wagner, S., and Beardsley, G. (1977). Social and private rates of return from industrial innovations. T*he Quarterly Journal of Economics*, 91(2): 221–240.

Martin, B. R., and Tang, P. (2007). The benefits from publicly funded research, SPRU Electronic Working Paper Series, Paper no. 161. University of Sussex, Brighton, UK: SPRU.

McComb, G., Lantz, V., Nash, K. and Rittmaster, R. (2006). *International valuation databases: Overview, methods and operational issues*, Ecological Economics, 60: 461-472.

Mendes, I. and Proença, I. (2005) 'Estimating the recreation value of ecosystems by using a travel cost method approach', Working Paper 2005/08, Department of Economics at the School of Economics and Management (ISEG), Technical University of Lisbon.

Midwest Research Institute (1971). 'Economic impact of stimulated technology activity'. Technical report, NASA.

Mincer, J. (1974), *Schooling, Experience and Earnings*. New York: Columbia University Press

Mitchell, R.C. and Carson, R.T. (1989) *Using Surveys to Value Public Goods: The Contingent Valuation Method*. Johns Hopkins University Press, Baltimore, MD

MMC (2005). Natural Hazard Mitigation Saves: An Independent Study to Assess the Future Savings from Mitigation Activities. Volume 2-Study Documentation. Washington DC: Multihazard Mitigation Council.

Moed, H.F., Burger, W.J.M., Frankfort, J.G., van Raan, A.F.J. (1985). The use of bibliometric data for the measurement of university research performance. *Res. Pol.* 14, 131–149.

Mrozek, J. R. and Taylor, L. O. (2002), What determines the value of life? a meta-analysis. *J. Pol. Anal. Manage.*, 21: 253–270.

Nederhof, A.J. (2006) 'Bibliometric monitoring of research performance in the social sciences and the humanities: a review', *Scientometrics*, Vol. 66, pp. 81-100.

Newell, R.G. and Pizer, W.A. (2004). 'Uncertain discount rates in climate policy analysis, *Energy Policy*, 32(4): 519-529.

NOAA (1993). 'Report on the NOAA panel on contingent valuation', Federal Register, vol. 58 n.10, Friday January 15, pp. 4602-4614.

Nordhaus, W. (2007) 'A review of the Stern Review on the Economics of climate change', *Journal of Economic Literature*, 45 (3): 686-702.

Nunes, P.A. (1999). 'Contingent valuation of the benefits of natural areas and its warm glow component', UK: Ph.D dissertation, Faculty of Economics and applied Economics.

OECD (1996). *Venture Capital and innovation*, OECD Publishing, Paris.

OECD (1999) Handbook of Incentive Measures for biodiversity, OECD Publishing, Paris.

OECD (2006) 'Quasi option value' in Cost Benefit Analysis and the Environment. Recent developments, Paris: OECD Publishing.

OECD (2007). 'Use of discount rates in the estimation of the costs of inaction with respect to selected environmental concerns', Working Party on National Environmental Policies, prepared by Hepburn, C., OECD.

OECD (2008). Report on Roadmapping of Large Research Infrastructures, OECD Publishing, Paris.

OECD (2009), Clusters, Innovation and Entrepreneurship, OECD Publishing, Paris.

OECD (2012). 'The value of Statisical Life: a Meta-Analysis' in Working Party on National Environmental Policies, Paris.

OECD (2014a). International distributed research infrastructures: issues and options. Global Science Forum report. OECD Publishing, Paris.

OECD (2014b). Report on the Impacts of Large Research Infrastructures on Economic Innovation and on Society. Case studies at CERN. Global Science Forum report. OECD Publishing, Paris.

OECD (2015). "Cost benefit analysis of investment projects", in *Government at a Glance 2015*, OECD Publishing, Paris.

Oskarsson, I. and Schläpfer, A. (2008). The performance of Spin-off companies at the Swiss Federal Institute of Technology Zurich, ETH transfer Publishing.

Pakes, A. (1986). "Patents as Options: Some Estimates of the Value of Holding European Patent Stocks", *Econometrica*, 54, No. 4

Pancotti, C., Battistoni, G., Genco, M., Livraga, M.A., Mella, P., Rossi, S. and Vignetti, S. (2015). 'The socio-economic impact of the National Hadrontherapy Centre for Cancer Treatment (CNAO): applying a cost-benefit analysis analytical framework', DEMM Working Paper n. 2015-05.

Pearce, D.W. (1993) *Economic Values and the Natural World*, London: Earthscan.

Pearce, D.W, Atkinson, G. and Mourato, S. (2006). *Cost-Benefit Analysis and the Environment*. Recent developments, Paris: OECD Publishing.

Picazo-Tadeo, A. and E. Reig-Martínez (2005). 'Calculating shadow wages for family labour in agriculture : An analysis for Spanish citrus fruit farms', Cahiers d' Economie et Sociologie Rurales, INRA Department of Economics, 75: 5-21.

Pigou, A. (1920). *The Economics of welfare*, London: Macmillen.

Poor, J.P. and Smith, J.M. (2004). Travel cost analysis of a cultural heritage site: the case of historic St. Mary's City of Maryland, *Journal of Cultural Economics* 28, 217–229.

Portney, P.R. (1994) The Contingent Valuation Debate: Why Economists Should Care. *Journal of Economic Perspectives* 8 (4): 3–17.

Potts, D. (2002) Project planning and analysis for development, London: Lynne Rienner.

Potts, D. (2012a). 'Semi-input-output methods of shadow price estimation: are they still useful?', in Weiss J. and Potts D. (eds.) *Current Issues in Project Analysis for Development*, Cheltenham (UK) and Northampton, Massachusetts (US): Edward Elgar.

Potts, D. (2012b). 'Shadow wage rates in a changing world' in Weiss, J. and D. Potts (eds.) *Current Issues in Project Analysis for Development*, Cheltenham (UK) and Northampton, Massachusetts (US): Edward Elgar.

Pouliquen, L.Y. (1970). 'Risk Analysis in Project Appraisal', World Bank Staff Occasional Papers Number 11, Baltimore: Johns Hopkins University Press.

Price, D.J.d.S. (1965). Networks of Scientific Papers. *Science*, 149, 510–515.

Psacharopoulos, G. (1994). 'Returns to investment in education: A global update', *World Development*, Elsevier, vol. 22(9): 1325-1343, September.

Psachoroupolos, J. (1995). 'The Profitability of Investment in Education: Concepts and Methods', The World Bank Human Capital Development and Operations Policy Working Paper No. 63.

Psacharopoulos, G., and Patrinos H. A. (2004). 'Returns to Investment in Education: A Further Update'. *Education Economics* 12(2): 111-134.

Quinet, E. (2007). 'Cost Benefit Analysis of Transport Projects in France', in Florio, M. (ed) *Cost Benefit Analysis and Incentives in Evaluation*, Edward Elgar.

Radicchi, F., Fortunato, S., Castellano, C. (2008). 'Universality of citation distributions: toward an objective measure of scientific impact'. *Proc. Natl. Acad. Sci. USA* 105(45): 17268–17272.

Reiling, S.D. and Anderson, M.W. (1980). 'The relevance of option value in benefit-cost analysis'. Technical Bulletin 101. Orono, ME: University of Maine.

Research Council UK. (2010). *Large Facilities Roadmap 2010*.

Reutlinger, S. (1970). 'Techniques for Project Appraisal Under Uncertainty', World Bank Staff Occasional Paper number 10, Baltimore: Johns Hopkins University Press.

Rice, D. P. (1967). Estimating the cost of illness. *American Journal of Public Health and the Nations Health*, 57(3): 424–440.

Rice, D. P., Hodgson, T. A., and Kopstein, A. N. (1985). The economic costs of illness: A replication and update. *Health Care Financing Review*, 7(1): 61–80.

Richer, J.R. (1995). 'Willingness to pay for desert protection'. *Contemporary Economic Policy* 13: 93–104.

Robert, C. and Casella, G. (2004). *Monte Carlo Statistical Methods*. 2nd ed. Springer-Verlag, New York.

Ruijgrok E.C.M. (2006). The three economic values of cultural heritage: a case study in the Netherlands. *Journal of Cultural Heritage*, 7(3): 206-213.

Sakakibara, M. (2010). 'An empirical analysis of pricing in patent licensing contracts'. *Industrial and Corporate Change,* 19: 927-945.

Saleh, I (2004). 'Estimating shadow wages for economic project appraisal', *The Pakistan Development Review*, 43(3): 253-266.

Salina, G. (2006). Dalla ricerca di base al trasferimento tecnologico: impatto dell'attività scientifica dell'istituto nazionale di fisica nucleare sull'industria italiana, *Rivista di cultura e politica scientifica*, No. 2/2006.

Sattout, E.J., Talhouk, S.N. and Caligari, P.D.S. (2007). Economics Value of Cedar Relics in Lebanon: An Application of Contingent Valuation Method for Conservation. *Ecological Economics* 61(2007): 315 – 322.

Sauter, R., and Volkery, A. (2013). Review of costs and benefits of energy savings, A report by the Institute for European Environmental Policy (IEEP) for the Coalition of Energy Savings. Task 1 Report. Brussels.

Savvides, S.C. (1994). Risk Analysis in Investment Appraisal, *Project Appraisal Journal* 9(1), March.

Schankerman, M. (1998), 'How Valuable is Patent Protection? Estimates by Technology Field', *RAND Journal of Economics*, 29(1): 77-107

Schankerman, M., Pakes, A. (1986), 'Estimates of the Value of Patent Rights in European Countries During the Post-1950 Period', *Economic Journal*, Vol. 96, No. 384.

Schuster, R.M, Cordell, H.K. and Phillips, B. (2005), 'Understanding the Cultural, Existence, and Bequest Values of Wilderness', December, *International Journal of Wilderness*, Vol. 11, No.3.

Schmied, H. (1975). 'A study of economic utility resulting from CERN contracts'. Technical report, European Organization for Nuclear Research, Geneva (Switzerland).

Schmied, H. (1982). 'Results of attempts to quantify the secondary economic effects generated by big research centers'. *Engineering Management*, IEEE Transactions on, EM-29(4):154–165.

Sharp, B. and Kerr, G. (2005) Option and Existence Values for the Waitaki Catchment. Report prepared for Ministry for the Environment, Wellington, New Zealand.

Sneed, K.A. and Johnson, D.K.N. (2007), 'Selling ideas: The determinants of patent value in an auction environment', Colorado College Working Paper 2007-05, June.

Squicciarini, M., H. Dernis and C. Criscuolo (2013), "Measuring Patent Quality: Indicators of Technological and Economic Value", OECD Science, Technology and Industry Working Papers, 2013/03, OECD Publishing.

Seppä, T. J., and Laamanen, T. (2001). Valuation of Venture Capital Investments: Empirical Evidence. *R&D Management*, 31(2): 215–230.

Serrano, C.J (2008). 'The Dynamics of the Transfer and Renewal of Patents', NBER Working Papers 13938, National Bureau of Economic Research, Inc.

Sidiropoulos, A., Katsaros, D. and Manolopoulos, Y. (2006). Generalized h-index for disclosing latent facts in citation networks. Available at arXiv:cs/0607066.

Simkin M.V, Roychowdhury V.P. A mathematical theory of citing. *J Am Soc Inf Sci Technol.* 2007 Sep; 58(11):1661–73.

Sorg, C.F. and Loomis, J.B. (1984) *Empirical Estimates of Amenity Forest Values: A Comparative Review*, General Technical Report RM-107. Fort Collins, CO: USDA Forest Service, Rocky Mountain Forest and Range Experiment Station.

Stevens, T.H., DeCoteau, N.E., Willis, C.E. (1991). 'Sensitivity of contingent valuation to alternative payment schedules'. *Land Economics* 73 (1): 140-148.

Sun, J., Min, C. and Li, J. (2015). A Vector for Measuring Obsolescence of Scientific Articles. Proceedings of ISSI 2015 Istanbul: 15th International Society of Scientometrics and Informetrics Conference, Istanbul, Turkey, 29 June to 3 July, 2015, Bogaziçi University Printhouse.

Technopolis (2011). The role and added value of large-scale research facilities. Final Report.

Technopolis (2013). *Big Science and Innovation*, 5 July 2013.

Teichmann, D. and Schempp, C. (2013). 'Calculation of GHG Emissions of Waste Management Projects'. JASPERS Staff Working Papers.

The Word Bank (2003) Building Safer Cities The Future of Disaster Risk. Washington, D.C

Thompson E., Berger M., Blomquist G. and Allen S. (2002). 'Valuing the Arts: A Contingent Valuation Approach', *Journal of Cultural Economics*, 26(2): 87-113.

Tietenberg, T. and Lewis, L. (2009). *Environmental and Natural Resource Economics*, Pearson, Boston, MA

Togridou, A., Hovardas, T. & Pantis, J.D. (2006). Determinants of visitors' willingness to pay for the National Marine Park of Zakynthos, Greece. *Ecological Economics*, 60(1), 308-319.

Treasury Board Secretariat, Canada. (2007). 'Canadian Cost-Benefit Analysis Guide: Regulatory Proposals (Interim)'. Ottawa, ON: Treasury Board Secretariat.

Turner, R.K. (1999). 'The Place of Economic Values in Environmental Valuation', in Bateman I.J. and Willis K.G. (eds.) *Valuing Environmental Preferences*. New York: Oxford University Press.

Trajtenberg M., Henderson R. and Jaffe A. (1997). University vs. Corporate Patents: A Window on the Basicness of Invention. *Economics of Innovation and New Technology* 5, 19-50.

Venables, A.J. (2007). 'Evaluating urban transport improvements: cost-benefit analysis in the presence of agglomeration and income taxation', *Journal of Transport Economics and Policy*, 41, 173-188.

Vesely, É. (2007). Green for green: The perceived value of a quantitative change in the urban tree estate of New Zealand. *Ecological Economics*, 63(2-3), 605-615.

Viscusi, W. and Aldy, J. E. (2003), 'The value of a statistical life: a critical review of market estimates throughout the world', *Journal of Risk and Uncertainty*, 27(1): 5-76.

Viscusi, W. (2014). 'The Value of Individual and Societal Risks to Life and Health'. In *Handbook of the Economics of Risk and Uncertainty*, MJ Machina and WK Viscusi (Eds), Elsevier.

Vock, M., van Dolen, W., and de Ruyter, K. (2013). 'Understanding Willingness to Pay for Social Network Sites', *Journal of Service Research*, 16 (3): 311-325.

Vose, D. (2008). *Risk Analysis: A Quantitative Guide*, Great Britain: John Wiley and Sons.

Walsh, R.G., Loomis, J.B., Gillman, R.A. (1984). 'Valuing option, existence, and bequest demands for wilderness'. *Land Economics* 60 (1), 14–29.

Weisbrod, B.A. (1964). Collective-Consumption Services of Individual-Consumption Goods. *Quarterly Journal of Economics* 78 (Aug.): 471-477.

Weitzman, M.L (2001) 'Gamma Discounting', *The American Economic Review*, 91(1): 260-271.

Weitzman, M.L. (2007). 'A review of the Stern Review on the economics of climate change', *Journal of Economic Literature*, 45(3): 703-724.

Westland, J.C. (2010). 'Critical mass and willingness to pay for social networks', Electronic Commerce Research and Applications, 9(1): 9-19.

Wiestra, E. (1996). 'On the Domain of Contingent Valuation'. Ph.D. Dissertation, Faculteit Bestuurskunde, Universiteit Twente. Twente University Press, Enschede.

Wissenschaftsrat (2013). Report on the Science driven Evaluation of Large Research Infrastructure Projects for the National Roadmap (Pilot Phase).

World Health Organisation (2006), Guidelines for conducting cost-benefit analysis of household energy and health interventions, by Hutton G. and Rehfuess E., WHO Publication.

Wright, M. and Robbie, K. (1996). Venture capitalists, unquoted equity investment appraisal and the role of accounting information. *Accounting and Business Research* 26(2): 153-168.

Zerbe, R.O. and Dively, D. (1994). *Benefit-cost Analysis in Theory and Practice*. The HarperCollins Series In Economics. New York: HarperCollins College Publishers.

Zhuang, L., Liang, Z., Lin, T. and De Guzman, F. (2007). 'Theory and practice in the choice of social discount rate for cost benefit analysis: A survey', ERD Working Paper n. 94, Asian Development Bank.